\documentclass[iop,tighten]{emulateapj}

\def\simgt{\lower 2pt \hbox{$\, \buildrel {\scriptstyle >}\over{\scriptstyle \sim}\,$}}
\def\simlt{\lower 2pt \hbox{$\, \buildrel {\scriptstyle <}\over{\scriptstyle \sim}\,$}}

\begin{document}

\title{X-ray Emission from Optically Selected Radio-Intermediate and
  Radio-Loud Quasars}

\author{B.~P.~Miller,$^{1,2,3}$ ~~W.~N.~Brandt,$^{2,4}$
  ~~D.~P.~Schneider,$^{2}$ ~~R.~R.~Gibson,$^{5}$
  ~~A.~T.~Steffen,$^{6}$ ~~and Jianfeng~Wu$^{2,4}$}

\footnotetext[1]{Department of Astronomy, University of Michigan, 500
  Church Street, Ann Arbor, MI 48109; {\it mbrendan@umich.edu}}

\footnotetext[2]{Department of Astronomy and Astrophysics, The
  Pennsylvania State University, 525 Davey Laboratory, University
  Park, PA 16802}

\footnotetext[3]{Department of Physics, The College of Wooster, 308
  East University Street, Wooster, OH 44691}

\footnotetext[4]{Institute for Gravitation and the Cosmos, The
  Pennsylvania State University, University Park, PA 16802}

\footnotetext[5]{Department of Astronomy, University of Washington,
  Box 351580, Seattle, WA 98195}

\footnotetext[6]{Department of Astronomy, University of Wisconsin
  Marathon County, 518 S. 7th Avenue, Wausau, WI 54401}

\begin{abstract}

We present the results of an investigation into the \hbox{X-ray}
properties of radio-intermediate and radio-loud quasars (RIQs and
RLQs, respectively). We combine large, modern optical (e.g., SDSS) and
radio (e.g., FIRST) surveys with archival \hbox{X-ray} data from {\it
  Chandra\/}, {\it XMM-Newton\/}, and {\it ROSAT\/} to generate an
optically selected sample that includes 188 RIQs and 603 RLQs. This
sample is constructed independently of \hbox{X-ray} properties but has
a high \hbox{X-ray} detection rate (85\%); it provides broad and dense
coverage of the ${\ell}-z$ plane, including at high redshifts (22\% of
objects have $z=2-5$), and it extends to high radio-loudness values
(33\% of objects have $R^{*}=3-5$, using logarithmic units). We
measure the ``excess'' \hbox{X-ray} luminosity of RIQs and RLQs
relative to radio-quiet quasars (RQQs) as a function of radio loudness
and luminosity, and parameterize the \hbox{X-ray} luminosity of RIQs
and RLQs both as a function of optical/UV luminosity and also as a
joint function of optical/UV and radio luminosity. RIQs are only
modestly \hbox{X-ray} bright relative to RQQs; it is only at high
values of radio-loudness ($R^{*}\simgt3.5$) and radio luminosity that
RLQs become strongly \hbox{X-ray} bright. We find no evidence for
evolution in the \hbox{X-ray} properties of RIQs and RLQs with
redshift (implying jet-linked IC/CMB emission does not contribute
substantially to the nuclear \hbox{X-ray} continuum). Finally, we
consider a model in which the nuclear \hbox{X-ray} emission contains
both disk/corona-linked and jet-linked components and demonstrate that
the \hbox{X-ray} jet-linked emission is likely beamed but to a lesser
degree than applies to the radio jet. This model is used to
investigate the increasing dominance of jet-linked \hbox{X-ray}
emission at low inclinations.

\end{abstract}


\section{Introduction}

\subsection{Radio-loud and radio-intermediate quasars}

Quasar emission is believed to result largely from accretion onto a
supermassive black hole (e.g., Lynden-Bell 1969). The bulk of the
optical/UV continuum in radio-quiet quasars (RQQs) is associated with
quasi-thermal emission originating in the accretion disk (e.g.,
Shields 1978), while the \hbox{X-ray} emission is postulated to arise
from Compton upscattering of disk photons occuring in a hot ``corona''
(e.g., Haardt \& Maraschi 1991). This scenario leads naturally to a
correlation between optical/UV and \hbox{X-ray} luminosity. Extensive
studies of RQQs (e.g., Avni \& Tananbaum 1986; Strateva et al.~2005;
Steffen et al.~2006; Just et al.~2007; Kelly et al.~2007; Green et
al.~2009) have found that the optical/UV-to-X-ray spectral slope
steepens (in the sense that objects become relatively less
\hbox{X-ray} luminous) with increasing optical/UV
luminosity. Intriguingly, most studies (see above references) find
that there does not appear to be significant evolution with redshift
in the spectral energy distributions of RQQs; despite the strong
evolution in the space density of quasars, these studies generally
find that RQQs in the early universe appear to have similar
optical/UV-to-X-ray spectral slopes to their local analogs.

Radio-loud quasars (RLQs) are often defined to be the subset of
quasars with a radio-loudness parameter satisfying $R^{*}\ge1$, where
$R^{*}$ is the logarithmic ratio of monochromatic luminosities (with
units of erg~s$^{-1}$~Hz$^{-1}$) measured at (rest-frame) 5~GHz and
2500~\AA~(e.g., Stocke et al.~1992)\footnote{Some authors measure at
  4400~\AA~instead, following Kellerman et al.~(1989); this method
  results in only a minor change ($\sim0.1$) in calculated $R^{*}$
  values. Note that many authors prefer to define $R^{*}$ in linear
  units.}; RQQs must minimally satisfy $R^{*}<1$ and often are found
to have $R^{*}<0$. RLQs comprise $\sim$10\% of quasars, with this
fraction apparently varying with both luminosity and redshift (e.g.,
Jiang et al.~2007 and references therein). The definitive physical
trigger for radio-loudness remains elusive, but RLQs generally have
more massive central black holes than RQQs (e.g., Laor 2000; Metcalf
\& Magliocchetti 2006; also, Shankar et al.~2010 find this to be
redshift-dependent), and it has also been suggested that RLQs host
more rapidly spinning black holes than do RQQs (e.g., Wilson \&
Colbert 1995; Meier 2001; but see also Garofalo et al.~2010). RLQs and
RQQs are typically treated as distinct populations, in part due to the
apparent relative scarcity of objects with $R^{*}\approx1$. The
appropriateness of this canonical separation has been questioned due
to the discovery of numerous quasars of intermediate radio-loudness
(e.g., White et al.~2000; Cirasuolo et al.~2003), which may outnumber
RLQs (e.g., White et al.~2007), but there does appear to be a genuine
bimodality of $R^{*}$ allowing fairly objective distinction between
RQQs and RLQs (e.g., Ivezi{\'c} et al.~2004; Zamfir et al.~2008). It
should be noted that RQQs are not necessarily radio-silent; for
example, Miller et al.~(1993) found the radio emission from
radio-detected RQQs to be dominated by a starburst-linked
component,\footnote{Weak radio emission from RQQs has also been
  suggested to be generated within magnetically heated coronae (Laor
  \& Behar~2008) or slow and dense disk winds (Blundell \&
  Kuncic~2007).} and interpreted radio-intermediate quasars (RIQs) as
being RQQs in which a low-power and mildly relativistic jet is viewed
at low inclinations (see also, e.g., Falcke et al.~1996; Zamfir et
al.~2008).

The observed properties of RLQs and their likely parent population of
radio galaxies are dependent upon orientation to the observer's line
of sight (e.g., Barthel 1989; Urry \& Padovani 1995). As with RQQs
(e.g., Antonucci 1993), there is believed to be an obscuring ``torus''
present in RLQs that blocks the broad-line region from view at large
inclinations, but RLQs are further complicated by significant
non-isotropic jet emission. Radio jets have been measured to have
relativistic bulk velocities on parsec scales from multi-epoch
high-resolution radio imaging of moving knots in the inner jet, and
the frequent lack of a detectable counterjet is consistent with
Doppler beaming (e.g., Worrall \& Birkinshaw 2006 and references
therein). RLQ jets terminate in hotspots within lobes, for which the
velocities are typically non-relativistic (e.g., Scheuer 1995) and so
this emission is relatively isotropic. Both the ratio of radio
core-to-lobe flux and the ratio of core radio-to-optical luminosities
are observed to depend upon orientation, and these results suggest
that both the lobe emission\footnote{The scatter within the
  correlation of core-to-lobe flux ratio to inclination is a factor of
  $\simeq5-10$ for a given inclination (Figure 1a of Wills \&
  Brotherton 1995); this scatter may reflect environmental effects,
  which can be sufficient to induce lobe asymmetries (e.g., Mackay et
  al.~1971; Gopal-Krishna \& Wiita 2000).} and the optical continuum
in RLQs are correlated with intrinsic unbeamed jet power (e.g., Wills
\& Brotherton 1995). The luminosities of narrow emission lines appear
to correlate directly with jet power, with the link plausibly coming
from a mutual underlying dependence upon accretion rate and/or
black-hole spin (e.g., Rawlings \& Saunders 1991; Willott et
al.~1999). Various unification models (e.g., Urry \& Padovani 1995;
Jackson \& Wall 1999) link narrow-line radio galaxies, broad-line
radio galaxies, RLQs, and blazars by decreasing inclination. Our focus
in the present study is restricted to broad-line quasars, but our
results are of potential relevance to radio galaxies and blazars in
the context of such unification schemes.

X-ray studies of RLQs strongly suggest that the nuclear \hbox{X-ray}
emission contains a significant jet-linked component. Zamorani et
al.~(1981) discovered that RLQs are more \hbox{X-ray} luminous than
are RQQs with comparable optical/UV luminosities, by a typical factor
of about three. Worrall et al.~(1987) used {\it Einstein\/} data to
show that the relative \hbox{X-ray} brightness is greater for RLQs
that are more radio-luminous or have flatter radio spectra, and found
no evidence for redshift evolution out to $z\sim3.5$ in the properties
of RLQs. Wilkes \& Elvis (1987) and Shastri et al.~(1993) uncovered
\hbox{X-ray} spectral flattening with increasing radio loudness and
radio core dominance in samples of quasars observed with {\it
  Einstein\/}. Brinkmann et al.~(1997) investigated a large sample of
{\it ROSAT\/}-detected RLQs and found that both lobe-dominated and
core-dominated RLQs show \hbox{X-ray} luminosity correlated with core
radio luminosity, with the \hbox{X-ray} luminosity of core-dominated
RLQs increasing more rapidly with increasing core radio luminosity
(e.g., their Figure 15). It is also noteworthy in the context of
unification schemes that FR~II (see Fanaroff \& Riley~1974 for
description of the FR~I and FR~II classes) radio-galaxy X-ray spectra
typically show both a dominant absorbed and a weaker unabsorbed
component, apparently linked with the disk/corona and jet,
respectively (e.g., Evans et al.~2006; Hardcastle et al.~2009).

\subsection{Aims of this work}

Recent wide-angle, overlapping surveys in the radio (e.g., Faint
Images of the Radio Sky at Twenty-cm, or FIRST; Becker et al.~1995)
and optical (e.g., the Sloan Digital Sky Survey, or SDSS; York et
al.~2000) may be matched (e.g., Ivezi{\'c} et al.~2002) to enable the
selection of large, well-defined samples of RLQs, for which
\hbox{X-ray} properties may be investigated. For example, Suchkov et
al.~(2006) present a catalog of SDSS Data Release Four (DR4) quasars
matched to FIRST sources as well as \hbox{X-ray} sources from pointed
{\it ROSAT\/} PSPC observations. Jester et al.~(2006a) matched a
subset of SDSS DR5 quasars to FIRST sources and \hbox{X-ray} sources
from the {\it ROSAT\/} All Sky Survey, finding radio loudness to be
dependent upon both optical and \hbox{X-ray} luminosity. The improved
capabilities of modern X-ray observatories such as {\it Chandra\/} and
{\it XMM-Newton\/} have substantially advanced understanding of RLQs.
For example, the high angular resolution and low background of {\it
  Chandra\/} enable the routine detection of \hbox{X-ray} emission
from the knots of large-scale RLQ jets (e.g., Worrall 2009 and
references therein), while the broad bandpass and high throughput of
{\it XMM-Newton\/} generate high signal-to-noise \hbox{X-ray} spectra
useful for quantifying differences between RQQs and RLQs (e.g., Page
et al.~2005; Young et al.~2009) or radio galaxies and RLQs (e.g.,
Belsole et al.~2006). Samples of SDSS quasars with \hbox{X-ray}
coverage by {\it Chandra\/} or {\it XMM-Newton\/} that include
subsamples matched to FIRST sources are presented and discussed by
Green et al.~(2009) and Young et al.~(2009).

Guided by the results described in $\S$1.1 (and taking advantage of
our large sample size, which permits finer categorization), we
consider three categories of quasars in this work: RQQs, RIQs, and
RLQs (rather than just RQQs and RLQs), where we define RIQs to consist
of objects with $1{\leq}R^{*}$$<2$; consequently, the objects we
classify as RLQs satisfy $R^{*}\ge2$. The goal of this study is to
quantify the optical-to-X-ray properties of RIQs and RLQs and to
investigate the physical origin of their \hbox{X-ray} emission. Such
an effort requires (1) consistent selection criteria that are unbiased
with respect to the \hbox{X-ray} properties we wish to investigate;
(2) a large sample of quasars spanning a broad range of radio
properties and possessing sensitive \hbox{X-ray} coverage; (3) radio
imaging capable of resolving extended sources (multifrequency radio
coverage to calculate or constrain spectral indices is also useful);
(4) a high fraction of \hbox{X-ray} detections along with proper
statistical consideration of \hbox{X-ray} limits; and (5) effective,
broad coverage of the luminosity-redshift plane to include the full
population of RIQs and RLQs, and to avoid degeneracies in regression
analysis and other biases. We generate a large sample of RIQs and RLQs
with archival \hbox{X-ray} coverage by matching the SDSS DR5 quasar
catalog of Schneider et al.~(2007) and the photometrically selected
quasars from Richards et al.~(2009) to FIRST and to high-quality
observations from {\it Chandra\/}, {\it XMM-Newton\/}, or {\it
  ROSAT\/}. We supplement this primary sample with additional RLQs
observed by {\it Einstein\/}, high-redshift RLQs, and low-luminosity
RLQs detected in deep multiwavelength surveys. The full sample enables
accurate parameterization of \hbox{X-ray} luminosity correlations
across a wide range of radio properties, notably including the
previously sparsely probed but well-populated RIQ regime. We are also
able to take advantage of recent advances in statistical methods
(e.g., Kelly 2007) in our analysis and of newly established
constraints on jet properties (e.g., Mullin \& Hardcastle 2009) in our
modeling. In addition, our use of modern accurate cosmological
parameters eliminates a source of systematic error present in some
earlier analyses of luminosity correlations.

The outline of this paper is as follows: in $\S$2 we describe the
selection methods used to generate our sample, in $\S$3 we discuss
characteristics of the RIQs and RLQs studied here, in $\S$4 we compare
the \hbox{X-ray} properties of RIQs and RLQs to those of RQQs, in
$\S$5 we parameterize \hbox{X-ray} luminosity in RIQs and RLQs as a
joint function of optical/UV and radio properties, in $\S$6 we
determine a plausible physical model for the spectral energy
distributions of RIQs and RLQs, and in $\S$7 we summarize our
results. We adopt a standard cosmology with $H_{0}$ = 70 km
s$^{-1}$~Mpc$^{-1}$, ${\Omega}_{M}$~=~0.3, and
${\Omega}_{\Lambda}$~=~0.7 (e.g., Spergel et al.~2007)
throughout. Radio, optical/UV, and \hbox{X-ray} monochromatic
luminosities ${\ell}_{\rm r}$, ${\ell}_{\rm uv}$, and ${\ell}_{\rm x}$
are expressed as logarithmic quantities with units of
erg~s$^{-1}$~Hz$^{-1}$ (suppressed hereafter), at rest-frame 5~GHz,
2500~\AA, and 2~keV, respectively. In these units the radio loudness
is $R^{*} = {\ell}_{\rm r}-{\ell}_{\rm uv}$ and the useful quantity
${\alpha}_{\rm ox}$, the optical/UV-to-X-ray spectral slope (e.g.,
Avni \& Tananbaum 1986), is ${\alpha}_{\rm ox} =
0.384\times({\ell}_{\rm x}-{\ell}_{\rm uv})$. Object names are
typically taken from the SDSS DR5 spectroscopic quasar catalog of
Schneider et al.~(2007) or from the SDSS DR6 photometric quasar
catalog of Richards et al.~(2009) and are J2000 throughout.

\section{Sample selection}

Our primary sample consists of 654 optically selected RIQs and RLQs
with SDSS/FIRST observations and high-quality X-ray coverage from {\it
  Chandra\/} (171), {\it XMM-Newton\/} (202), or {\it ROSAT\/}
(281). The primary sample is split nearly evenly between spectroscopic
(312) and high-confidence photometric (342) quasars. Most (562) of the
primary sample objects possess serendipitous off-axis \hbox{X-ray}
coverage, while the remainder (92) were targeted in the observations
used in our sample. The \hbox{X-ray} detection fraction for the
primary sample is 84\%; the detection fraction for those objects with
{\it Chandra\/}/{\it XMM-Newton\/}/{\it ROSAT\/} coverage is
95\%/92\%/70\% (typical {\it ROSAT\/} observations are comparatively
less sensitive and have higher background). Two additional samples of
spectroscopic SDSS quasars are constructed to examine the influence of
SDSS targeting selection flags: a ``QSO/HIZ'' sample that contains 200
SDSS/FIRST RIQs and RLQs targeted as quasars from their SDSS colors
(this is entirely a subsample of the spectroscopic primary sample),
and a ``FIRST'' sample that contains 180 SDSS/FIRST RIQs and RLQs
targeted as quasars based on their radio emission; there is
considerable overlap between these two groups, but neither contain the
quasars targeted by SDSS as ``serendipitous'' that we also include in
the primary spectroscopic sample.

We extend the primary sample of spectroscopic/photometric RIQs and
RLQs with several supplemental samples to increase coverage of the
${\ell}-z$ plane: 93 luminous RLQs with {\it Einstein\/} coverage from
Worrall et al.~(1987), 13 high-redshift RLQs studied by Bassett et
al.~(2004) and Lopez et al.~(2006), and 31 low-luminosity RLQs
selected from deep multiwavelength surveys (see $\S$2.2.3 for details
and references) including the Extended {\it Chandra\/} Deep
Field--South (\hbox{E-CDF-S}) and the {\it Chandra\/} Deep
Field--South (\hbox{CDF-S}), the {\it Chandra\/} Deep Field--North
(CDF-N), and the Cosmic Evolution Survey (COSMOS). These supplemental
samples are combined with the primary sample to produce the full
sample. Almost all analyses are performed separately on the primary
sample as well as on the full sample.

The full sample consists of 791 quasars with $R^{*}\ge1$ (with an
X-ray detection fraction of 85\%), of which 188 are RIQs with
$R^{*}<2$ and 603 are RLQs with $R^{*}\ge2$. The sky coverage of the
full sample is shown in Figure 1. Properties for objects in the
primary sample are provided in Table 1, properties for objects in the
deep-fields sample are provided in Table 2, and characteristics of the
various samples are provided in Table 3. In the remainder of this
section, we provide details about the selection of all the samples
used throughout this paper.

\begin{figure}
\includegraphics[scale=0.47]{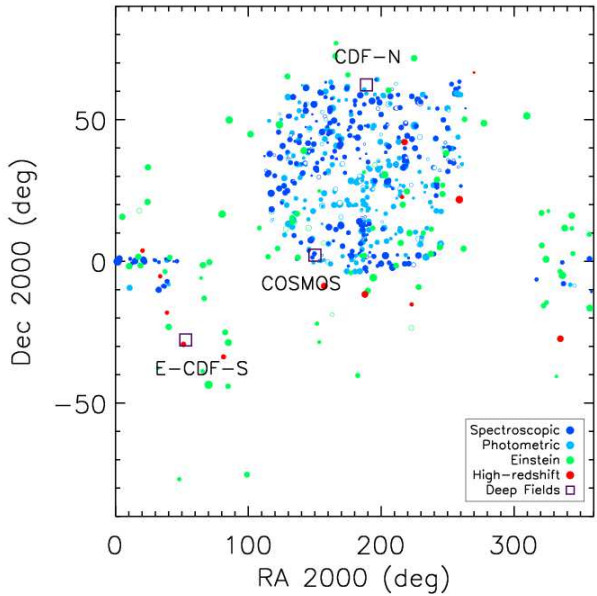} \figcaption{The sky coordinates of
  objects in the full sample of 791 RIQs and RLQs. The filled symbols
  are \hbox{X-ray} detections, and the open symbols are \hbox{X-ray}
  upper limits. The symbol size is linearly proportional to $R^{*}$;
  larger symbols correspond to quasars with greater radio loudness
  values. Primary sample SDSS spectroscopic/photometric quasars
  ($n=654$) are plotted in blue/cyan, and the supplemental samples of
  {\it Einstein\/} ($n=93$) and high-redshift ($n=13$) RLQs are
  plotted in green and red, respectively. The deep-field sample
  ($n=31$) is indicated by the locations of the COSMOS, E-CDF-S, and
  CDF-N surveys (marked with purple squares and labels; the square
  size does not indicate the solid-angle coverage of these surveys).}
\end{figure}

\subsection{Primary sample}

\subsubsection{Spectroscopic sample}

The spectroscopic sample of 312 RIQs and RLQs is drawn from the SDSS
DR5 Quasar Catalog of Schneider et al.~(2007). The sky area covered by
DR5 spectroscopic observations is 5740 deg$^{2}$ near the north
Galactic cap (Adelman-McCarthy et al.~2007). The Schneider et
al.~(2007) quasar catalog includes quasars targeted for matching a
variety of (often overlapping) criteria (see Schneider et al.~2007 and
Richards et al.~2002 for details). Of the 77429 objects in the quasar
catalog, 51577 were targeted based on quasar-like optical colors and
have BEST target flags of ``QSO'' or ``HIZ'' set (see Schneider et
al.~2007 for description of these parameters). Most of the remaining
quasars were targeted as ``serendipitous'' objects based on possessing
non-stellar optical colors. A smaller number of quasars were targeted
based on their radio (FIRST) or X-ray (RASS) emission and/or were
photometrically categorized as stars or galaxies. Quasars targeted as
``QSO'' or ``HIZ'' or ``serendipitous'' were considered for inclusion
in the optically-selected spectroscopic sample; matching to the FIRST
radio survey and then to archival high-quality \hbox{X-ray} coverage
provides an initial list of 345 such RIQs and RLQs. As described in
Appendix A, we remove from this initial list 22 BAL quasars, 8 highly
reddened quasars, and 3 GHz-peaked spectrum sources. The remaining 312
objects constitute our spectroscopic sample of optically-selected RIQs
and RLQs.

The radio properties of these quasars are determined from the 1.4 GHz
FIRST survey, which has a resolution of $\sim$5$''$, a 5$\sigma$
limiting flux density of $\sim$1~mJy, and 9030 deg$^{2}$ of sky
coverage, much of which overlaps the SDSS area (Becker et
al.~1995). Objects were retained as RIQs or RLQs if the summed
luminosity of their constituent components satisfied ${\ell}_{\rm
  r}>31.0$ (motivated by, e.g., Zamfir et al.~2008, who define
$31<{\ell}_{\rm 1.4}<31.6$ as RIQs and ${\ell}_{\rm 1.4}>31.6$ as
RLQs) along with $R^{*}>1.0$ (with ${\ell}_{\rm r}$ and $R^{*}$
defined as in $\S$1.2).  Although the requirements of optical
selection and a joint SDSS/FIRST detection necessarily limit the
completeness of our sample, the depths of the SDSS and FIRST surveys
are well matched for detecting RIQs and RLQs. An $m_{i}$=19.1 quasar
(the limit for $z<3$ candidate ``QSO'' SDSS spectroscopic targeting)
with $R^{*}=1$ and a typical spectral slopes (${\alpha}_{\rm r}=-0.5$)
would have a 1.4 GHz flux density of $\simeq$0.9 mJy, near the FIRST
point-source detection limit.  Details of the optical screening, radio
matching, and selection of \hbox{X-ray} observations are provided in
Appendix A.

In addition to the optically-selected (``QSO/HIZ'' or
``serendipitous'') spectroscopic sample of 312 RIQs and RLQs, we
construct and separately analyze a slightly smaller sample of 200
``QSO/HIZ'' targeted spectroscopic quasars, and we also generate a
sample of 180 quasi-radio-selected ``FIRST'' targeted spectroscopic
quasars. Such objects were targeted by the SDSS on the (not
necessarily exclusive) basis of being likely optical counterparts to
FIRST radio sources. This does not constitute a true radio-selected
sample because the SDSS retains optical magnitude limits for FIRST
sources and because lobe-dominated radio sources without a FIRST core
component will not be targeted by SDSS as FIRST sources, but this
approach provides a useful basis for broad comparison to RIQs and RLQs
targeted on the basis of their optical colors. There is considerable
overlap between the categories of ``QSO/HIZ'' and ``FIRST'' targeted
objects; 163 of the 180 objects (91\%) with the ``FIRST'' flag set
also have either the ``QSO'' or ``HIZ'' flags set. The median
properties of these samples are similar, with median
${\Delta}(g-i)=0.06/0.08$, median ${\ell}_{\rm r}=33.14/33.16$, median
$R^{*}=2.50/2.41$, and median ${\alpha}_{\rm ox}=-1.41/-1.43$ for the
objects selected (as ``QSO/HIZ'')/(as ``FIRST''). These results do not
mandate that a complete radio-selected sample of RIQs and RLQs would
have properties consistent with those of color-selected SDSS RIQs and
RLQs (see, e.g., McGreer et al.~2009), but they do indicate that there
is substantial overlap in these selection methods.

\begin{figure}
\includegraphics[scale=0.47]{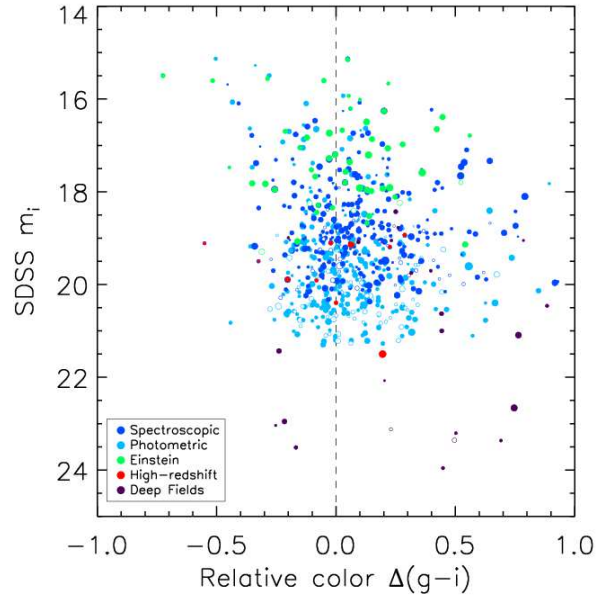} \figcaption{Color-magnitude plot
  of RIQs and RLQs, where the relative color ${\Delta}(g-i)$ is the
  $g-i$ color (corrected for Galactic extinction) for a given object
  less the median quasar color at that redshift. Bluer/redder objects
  have negative/positive values of ${\Delta}(g-i)$. The filled symbols
  are \hbox{X-ray} detections and the open symbols are \hbox{X-ray}
  upper limits. Larger symbols correspond to more radio-loud
  quasars. Primary sample SDSS spectroscopic/photometric quasars are
  plotted in blue/cyan, and the supplemental samples of {\it
    Einstein\/}, high-redshift, and deep-field RLQs are plotted in
  green, red, and purple, respectively. The ${\Delta}(g-i)$ and
  $m_{\rm i}$ values for the deep field RLQs have been calculated by
  tranforming $UBVRI$ magnitudes to $ugriz$ equivalents.}
\end{figure}

\subsubsection{Photometric sample}

The photometric sample of 342 RIQs and RLQs is constructed from a
parent population of over 1,000,000 photometric SDSS sources
identified as potential quasars through the nonparametric Bayesian
classification conducted by Richards et al.~(2009) on unresolved SDSS
DR6 objects. The efficiency of the photometric catalog at excluding
non-quasar contaminants within the list of candidates is high (for
example, it is estimated at 97\% within a subsample of $\sim$500,000
robust UV-excess candidates; it is lower for high-redshift
candidates). We consider only the most likely quasar candidates by
requiring the {\it good\/} flag to be $\ge1$ (this measure is
determined based on several metrics; see Richards et al.~2009 for
details). Our analysis requires reliable redshifts and luminosities,
so we discard those sources with more uncertain photometric redshifts
(probability $<0.5$ of lying within the given range). Our minimum
radio loudness and radio luminosity requirements improve efficiency
still further, as only a small fraction of the non-quasars in the
photometric sample would be expected to display sufficient radio
emission to pass these cuts; we expect that non-quasar contamination
in the matched photometric SDSS/FIRST sample is negligible. By
utilizing photometrically selected quasars, it is possible to expand
the luminosity coverage of the primary sample. Over half of the
Richards et al.~(2009) catalog consists of objects fainter than
$m_{\rm i}=20.1$, which is already a full magnitude fainter than the
SDSS limit for spectroscopic targeting of $z<3$ ``QSO'' candidates
(SDSS spectroscopic targeting of ``HIZ'' candidates is limited to
objects with $m_{\rm i}\le20.2$). The classification scheme is
calibrated with spectroscopically confirmed SDSS quasars, and
consequently the optical properties of these photometrically
identified quasars are expected to be consistent with those selected
from the DR5 Quasar Catalog. This can be verified from Figure 2, which
shows that the relative colors of the photometric RIQs and RLQs are
distributed similarly to those of the spectroscopic sample, but that
the photometric sample extends to fainter magnitudes.

The matching to radio sources and determination of X-ray coverage for
the photometric quasars is identical to the procedure described for
the spectroscopic quasars in Section 2.1.1 and Appendix A. This
process produces a candidate list of 427 photometric RIQs and
RLQs. However, 15 of these objects have SDSS spectroscopic redshifts
obtained on an MJD of less than 53535; these objects were available
for inclusion in the Schneider et al.~(2007) DR5 quasar catalog but
were deliberately not included therein, and are therefore not
appropriate for our sample.\footnote{We verified that these objects
  were properly excluded from the Schneider et al.~(2007) catalog and
  from our study. Of these 15 objects, 12 (082324.75+222303.2,
  100656.46+345445.1, 101858.54+591127.8, 105829.60+013358.8,
  110021.06+401928.0, 112211.80+431649.7 121026.59+392908.6,
  124141.38+344031.0, 131106.47+003510.0, 132833.56+114520.5
  160740.59+254115.7, 162625.85+351341.4) are included as BL Lacs in
  the DR5 catalog of Plotkin et al.~(2008); the other three
  (030055.97$-$002206.5, 123251.42+123110.9, 133925.47$-$002705.5)
  have non-QSO SDSS spectra (classified by the SDSS pipeline as
  ``Unknown'' or ``Galaxy'').}  (Recall that no known DR5
spectroscopic quasars are permitted in our photometric sample, since
spectroscopic data are preferred.) After excluding objects rejected
from the DR5 Schneider et al.~(2007) quasar catalog, and additionally
culling six GHz-peaked spectrum objects, the updated candidate list of
photometric RIQs and RLQs contains 406 objects.

We perform an additional check for those RIQs and RLQs with
photometric redshifts of $z_{\rm phot}\ge1.9$ in order to improve
further sample fidelity. To our knowledge, the only significant set of
systematically erroneous photometric redshifts within the Richards et
al.~(2009) catalog, as established via cross-checking with SDSS
spectroscopic data, consists of a small fraction of low-redshift
($z_{\rm spec}\le1$) quasars assigned photometric redshifts of $z_{\rm
  phot}\sim2$. (The additional and unavoidable effect of increasing
redshift uncertainty at very faint magnitudes is difficult to quantify
in the absence of spectroscopic coverage, and we do not consider it
here.) While such inaccuracies are atypical, it is possible to
identify many of the low-redshift quasars with $z_{\rm phot}\sim2$
through matching to UV observations carried out by the Galaxy
Evolution Explorer (GALEX; Martin et al.~2005). We make use of both
redshift-dependent color-color separation (D.~W.~Hogg 2009, personal
communication) and FUV/NUV (far/near UV;
$\sim$1350--1750/1750-2750~\AA) band SDSS detection rates (Trammell et
al.~2007) to find and discard 34 RIQs and RLQs for which the $z_{\rm
  phot}\sim2$ photometric redshift is likely inaccurate (and retain
another 9 for which GALEX data correctly indicates an inaccurate
photometric redshift but for which we are able to use available
spectroscopic redshifts). We conservatively also discard a further 25
objects with $z_{\rm phot}>1.9$ that lack GALEX coverage.  Appendix B
contains details of the use of GALEX data to assess the accuracy of
objects with $z_{\rm phot}>1.9$. The updated candidate list of
photometric RIQs and RLQs contains 347 ($=406-34-25$) objects, within
which the remaining fraction with this type of redshift
misidentification is only $\sim1.5$\%.

We found 82 (out of 347) photometric RIQs or RLQs with SDSS
spectroscopic redshifts obtained on an MJD of greater than 53535;
these objects were not available for inclusion in the Schneider et
al.~(2007) DR5 quasar catalog, and thus provide a ``blind test'' of
our selection methodology above. After examination of the SDSS
spectra, only two objects\footnote{085448.87+200630.7 and
  133818.26+222156.4 have ``Unknown'' SDSS pipeline spectral type and
  $z_{\rm phot}/z_{\rm spec}=0.415/4.517$ and 2.035/0.392,
  respectively, but the spectroscopic redshift is highly uncertain.}
did not show obvious broad lines. Two non-quasar object from 82
photometric candidate RIQs and RLQs with SDSS spectra corresponds to
2.4\%, suggesting that non-quasar contamination of our optical/radio
matched sample is quite low, at least for brighter objects. Another
three objects\footnote{080447.96+101523.7, 141651.49+185014.1, and
  155259.18+203107.9 appear to be (DR7) BAL quasars; since, as
  discussed in Appendix A, BAL quasars are typically \hbox{X-ray}
  absorbed, these three objects are not appropriate for our sample.}
show BAL features. Three BAL quasars from 82 photometric RIQs and RLQs
with SDSS spectra corresponds to 3.7\%, suggesting that BAL
contamination of our optical/radio matched sample is quite low, at
least for brighter objects. This fraction is slightly lower than the
typical fraction of RIQs and RLQs with BALs (e.g., Shankar et
al.~2008), perhaps because the photometric color-selection is less
efficient for BAL quasars with redder colors (with such a tendency
reinforced by the requirement that ${\Delta}(g-i)<1$). There are then
342 ($=347-2-3$) RIQs and RLQs in the photometric sample.

The photometric redshifts for the remaining 77 photometric RIQs and
RLQs with SDSS spectra were replaced with their spectroscopic
redshifts. Photometric redshifts for an additional 25 photometric RIQs
and RLQs were replaced with spectroscopic redshifts listed in the
NASA/IPAC Extragalactic Database
(NED\footnote{http://nedwww.ipac.caltech.edu/}). The ratio of the
absolute value of the difference between photometric and spectroscopic
redshifts to the spectroscopic redshift was checked to assess redshift
accuracy. The 9 objects already identified as having substantially
inaccurate photometric redshifts by the prior process of matching to
GALEX data are not included in this comparison. There are only
1/7/24 objects for which this ratio exceeds 0.8/0.2/0.1, and these
substantial/modest/tiny redshift errors are relatively random (the
median/mean/standard-deviation of $z_{\rm phot}-z_{\rm
  spec}=-0.02/-0.30/0.59$ for the 24 objects with $|z_{\rm
  phot}-z_{\rm spec}|/z_{\rm spec}>0.1$, and $0.01/-0.08/0.33$ for all
93 objects compared). These 1/7/24 objects correspond to percentages
of 1.1\%/7.5\%/25.8\% out of the 93 redshifts checked; applying these
percentages to the 240 ($=342-93-9$) photometric RIQs and RLQs lacking
spectroscopic coverage suggests substantial/modest/tiny redshift
errors in 0.8\%/5.3\%/18.1\% of the full photometric sample. Not only
are the percentages of errors small and relatively random, but the
impact on the luminosities for objects with modestly incorrect
photometric redshifts is only $\sim$0.2 (expressed in logarithmic
units), less than the intrinsic scatter; this should have no
appreciable impact on our analysis. The luminosities for the
photometric RIQs and RLQs with SDSS spectra or NED $z_{\rm spec}$
values are recalculated using the spectroscopic redshifts. In no case
did this cause the radio luminosity of a previously accepted RIQ or
RLQ to drop below the minimum selection cutoff values of ${\ell}_{\rm
  r}=31$. Median properties of the 342 objects in the final
photometric sample are presented in Table~3.

\subsection{Supplemental samples}

We supplement the primary sample with additional RIQs and RLQs chosen
to increase coverage of the ${\ell}-z$ plane, including 93 objects
previously observed by {\it Einstein\/}, 13 high-redshift objects
(primarily targeted by {\it Chandra\/}), and 31 objects selected from
deep-field surveys. The properties of the deep-field objects are
presented in Table~2, and Table~3 includes the properties of the
supplemental samples. The selection methods for these additional RIQs
and RLQs are by necessity not identical to those employed to obtain
our primary sample, but the optical colors of the supplemental sample
RLQs are similar to those of the primary sample, as can be seen in
Figure~2. The additional ${\ell}-z$ coverage provided by the
supplemental samples (Figure~3) helps ensure consideration of the full
population of RIQs and RLQs and also considerably reduces the
${\ell}-z$ degeneracy when performing statistical analyses below (but
we conduct most fitting on both the full and primary-only samples).

\subsubsection{Einstein sample}

To increase population of the high-luminosity region of the ${\ell}-z$
plane, we include the RLQs with {\it Einstein\/} observations analyzed
by Worrall et al.~(1987), as this sample includes many luminous RLQs
that even today do not have high-quality \hbox{X-ray} coverage from
other observatories. Their sample of 114 RLQs includes objects from
both the North and South celestial hemispheres and has an \hbox{X-ray}
detection rate of 89\%. Their sample was primarily radio-selected at
both high and low frequencies and consequently includes a mix of
compact and extended radio sources, and their RLQs tend to have higher
radio, optical, and \hbox{X-ray} luminosities (and also radio-loudness
values) than the objects in our primary sample. We take the radio,
optical, and \hbox{X-ray} luminosities from Worrall et al.~(1987) but
translate their values to our adopted cosmology. We discard three BAL
quasars, one compact steep spectrum source, and one GHz-peaked
spectrum source from the {\it Einstein\/} sample. The relative colors
of the 56 {\it Einstein\/} non-BAL quasars with SDSS magnitudes (of
which 44 also have SDSS spectra) are plotted on Figure 2 and are
similar to the relative colors of the primary sample; a
Kolmogorov-Smirnov (KS) test gives a probability $p=0.36$ that the two
samples are not inconsistent with arising from the same underlying
distribution. Note that 3C~273 is too bright to fit on this plot, and
also too bright to be targeted for spectroscopy by the SDSS. We
independently find 16 objects from the full Worrall et al.~(1987)
sample in our primary sample (10 spectroscopic, 6 photometric) and
omit these objects from the {\it Einstein\/} sample to avoid
duplication. Many of the other {\it Einstein\/} objects with SDSS
coverage do not appear in our primary sample for various reasons
(e.g., they were targeted by SDSS based on radio rather than optical
properties, or they lack FIRST coverage, or most commonly they lack
recent high-quality \hbox{X-ray} observations). The {\it Einstein\/}
sample is then made up of 93 RLQs. Of these, 11 were undetected by
{\it Einstein\/}, but three of these are detected in shallow {\it
  ROSAT\/} observations, and we use these ${\ell}_{\rm x}$
measurements rather than the {\it Einstein\/} limits.

\subsubsection{High-redshift sample}

To increase population of the high-redshift region of the ${\ell}-z$
plane, we include the 15 high-redshift RIQs and RLQs tabulated by
Bassett et al.~(2004) and the 6 high-redshift RLQs observed by Lopez
et al.~(2006). These high-redshift objects were selected based on
radio flux as well as redshift and were typically (15/21) targeted by
{\it Chandra\/} in ``snapshot'' observations of $\sim$5~ks (6/21 were
observed instead by {\it ROSAT\/}; all these are from Bassett et
al.~2004); all are \hbox{X-ray} detected. The Lopez et al.~(2006)
objects have Southern declinations and are therefore unavailable to
the SDSS/FIRST surveys. Most (11/15) of the Bassett et al.~(2004)
objects have SDSS coverage, and most (7/11) of these have SDSS spectra
and are identified as SDSS quasars. The relative colors of the Bassett
et al.~(2004) RLQs for which we have SDSS magnitudes are plotted in
Figure 2. One object with ${\Delta}(g-i)>1$ is discarded and not
shown.\footnote{Additionally, the redshift for 091316.55+591921.6
  ($z=5.122$) is too high to obtain a reliable relative color and so
  ${\Delta}(g-i)$ has been set to zero for this object.} We
independently find 7 of the 15 objects from Bassett et al.~(2004) in
our primary sample and for consistency we use our measurements in our
analysis of these objects. The high-redshift sample then consists of
13 ($=15-1-7+6$) objects, of which 12 are RLQs and one is an RIQ.

Some of the high-redshift RLQs have particularly large radio-loudness
values, with five having $R^{*}>3.5$. These objects are referred to as
``blazars'' by Bassett et al.~(2004) and Lopez et al.~(2006) and are
likely viewed at lower inclinations than most of our primary sample
objects (though all of the high-redshift objects are broad-line
quasars and not BL Lacs). The relatively large fraction of objects
with extreme radio-loudness values within the high-redshift sample
should not necessarily be taken to be representative of high-redshift
RLQs.

\subsubsection{Deep-fields sample}

To increase population of the low-luminosity region of the ${\ell}-z$
plane, we include RIQs and RLQs identified from deep-field surveys;
properties of these objects are given in Table~2. We select 16 objects
by optical color (of which 14 have X-ray detections) and include a
further 15 \hbox{X-ray} detected objects known to possess broad-line
optical spectra.

We apply a color selection technique which approximates that of the
primary sample when searching for lower luminosity RIQs and RLQs in
deep surveys. Our general procedure is to utilize the Vanden Berk et
al.~(2000) color-color selection method to identify potential quasars
from large catalogs of objects with $UBVRI$ photometry. This set of
color cuts primarily selects for $z<2$ UV-excess objects, but also
includes additional color cuts designed to identify potential quasars
at higher redshift ($z=2-4$). Optical catalogs for the COSMOS region
are described in Ilbert et al.~(2008, 2009); for the E-CDF-S they
include \hbox{COMBO-17} (Wolf et al.~2004) and MUSYC (Gawiser et
al.~2006); for the CDF-N they include the Hawaii survey (Capak et
al.~2007). We converted $UBVRI$\footnote{For COSMOS only, we use
  available $u^{*}B_{\rm j}V_{\rm j}r^{+}i^{+}$ magnitudes for color
  selection, then convert the more precise $u^{*}g^{+}r^{+}i^{+}z^{+}$
  to SDSS $ugriz$ by subtracting median differences. For the somewhat
  brighter broad-line selected COSMOS objects (described below) we use
  SDSS magnitudes directly.} to SDSS $ugriz$ magnitudes following the
transformations given by Jester et al.~(2005; see their Table 1) as
calculated for $z\le2.1$ quasars and use the $ugriz$ magnitudes to
calculate ${\Delta}(g-i)$ (discarding any heavily reddened objects
with ${\Delta}(g-i)\simgt1$) and ${\ell}_{\rm uv}$. Accurate
photometry is important for calculating colors, luminosities, and
photometric redshifts, and so we additionally require $m_{\rm i}<24$
within the {\it Chandra\/} Deep Fields; since COSMOS has shallower
\hbox{X-ray} coverage, we require $m_{\rm i}<22.5$ for this survey to
maintain a reasonable \hbox{X-ray} detection fraction. These magnitude
limits are factors of $\sim$90 and $\sim$20 deeper than the $m_{\rm
  i}<19.1$ limit for SDSS targeting of $z<3$ ``QSO'' objects. These
optical selection criteria do not discriminate with respect to
\hbox{X-ray} properties. The majority of the selected deep-field RIQs
and RLQs have spectroscopic redshifts (see Table~2 for references),
and the remainder have accurate photometric redshifts (references in
Table~2) that have been derived including UV or IR data where
available.

The resulting optically-selected quasar candidates are then matched to
radio catalogs, and non-radio-loud objects are removed from further
consideration. This step significantly improves the efficiency of the
candidate list at excluding non-quasar contaminants. The COSMOS,
E-CDF-S, and CDF-N surveys have highly sensitive radio coverage, with
detection limits better than $\sim$50~$\mu$Jy at 1.4~GHz. The VLA
1.4~GHz radio catalogs used for the COSMOS, E-CDF-S and CDF-S, and
CDF-N fields are presented in Schinnerer et al.~(2007), Miller et
al.~(2008) [which includes many objects also given in Kellerman et
  al.~(2008)], and Biggs \& Ivison (2006), respectively. Luminous
starburst galaxies make up an increasing fraction of the radio-source
population at low radio fluxes and luminosities (e.g., Windhorst et
al.~1985; Barger et al.~2007), and so we also impose radio-selection
criteria designed to screen out starbursts. We require ${\ell}_{\rm
  r}>31.0$ as for the primary sample and further impose a more
stringent requirement of $R^{*}>1.3$ (e.g., see Figure 8 of Barger et
al.~2007) upon deep-field candidates, thereby decreasing potential
starburst contamination of the sample with the tradeoff of omitting
some genuine radio-intermediate deep-field quasars.

We next match to \hbox{X-ray} catalogs, with any \hbox{X-ray} limits
estimated from sensitivity maps. We make use of \hbox{X-ray}
point-source catalogs based on {\it Chandra\/} observations of the
\hbox{E-CDF-S}, \hbox{CDF-S}, \hbox{CDF-N}, and COSMOS as presented in
Lehmer et al.~(2005), Luo et al.~(2008), Alexander et al.~(2003), and
Elvis et al.~(2009), respectively. Maximum {\it Chandra\/} effective
exposure times are $\sim$250~ks for the \hbox{E-CDF-S}, $\sim$2~Ms for
the \hbox{CDF-S} and the \hbox{CDF-N}, and $\sim$160~ks for COSMOS. We
use a matching radius around the optical position of $2.5''$, which is
large enough to account for joint uncertainties in position but
sufficiently small that no spurious matches are likely (see above
references).

As we are interested in characterizing the fundamental \hbox{X-ray}
emission properties of RIQs and RLQs, it is helpful to identify and
remove objects with heavy intrinsic \hbox{X-ray} absorption. This is
important for the low-luminosity deep-field sample, since the fraction
of obscured AGNs is large at low \hbox{X-ray} luminosities and
decreases to high \hbox{X-ray} luminosities (from $\approx80$\% for a
2--10~keV luminosity of $10^{42}$~erg~s$^{-1}$ to $\approx20$\% at
$10^{45}$~erg~s$^{-1}$; e.g., Hasinger 2008, and see also discussion
in Brandt \& Alexander 2010). Removing objects with strong optical
reddening (as we do for both the primary and the deep-field samples)
can also remove many objects with \hbox{X-ray} absorption, but for the
deep-field sample we take the additional step of considering
\hbox{X-ray} spectral shape (but not \hbox{X-ray} luminosity) as a
selection criterion, as measured using the \hbox{X-ray} hardness ratio
[defined as $HR = (H-S)/(H+S)$, where $H$ and $S$ are the
  \hbox{2--8~keV} and \hbox{0.5--2~keV} counts, respectively]. We
screen out sources that are likely absorbed by requiring that the
hardness ratio satisfy $HR<0$; this would correspond to a power-law
slope of $\Gamma\simeq1$ for no intrinsic absorption, $\simeq2\sigma$
from the $\Gamma\simeq1.55$ typical of RLQs (e.g., Page et
al.~2005). After application of the hardness-ratio cut, this
color-selection method yields 16 deep-field RIQs and RLQs, of which 14
have X-ray detections; 9 are from the E-CDF-S, 4 from the CDF-N, and 3
from COSMOS. In Appendix C, we briefly comment on interesting aspects
of some of these RIQs and RLQs. Most (12/16) of the deep-field quasars
selected in this manner are RLQs with $R^{*}>2$.

In addition to the color selection, we also employ optical spectra for
selection, accepting RIQs or RLQs which are known to have broad
emission-line spectra. Optical spectroscopy specifically targeting
X-ray sources is available in the CDF-S (based on 1~Ms sources;
Giacconi et al.~2002), the CDF-N (based on 2~Ms sources; Alexander et
al.~2003), and COSMOS (based on {\it XMM-Newton\/} detections;
Cappelluti et al.~2009), and is presented in Szokoly et al.~(2004),
Barger et al.~(2003) as well as Trouille et al.~(2008), and Trump et
al.~(2009), respectively. We include 15 RIQs and RLQs selected based
on broad-line emission. Of these, one is from the CDF-N and two are
from the CDF-S; these three objects have radio luminosities of
${\ell}_{\rm r}<31$ and so were not available for consideration as
color-selected objects, but they do satisfy the Vanden Berk et
al.~(2000) color-color selection method for quasar candidates, as well
as the hardness ratio cut, and so would otherwise have been expected
to be selected through this process. The remaining 12 RIQs and RLQs
are from the {\it XMM-Newton\/} COSMOS survey, which is shallower than
the {\it Chandra\/} COSMOS survey but covers a wider area; many of
these broad-line objects do not fall within the {\it Chandra\/}
coverage and/or have radio luminosities of ${\ell}_{\rm r}<31$, and so
were not considered for color selection. Again, all of these
broad-line objects would have been identified as quasar candidates
based on their optical colors, and all further satisfy the hardness
ratio cut.

In principle this manner of selecting broad-line objects known to be
X-ray sources might introduce a bias toward \hbox{X-ray} bright
sources, were similarly optically-bright broad-line objects with
\hbox{X-ray} limits not also considered. However, we are unaware of
any broad-line objects present in the extensive optical surveys of the
\hbox{E-CDF-S} or \hbox{CDF-N} which lack \hbox{X-ray} detections, and
in any event the RIQs and RLQs selected based on optical spectroscopy
could generally have had lower \hbox{X-ray} counts by factors of
$\sim$10 and still been detected, suggesting that their relative
\hbox{X-ray} brightness is not atypical for their optical/UV
luminosities. This is supported by the observation that the
${\alpha}_{\rm ox}$ values of the broad-line selected objects are
similar to those of the color-selected objects.

The complete deep-field sample then consists of 31 RIQs and RLQs, of
which 29 are \hbox{X-ray} detected. Of these, 16 were selected by
optical color (of which 14 have \hbox{X-ray} detections) and 15 were
selected based on possessing broad-line optical spectra.

\section{Sample characteristics}

Characteristics of the various samples are presented in Tables~3 and 4
and illustrated in Figures 1--6. The sky coverage of the full sample
is indicated in Figure~1, a color-magnitude diagram is provided as
Figure~2, the redshift distribution is plotted versus optical/UV
luminosities in Figure~3 and versus \hbox{X-ray} luminosities in
Figure~4, the radio characteristics of the primary sample are given in
Figure~5, and the \hbox{optical/UV-to-X-ray} spectral slope as a
function of optical/UV luminosity is shown in Figure~6.

\begin{figure*}
\includegraphics[scale=0.90]{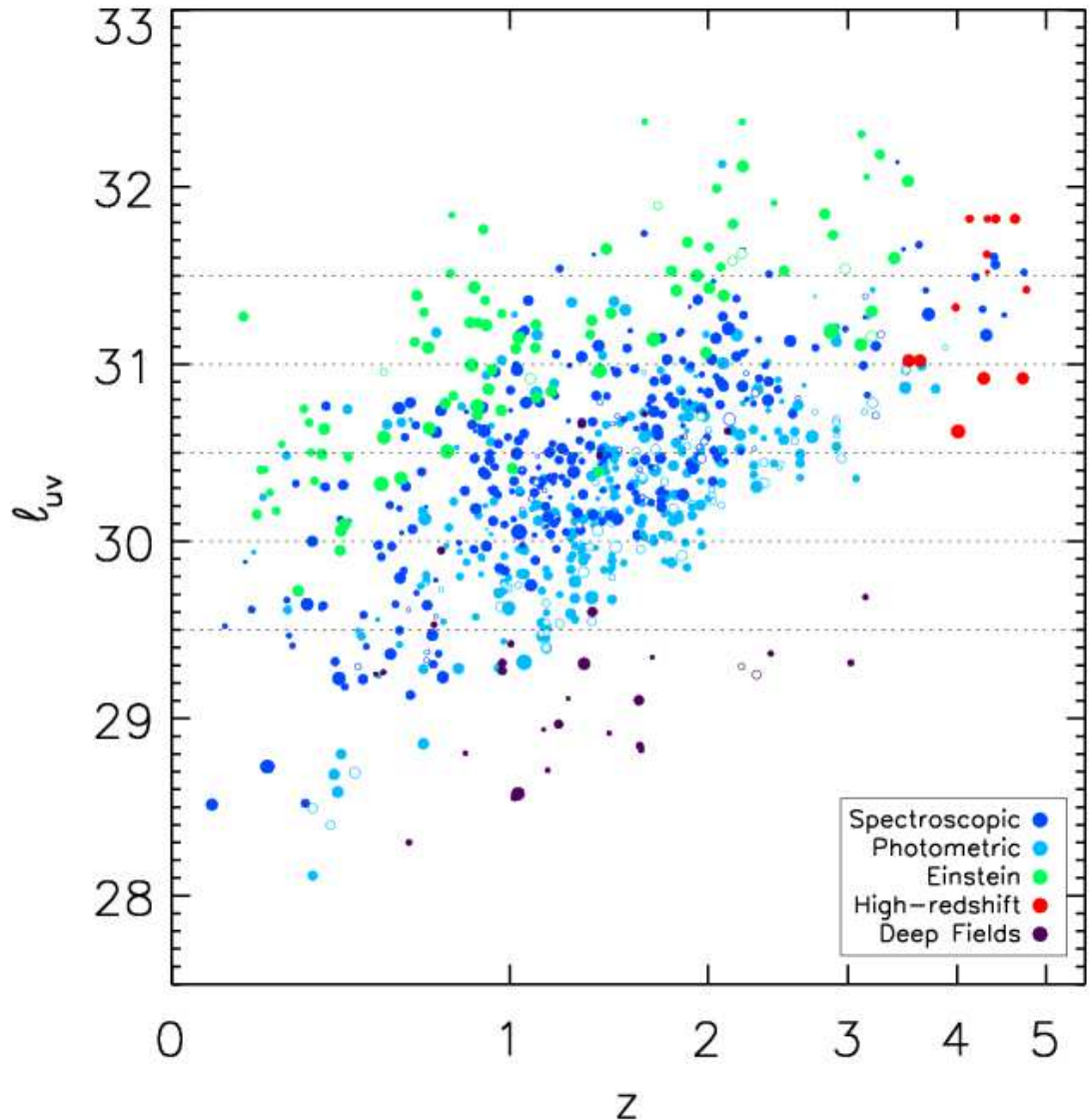}
\figcaption{Sample coverage of the ${\ell}_{\rm uv}-z$ plane, with
  optical/UV luminosity as measured at rest-frame 2500~\AA. The filled
  symbols are X-ray detections, and the open symbols are X-ray upper
  limits. Larger symbols correspond to more radio-loud
  quasars. Primary sample SDSS spectroscopic/photometric quasars are
  plotted in blue/cyan, and supplemental sample {\it
    Einstein\/}/high-redshift/deep fields RLQs are plotted in
  green/red/purple. The dotted lines define luminosity bins within
  which various properties are summarized in Table 4.}
\end{figure*}

\subsection{Optical/UV luminosities}

Optical/UV monochromatic luminosities for all primary sample objects
are calculated at rest-frame 2500~\AA~from SDSS photometric $ugriz$
PSF magnitudes (corrected for Galactic extinction) via comparison to a
redshifted unobscured composite quasar spectrum taken from Vanden Berk
et al.~(2001). This method accounts (in a statistical sense) for the
flux from emission lines as well as typical breaks in the continuum
slope. The error in this method due to differences in emission-line
strength or spectral shape in a particular object is expected to be
less than the inherent uncertainty ($\sim$30\%; e.g., Vanden Berk et
al.~2004; Kaspi et al.~2007) due to typical quasar optical
variability. Only the AGN power-law component is included in the
luminosity; a typical iron-emission ``bump'' near 2500~\AA~(e.g.,
Wills et al.~1985; Vanden Berk et al.~2001) is subtracted, as is the
contribution from a typical RLQ host galaxy (${\ell}_{\rm uv}=28.2$
based on an old stellar population as in, e.g., Coleman et
al.~1980). We verify the accuracy of this approach through comparison
to $\sim$100 RQQs for which Strateva et al.~(2005) calculated
luminosities by fitting SDSS spectra (after dereddening and correcting
for fiber inefficiencies, and also subtracting host-galaxy emission)
and find close agreement (mean difference of ${\ell}_{\rm
  uv}-{\ell}_{\rm uv,s}=-0.03$ with a standard deviation of 0.12).  By
using photometric rather than spectroscopic data to compute optical/UV
luminosities, we can treat the SDSS spectroscopic and photometric
samples in a consistent fashion.

Optical/UV luminosities for the {\it Einstein\/} RLQs are taken from
Worrall et al.~(1987), corrected to our adopted cosmology. Optical/UV
luminosities for the high-redshift RLQs are taken from Bassett et
al.~(2004) and Lopez et al.~(2006). Optical/UV luminosities for the
deep field objects are calculated from SDSS $ugriz$ equivalents,
determined as described in $\S$2.2.3. The full sample (Figure 3)
achieves dense coverage of the ${\ell}-z$ plane, a wide span (almost
five decades) in luminosity coverage ($\sim$2.5 decades at a given
redshift even without considering deep-field RIQs and RLQs, and
$\sim$3.5 including deep-field objects), and coverage to $z\approx5$.

\subsection{Radio luminosities and radio loudness}

Radio monochromatic luminosities are calculated at rest-frame 5~GHz
through extrapolation of the observed 1.4~GHz flux densities. It is
desirable to treat the entire sample in a uniform fashion, but many
sources lack multi-frequency radio measurements, or are
multi-component sources with multi-frequency radio flux densities
obtained at an angular resolution insufficient to distinguish
${\alpha}_{\rm r}$ for the core from any extended emission. Therefore,
we do not use individual ${\alpha}_{\rm r}$ values calculated for a
particular source (see $\S$3.3) to determine the radio luminosity of
that source, but instead assume a radio spectral index\footnote{The
  radio spectral index is defined as ${\alpha}_{\rm r} = \log{(S_{\rm
      h}/S_{\rm l})}/\log{({\nu}_{\rm h}/{\nu}_{\rm l})}$, with
  ${\nu}_{\rm h}, {\nu}_{\rm l}$ and $S_{\rm h}, S_{\rm l}$ the high,
  low frequencies (e.g., 5 and 1.4 GHz) and corresponding flux
  densities.} of ${\alpha}_{\rm r} = -0.9$ for lobe emission and
${\alpha}_{\rm r} = -0.3$ for core emission. These values are
approximately the mean ${\alpha}_{\rm r}$ for lobe-dominated and
core-dominated sources observed within the primary sample,
respectively (using ${\alpha}_{\rm r}$ calculated from low-frequency
radio measurements for the lobe-dominated sources). In any event,
alternative choices of ${\alpha}_{\rm r}$ produce only small changes
in ${\ell}_{\rm r}$ since it is only necessary to extrapolate over a
small frequency range. The total ${\ell}_{\rm r}$ is the sum of the
core and lobe components, and we also provide ${\ell}_{\rm r,core}$ in
Table 1. The radio monochromatic luminosities within the full sample
span over four decades, with a median ${\ell}_{\rm r} = 32.95$. The
median radio monochromatic luminosity within the primary sample is
slightly lower, with ${\ell}_{\rm r} = 32.83$. This difference chiefly
reflects the high radio luminosities of the supplemental {\it
  Einstein\/} RLQ sample, which may be due to some targets being radio
selected. The radio luminosities within the deep-field sample are
lower, with a median ${\ell}_{\rm r} = 31.50$ (recall that we permit
${\ell}_{\rm r} < 31$ for broad-line selected deep-field objects).

Radio loudness in our sample is correlated with radio luminosity
(Figure 5), signifying the influence of beamed jet emission that comes
to dominate the radio emission measured by FIRST for objects at low
inclinations (or with intrinsically powerful jets) but apparently does
not similarly dominate the optical/UV luminosity (as is also suggested
by the SDSS broad emission-line equivalent widths, which do not
suggest significant dilution by a featureless jet-linked continuum for
these sources). The median radio loudness for the full sample is
\hbox{$R^{*}=2.59$}; it is notably higher for the {\it Einstein\/}
sample (median $R^{*}=3.44$) and lower for the deep-field objects
(median $R^{*}=2.06$, with the lowest permitted value being
$R^{*}=1.3$).

\begin{figure}
\includegraphics[scale=0.47]{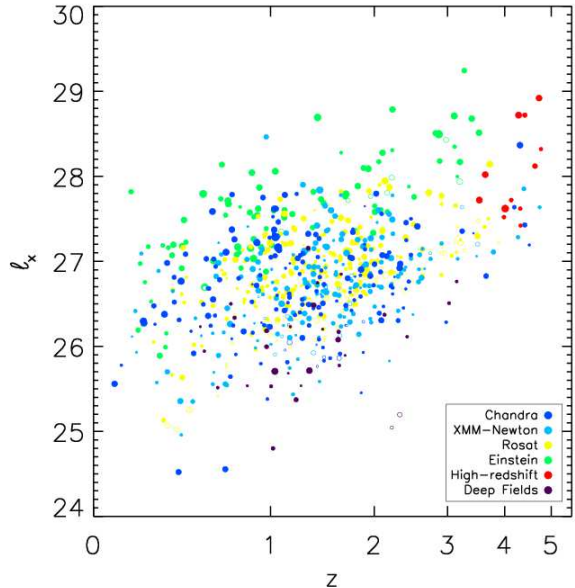}
\figcaption{Sample coverage of the ${\ell}_{\rm x}-z$ plane, with
  X-ray luminosity as measured at rest-frame 2~keV. The filled
  symbols are X-ray detections, and the open symbols are X-ray upper
  limits. Larger symbols correspond to more radio-loud
  quasars.}
\end{figure}

\subsection{Radio spectral shapes and morphologies}

Although we do not use individually measured radio spectral slopes to
calculate luminosity, we do distinguish between flat-spectrum and
steep-spectrum radio sources when radio spectral information is
available. Following Worrall et al.~(1987), objects with
${\alpha}_{\rm r} < -0.5$ are classified as steep spectrum, while
objects with ${\alpha}_{\rm r} \ge -0.5$ are classified as flat
spectrum. Flat-spectrum RLQs are \hbox{X-ray} brighter than their
steep-spectrum counterparts (e.g., Worrall et al.~1987), so we also
conduct analyses of \hbox{X-ray} luminosity correlations on separate
subsamples of flat-spectrum and steep-spectrum RLQs. The {\it
  Einstein\/} RLQs are identified as either flat-spectrum or
steep-spectrum in Worrall et al.~(1987), and we use their
classification; we do not identify sources from the other supplemental
samples as either flat-spectrum or steep-spectrum. In light of the
difficulties inherent in comparing fluxes obtained at widely differing
angular resolutions, spectral indices for primary sample objects with
multifrequency radio data are computed in the observed frame from the
source flux densities (summed over all components for cases of
resolved extended radio emission).

\begin{figure}
\includegraphics[scale=0.47]{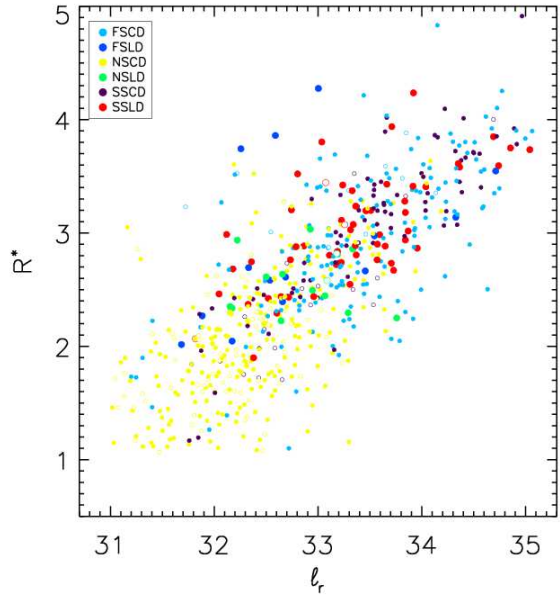} \figcaption{Radio loudness
  ($R^{*}$) versus radio luminosity for the 654 RIQs and RLQs in the
  SDSS primary sample. Symbol color identifies objects as
  flat-spectrum core-dominated (FSCD; cyan), flat-spectrum
  lobe-dominated (FSLD; blue), unmeasured radio spectrum
  core-dominated (NSCD; yellow), unmeasured radio spectrum
  lobe-dominated (NSLD; green), steep-spectrum core-dominated (SSCD;
  purple), and steep-spectrum lobe-dominated (SSLD; red). Symbols are
  slightly larger for lobe-dominated objects and open for \hbox{X-ray}
  limits. Radio loudness and luminosity are correlated within this
  sample, albeit with large scatter. Our sample shows the well-known
  tendencies for the most radio-loud objects to be primarily
  core-dominated and flat-spectrum, and for lobe-dominated objects to
  be mostly steep-spectrum; see also discussion in $\S$3.2 and
  $\S$3.3.}
\end{figure}

To find ${\alpha}_{\rm r}$ for objects with multifrequency radio
coverage, counterparts for all primary sample objects were sought in
the Green Bank 6 cm (5~GHz) catalog (Gregory et al.~1996), which has a
flux-density limit of $\sim$18 mJy and covers
$0^{\circ}<{\delta}<75^{\circ}$, in the Texas 82 cm (365 MHz) catalog
(Douglas et al.~1996), which has a flux-density limit of $\sim$250 mJy
and covers $-35^{\circ}<{\delta}<70^{\circ}$, and in the Westerbork 92
cm (325 MHz) Northern Sky Survey (Rengelink et al.~1997), which has a
flux-density limit of $\sim$18 mJy and covers
${\delta}>+30^{\circ}$. All of these surveys have significantly lower
angular resolution than does the FIRST survey, so components resolved
by FIRST may be blended in these catalogs. Any sources within $90''$
of the optical position are presumed to be associated with the quasar,
unless a FIRST background source is known to be present within the
field; for such instances the relative positions and fluxes have been
evaluated on a case-by-case basis and the multi-frequency radio data
discarded if deemed background contaminated.\footnote{Because we use
  the higher-resolution FIRST survey to screen for potential
  background sources, the effective matching radius is less than
  $90''$. The false-match probability is low even without FIRST
  screening: artificially shifting the declination by one degree and
  rematching returns 5.6\% as many matches within 90$''$.  However,
  most of these shifted matches are sufficiently distant from the core
  that they would be subject to screening as potential background
  objects as gauged by FIRST data.}

Green Bank data are prioritized when calculating ${\alpha}_{\rm r}$
since we are most interested in assessing the relationship of X-ray
emission to the radio core (rather than extended) emission. The
uncertainty on ${\alpha}_{\rm r}$ is then often dominated by the error
in the Green Bank flux measurements, which is generally $\sim$10\%. A
quasar with a typical FIRST radio flux and a borderline ${\alpha}_{\rm
  r} = -0.5$ radio spectral slope would have an uncertainty on
${\alpha}_{\rm r}$ of $\pm0.08$. There are 43 objects with
$-0.58\le{\alpha}_{\rm r}\le-0.42$, which is 11\% of the total number
of objects with determined ${\alpha}_{\rm r}$ values, so for most
objects the classification as flat or steep spectrum is
secure. Objects with both Green Bank and low-frequency measurements
were considered more closely. Sources with concave spectra are
presumed to arise from the emerging dominance of a flat-spectrum core
and are retained. However, as described in Appendix~A, the 9 objects
with convex spectra are identified as possible GPS sources, which have
properties not shared by RLQs in general and are thus not included in
the primary sample.

All primary sample objects are also classified as either
core-dominated or lobe-dominated, with core-dominated objects having a
core radio monochromatic luminosity (at 5 GHz) greater than half the
total radio luminosity.\footnote{Similar but slightly differing
  definitions of core-dominated are sometimes used; for example, Wills
  et al.~(1995) define core/lobe-dominated sources to be those for
  which the ratio of core/lobe emission at rest-frame 6~cm is
  greater/less than unity.} As a consequence of performing radio
selection at an observed frequency of 1.4~GHz (rather than at, for
example, a lower frequency such as 159~MHz or 178~MHz, at which the 3C
and 4C surveys, respectively, were carried out), core-dominated,
likely low-inclination sources are over-represented relative to their
presumed parent radio population. (Recall, however, that extremely
beamed objects are mostly excluded from our sample by the optical
selection criteria; for example, objects with featureless optical
spectra are not included in the SDSS DR5 quasar catalog.) The primary
sample lobe-dominated RIQs and RLQs are more likely to have steep
radio spectra, whereas the core-dominated objects are more likely to
have flat radio spectra (Figure~5). The primary sample RLQs with
particularly large radio-loudness values ($R^{*}>3.5$) tend to be
core-dominated and have flat radio spectra. Such well-known trends are
likely in large part a consequence of core small-scale jet emission
gradually overwhelming (steep-spectrum) lobe emission as inclination
decreases (e.g., Orr \& Browne 1982; see also $\S$6).

\subsection{X-ray luminosities}

X-ray counts were measured for all primary sample objects using the
IDL {\it aper\/} task. \hbox{X-ray} images for objects observed with
{\it ROSAT\/} or {\it XMM-Newton\/} were downloaded from
HEASARC\footnote{High Energy Astrophysics Science Archive Research
  Center: http://heasarc.gsfc.nasa.gov/} along with exposure maps and,
for {\it ROSAT\/}, background images. For objects observed with {\it
  Chandra\/}, the CIAO task {\it dmcopy\/} was used to produce
full-band images from the pipeline-processed level 2 event
files. Source counts were extracted from $\sim$90\% encircled-energy
apertures.\footnote{The $\sim$90\% encircled-energy apertures were
  determined for {\it Chandra\/} as in Luo et al.~(2008), for {\it
    XMM-Newton\/} as in
  http://xmm.esa.int/external/xmm\_user\_support/documentation/ uhb/node18.html,
  and for {\it ROSAT\/} as in
  http://www.mpe.mpg.de/ xray/wave/rosat/doc/tech\_reports/rosat\_psf.ps.}
The energy coverage of the utilized images is \hbox{0.4--2.4~keV} for
{\it ROSAT\/}, \hbox{0.5--8~keV} for {\it Chandra\/}, and
\hbox{0.2--12~keV} for {\it XMM-Newton\/}. Background counts were
determined from the provided background image for {\it ROSAT\/}
observations and as the median of 8 nearby non-overlapping apertures
for {\it Chandra\/} and {\it XMM-Newton\/} observations.

Source detection is evaluated by comparison of the observed aperture
counts to the 95\% confidence upper limit for background alone. Where
the number of background counts is less than 10 we use the Bayesian
formalism of Kraft et al.~(1991) to determine the limit; else, we use
Equation 9 from Gehrels (1986). If the aperture counts exceed the 95\%
confidence upper limit we consider the source detected and calculate
the net counts by subtracting the background from the aperture counts
and then dividing by the encircled-energy fraction; else, the source
is undetected and the 95\% confidence upper limit is used. Counts were
converted to count rates using the furnished or generated exposure
maps. Count rates were converted to observed-frame 2 keV flux
densities with
PIMMS,\footnote{http://cxc.harvard.edu/toolkit/pimms.jsp} in all cases
assuming Galactic absorption and a power-law spectrum\footnote{RLQ
  X-ray spectra can generally be fitted with a single power law (e.g.,
  Belsole et al.~2006), although some high-redshift RLQs do show
  evidence of intrinsic absorption (e.g., Cappi et al.~1997; Yuan et
  al.~2006) or spectral curvature (e.g., Tavecchio et al.~2007).} with
$\Gamma=1.5$ (alternate reasonable choices for $\Gamma$ have only a
few percent impact upon the calculated \hbox{X-ray} fluxes), which
were then used to determine bandpass-corrected rest-frame 2~keV
monochromatic luminosities.

Where available, data from the {\it Chandra\/} Source Catalog (Evans
et al.~2008) or the {\it XMM-Newton\/} Serendipitous Source Catalog
(Watson et al.~2009) were used in preference to our own
measurements. The \hbox{X-ray} luminosities were calculated from
catalog broad-band fluxes as given at 0.5--7 keV for {\it Chandra\/}
and 0.5--4.5 keV for {\it XMM-Newton\/}, with matching performed using
radii of 3$''$ and 10$''$, respectively. It is not possible for this
project to rely exclusively on catalogs, however, as large ``blind
search'' source catalogs must utilize more conservative detection
thresholds. In addition, it would be difficult to determine accurate
upper limits based on a catalog non-detection. Our calculated
luminosities are in good agreement with those derived from catalog
data (median/mean/standard deviation of the difference in ${\ell}_{\rm
  x}$ of 0.037/0.048/0.201 and 0.043/0.079/0.186 for {\it Chandra\/}
and {\it XMM-Newton\/}, respectively), indicating that our
measurements (detections or upper limits) for those objects not
included in catalogs should also be accurate.

The \hbox{X-ray} data for the primary sample of 654 RIQs and RLQs are
based for 171/202/281 objects on {\it Chandra\/}/{\it
  XMM-Newton\/}/{\it ROSAT\/} observations. The \hbox{X-ray} detection
rates for {\it Chandra\/}/{\it XMM-Newton\/}/{\it ROSAT\/} are
95\%/92\%/70\%. \hbox{X-ray} data for 115 objects detected with {\it
  Chandra\/} were taken from the {\it Chandra\/} Source Catalog, and
\hbox{X-ray} data for 160 objects detected with {\it XMM-Newton\/}
were taken from the {\it XMM-Newton\/} Serendipitous Source
Catalog. In a few cases these source catalogs provide total
\hbox{X-ray} coverage deeper than our single-observation photometry,
and our sample is improved by making use of these additional data. No
objects are discarded based on non-inclusion in source catalogs. Many
of the sources which are not included in the utilized source catalogs
but are detected by our photometry are associated with observations
not included in the source catalogs, either due to the observation
date falling outside of the range covered by the source catalogs or
some observation parameter (e.g., subarray type) not satisfying the
requirements for inclusion.

\begin{figure}
\includegraphics[scale=0.47]{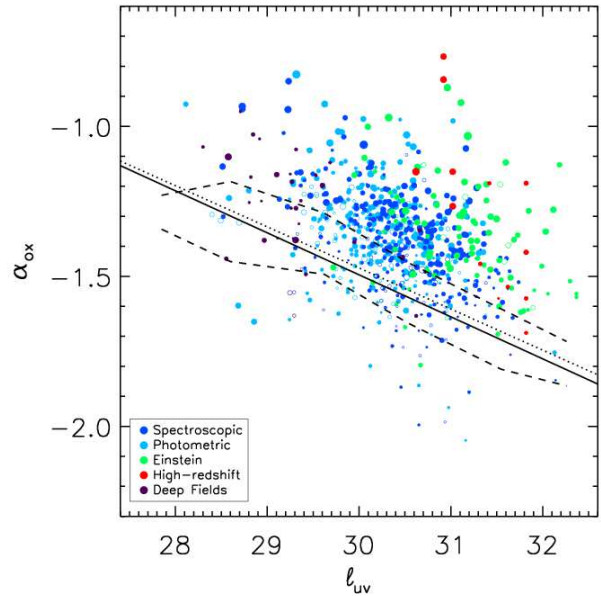} \figcaption{Optical/UV-to-X-ray
  spectral slope ${\alpha}_{\rm ox}$ as a function of ${\ell}_{\rm
    uv}$; symbols are coded as in Figure 3. The solid/dotted line is
  the best-fit linear relation for RQQs from Just et
  al.~(2007)/Steffen et al.~(2006), while the dashed lines show the
  25th and 75th percentiles for RQQs. Our sample shows the well-known
  tendency for RLQs to be X-ray bright relative to RQQs of comparable
  optical/UV luminosities.}
\end{figure}

\begin{figure*}
\includegraphics[scale=0.84]{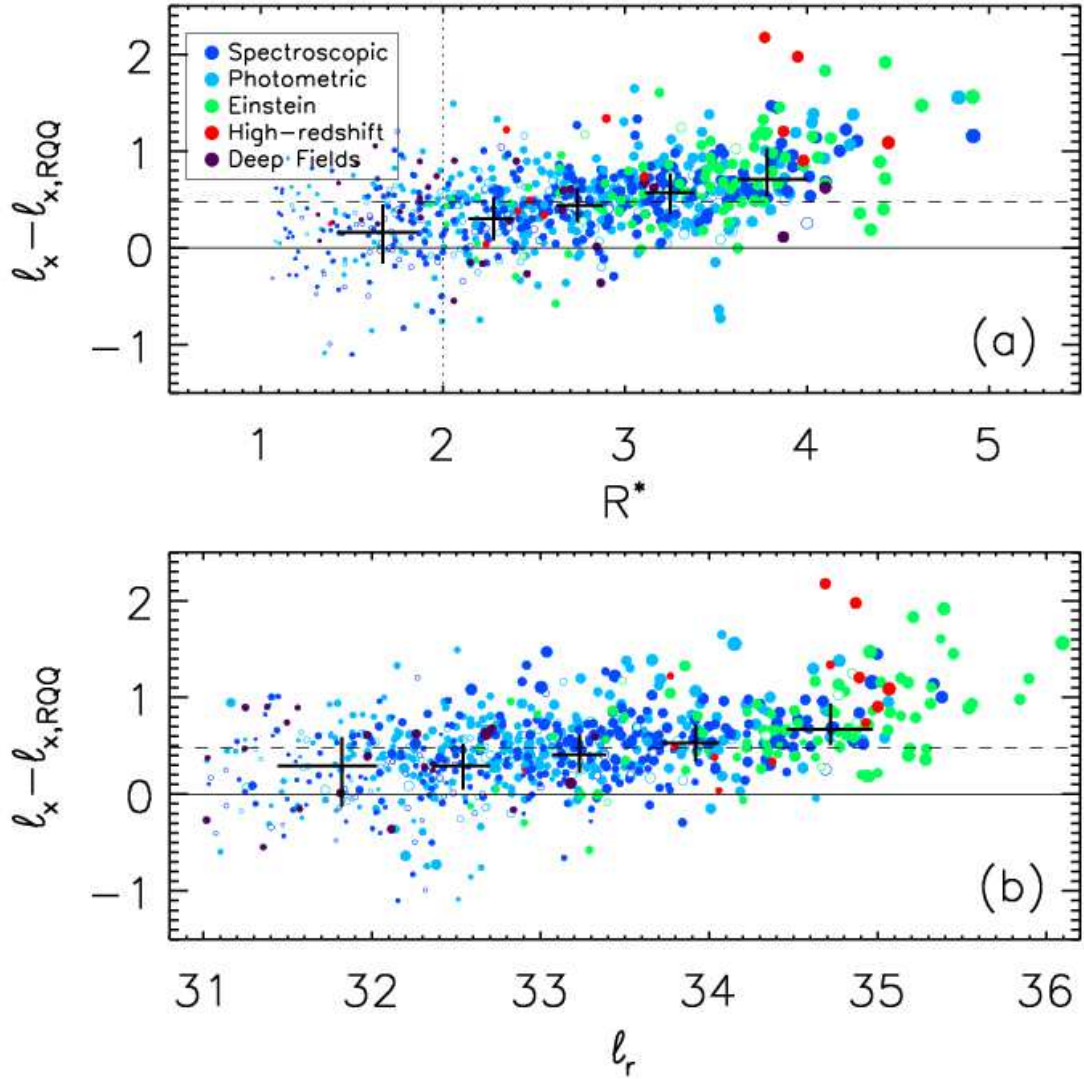} \figcaption{X-ray ``excess'' in
  RIQs and RLQs relative to comparable RQQs plotted as a function of
  radio loudness (top) and radio luminosity (bottom). The solid line
  corresponds to a linear multiplicative ratio of one and the dashed
  line to a ratio of three. The black crosses show median values
  within bins; the arms show the 25th--75th percentile range in each
  bin (values given in Table 5). Mean values are similar to medians
  and the error on the mean is much smaller than the interquartile
  range shown.  Symbols are coded as in Figure~3.}
\end{figure*}

Almost all of the {\it ROSAT\/} coverage is serendipitous; only 6/281
(2.1\%) sources are found within $3'$ of the aimpoint of the best {\it
  ROSAT\/} observation. The SDSS and FIRST surveys served as a source
of targets for many {\it XMM-Newton\/} and {\it Chandra\/}
observations; 34/202 (16.8\%) sources are found within $2'$ of the
aimpoint of the best {\it XMM-Newton\/} observation, and 58/171
(33.9\%) sources are found within $1'$ of the aimpoint of the best
{\it Chandra\/} observation. Some of these targeted objects also have
serendipitous coverage and would be in our sample even absent the
targeted observations: four of the objects targeted by {\it
  XMM-Newton\/} have off-axis {\it Chandra\/} coverage and two of the
objects targeted by {\it Chandra\/} have off-axis {\it ROSAT\/}
coverage. In total, then, 92/654 (14.1\%) of the RIQs and RLQs in our
primary sample were targeted for \hbox{X-ray} observations (of which
91/92 have \hbox{X-ray} detections). The analysis of luminosity
correlations below was carried out including targeted objects to
increase the size of the primary sample, but results are also provided
for an ``Off-axis'' sample of sources with only serendipitous
\hbox{X-ray} coverage and for the ``Targeted'' sample
exclusively. Differences between the properties of the ``Off-axis''
and ``Targeted'' subsamples are discussed in $\S$5.4.2.

There may be cases in which the extraction region used to calculate
the \hbox{X-ray} luminosity includes both nuclear and kpc-scale X-ray
jet emission. Many \hbox{X-ray} jets have been discovered by {\it
  Chandra\/} (e.g., Sambruna et al.~2004; Marshall et al.~2005), and
some would not be resolvable with {\it ROSAT\/} or even {\it
  XMM-Newton\/}. Additionally, in some instances the inner knot(s) of
an \hbox{X-ray} jet might lie inside the {\it Chandra\/} extraction
region. Generally even the inner knots of RLQs with \hbox{X-ray} jets
are observed to be only a few percent as bright as the \hbox{X-ray}
core\footnote{The \hbox{X-ray} core itself may contain an unresolved
  component of \hbox{X-ray} emission linked to the pc-scale radio jet,
  of course, but we desire such a component be included for our
  analysis and in any case it would be impossible to exclude from
  simple photometry.} (e.g., Marshall et al.~2005), which would not
significantly change the calculated ${\ell}_{\rm x}$ values. The XJET
catalog\footnote{http://hea-www.harvard.edu/XJET/} provides a useful
listing of $\sim$100 objects with known extended \hbox{X-ray} jet or
lobe emission. Out of the primary sample of 654 RIQs and RLQs, only 15
are listed in the XJET catalog. We examined \hbox{X-ray} images of all
15 objects and in no case did it appear likely that the extended
\hbox{X-ray} emission could significantly increase the measured
\hbox{X-ray} luminosity.

\begin{figure}
\includegraphics[scale=0.48]{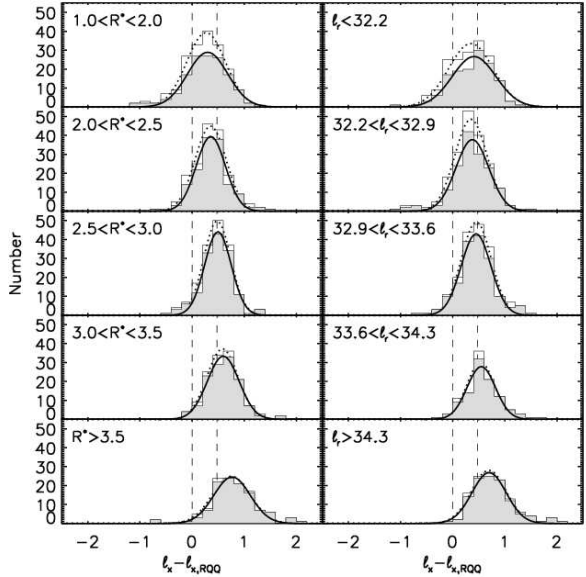} \figcaption{\small Histograms of
  ${\ell}_{\rm x}-{\ell}_{\rm x,RQQ}$ (defined as in $\S$4) plotted in
  (increasing downward) bins of radio-loudness (left) and radio
  luminosity (right); the bins are identical to those in Table~5. The
  filled histograms contain \hbox{X-ray}-detected objects and the open
  histograms include upper limits. The vertical dashed lines
  correspond to linear multiplicative factors of 1 and 3 for the ratio
  of the X-ray luminosity of RIQs and RLQs to comparable RQQs. The
  solid/dotted curves are Gaussians fit to detected/all objects to
  investigate log-normality; there appears to be a tail of objects
  with high X-ray luminosities in the maximum radio-loudness and
  luminosity bins.}
\end{figure}

\section{Comparison of the X-ray emission from RQQs, RIQs, and RLQs}

Several previous studies have noted the tendency for RLQs to be more
\hbox{X-ray} bright than RQQs of comparable optical/UV luminosity
(e.g., Zamorani et al.~1981; see also discussion and references in
$\S$1). We confirm that general result, and also quantify the degree
to which RIQs and RLQs differ in \hbox{X-ray} brightness from RQQs as
a function of radio loudness and luminosity. Our large and carefully
constructed sample permits relatively fine-grained binning for such
measurements.

The necessity of controlling for optical/UV luminosity when comparing
RIQs and RLQs to RQQs is driven by the well-known steepening in RQQs
of optical/UV-to-\hbox{X-ray} spectral slope with increasing
optical/UV luminosity (e.g., ${\alpha}_{\rm ox} =
-0.140\times{\ell}_{\rm uv}+2.705$ as given in Equation 3 of Just et
al.~2007; other studies as listed in $\S$1.1 typically find similar
results). As RLQs are \hbox{X-ray} brighter than RQQs at a given
${\ell}_{\rm uv}$, they have less negative values of ${\alpha}_{\rm
  ox}$. The median optical/UV-to-X-ray spectral slope for the full
sample of RIQs and RLQs is ${\alpha}_{\rm ox} = -1.40$, and for
RIQs/RLQs separately it is ${\alpha}_{\rm ox} = -1.50/-1.37$.  The
sample of RQQs we use for comparison (from Steffen et al.~2006) has a
median ${\alpha}_{\rm ox}=-1.52$.

The ${\alpha}_{\rm ox}$ and ${\ell}_{\rm uv}$ values for the complete
sample of RIQs and RLQs are plotted in Figure~6, along with the RQQ
${\alpha}_{\rm ox}({\ell}_{\rm uv})$ relations from Just et al.~(2007)
and Steffen et al.~(2006), and the 25th and 75th percentiles for
${\alpha}_{\rm ox}$ for RQQs within each ${\ell}_{\rm uv}$ decade
(from Table 5 of Steffen et al.~2006). The tendency of RQQs at lower
luminosities to fall below the best-fit ${\alpha}_{\rm ox}({\ell}_{\rm
  uv})$ relation appears to be genuine (e.g., Steffen et al.~2006;
Maoz et al.~2007); a larger sample is required to determine whether a
qualitatively similar effect may apply to RIQs and RLQs. Like RQQs,
RIQs and RLQs also show an anti-correlation between ${\alpha}_{\rm
  ox}$ and ${\ell}_{\rm uv}$ (a Spearman test gives $\rho=-0.31$, with
a probability less than $p=5\times10^{-5}$ that no correlation is
present), albeit with a systematic offset toward less negative values
of ${\alpha}_{\rm ox}$. Figure~6 also indicates that the degree to
which RIQs and RLQs are brighter in X-rays than comparable RQQs is
dependent upon radio loudness (the larger points, representing more
radio-loud objects, are generally further above the RQQ ${\alpha}_{\rm
  ox}({\ell}_{\rm uv})$ relation); we now quantify this dependence.

\begin{figure*}
\includegraphics[scale=0.85]{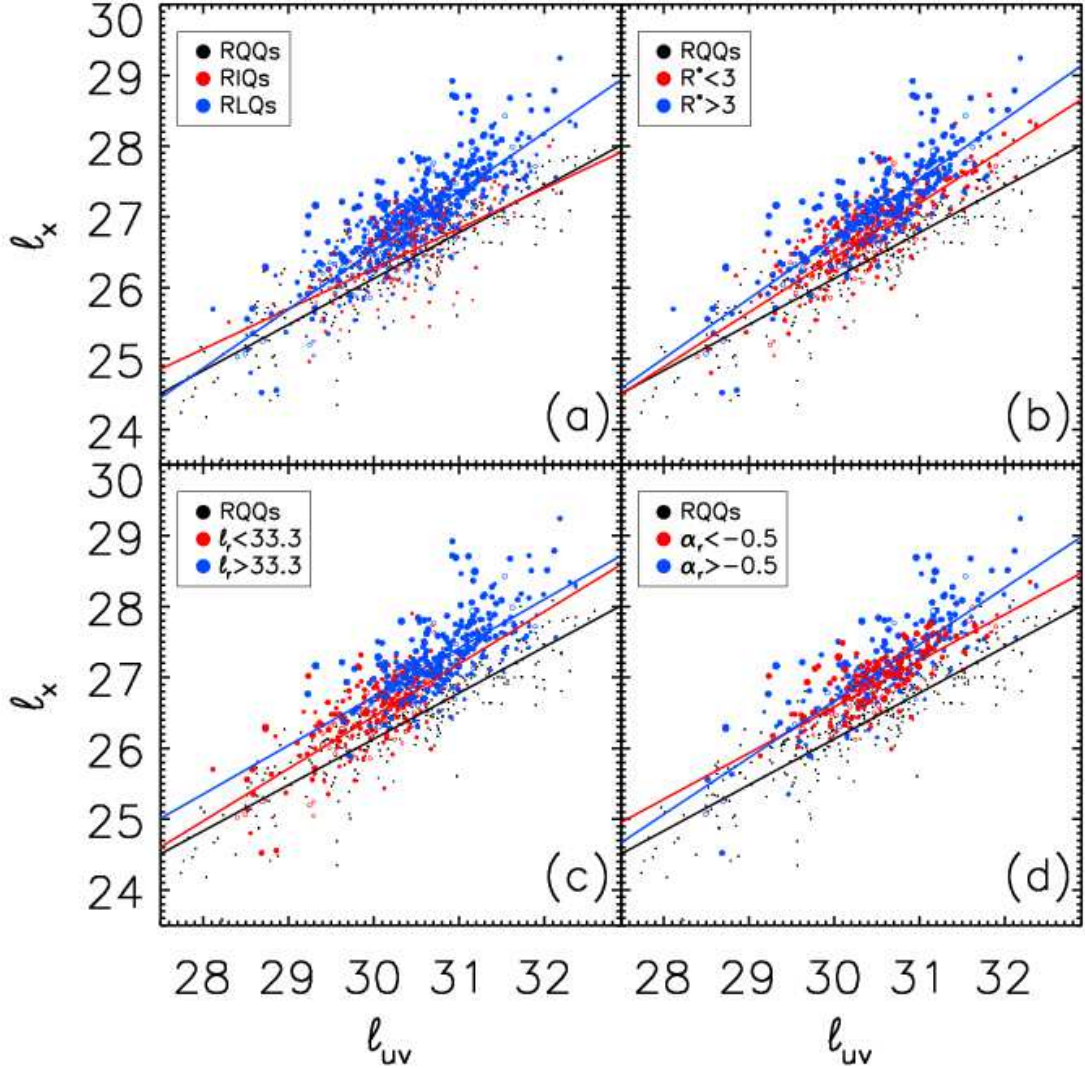} \figcaption{X-ray
  luminosity as a function of optical/UV luminosity for RIQs and RLQs
  (a) and for various sub-samples of RLQs (b, c, d). The model
  considered is ${\ell}_{\rm x} = a_{\rm 0} + b_{\rm
    uv}{\times}{\ell}_{\rm uv}$ (where ${\ell}_{\rm x}$ and
  ${\ell}_{\rm uv}$ have been normalized prior to fitting as described
  in $\S$5). Filled symbols are X-ray detections and larger symbols
  are more radio-loud throughout. Solid lines show best-fit linear
  models. }
\end{figure*}

The ``excess'' X-ray luminosity from RLQs may be defined as
${\ell}_{\rm x}-{\ell}_{\rm x,RQQ}$, where we take the expected X-ray
luminosity for RQQs to be ${\ell}_{\rm x,RQQ} = 0.709\times{\ell}_{\rm
  uv}+4.822$ (Equation~7 from Just et al.~2007) and use ${\ell}_{\rm
  uv}$ as measured for RIQs and RLQs.\footnote{This probes the
  relationship between radio and jet-linked \hbox{X-ray} emission,
  presuming the optical/UV luminosity in these RIQs and RLQs is
  disk-dominated and that the disk/corona in RIQs and RLQs displays a
  similar dependence of \hbox{X-ray} emission upon ${\ell}_{\rm uv}$
  as in RQQs. The first presumption is consistent with the apparently
  undiluted equivalent widths of the broad emission lines in the
  primary sample of RIQs and RLQs; see $\S$6 for further discussion of
  these points.}  Figure~7 shows excess X-ray luminosity for the
complete sample of RIQs and RLQs as a function of radio loudness and
of radio luminosity, along with median values and the 25th--75th
percentile range within bins of $R^{*}$ and ${\ell}_{\rm r}$. The
median values and 25th and 75th percentiles plotted in Figure~7 are
listed in Table~5. The mean values are consistent with the median
values, and the error on the mean is much less than the 25th--75th
percentile range. When expressed in linear units, the multiplicative
factor by which the \hbox{X-ray} luminosity for RIQs and RLQs exceeds
that of RQQs ranges (25th--75th percentiles) from 0.7--2.8 for RIQs
through the canonical $\sim$3 for RLQs (e.g., Zamorani et al.~1981) to
3.4--10.7 for extremely radio-loud ($R^{*}>3.5$) RLQs, with a
qualitatively similar increase in excess \hbox{X-ray} luminosity with
increasing radio luminosity. Figure~8 shows the distribution of
${\ell}_{\rm x}-{\ell}_{\rm x,RQQ}$ (the shaded histogram is detected
objects and the open histogram includes \hbox{X-ray} upper limits)
within the same bins of $R^{*}$ (left) and ${\ell}_{\rm r}$ (right) as
used to construct Table~5. The distribution of excess \hbox{X-ray}
luminosity is reasonably well-characterized as log-normal (see
overplotted Gaussians), with KS test probabilities $p>0.13$ in all
cases. There appears to be a tail to brighter \hbox{X-ray} luminosity
within the highest radio-loudness and luminosity bins (e.g., the
percentage of objects above 1.645$\sigma$ from the mean is not 5\% but
rather 10\%/11\% within the most radio-loud/luminous bin, calculated
using the fitted values for $\sigma$ and $\mu$). Similar results hold
for the primary sample only (Table 6), with the mean values of
${\ell}_{\rm x}-{\ell}_{\rm x,RQQ}$ consistent\footnote{In only 2/10
  cases (the highest radio-loudness and luminosity bins) is the
  difference in means greater than the error; the mean/standard
  deviation of the difference in means is 0.02/0.04.} between the full
and primary samples across $R^{*}$ and ${\ell}_{\rm r}$ bins.

The excess \hbox{X-ray} luminosity may also be fit directly as a
function of radio loudness or luminosity, for example as
${\ell}_{x}-{\ell}_{\rm x,RQQ} = a + b\times{R^{*}}$ or
${\ell}_{x}-{\ell}_{\rm x,RQQ} = a + b\times{\ell}_{\rm r}$, where $a$
and $b$ are fitting constants. We carry out such a fit using the IDL
code of Kelly (2007), which utilizes Bayesian techniques that
incorporate both uncertainties and upper limits. The best-fit models
for the full sample are ${\ell}_{x}-{\ell}_{\rm x,RQQ} =
(-0.354\pm0.050) + (0.289\pm0.018)\times{R^{*}}$ and
${\ell}_{x}-{\ell}_{\rm x,RQQ} = (0.450\pm0.015) +
(0.177\pm0.014)\times({\ell}_{\rm r}-33.3)$. Excess \hbox{X-ray}
luminosity is more strongly dependent upon radio-loudness than radio
luminosity. Flat-spectrum RLQs may have excess \hbox{X-ray} luminosity
more strongly correlated with radio properties than do steep-spectrum
RLQs, with flat/steep spectrum RLQs having coefficients of
\hbox{$(0.352\pm0.039)/(0.274\pm0.040){\times}R^{*}$} and
\hbox{$(0.204\pm0.027)/(0.088\pm0.029)\times({\ell}_{\rm r}-33.3)$}.
The large amount of scatter in these relations prevents productive
consideration of more complex models, but Figures~7 and 8 do suggest
that these linear fits (to log quantities) may not adequately capture
the apparent slow rise in \hbox{X-ray} excess at low radio-loudness or
luminosity and the more rapid increase at higher $R^{*}$ or
${\ell}_{\rm r}$ values.

\begin{figure}
\includegraphics[scale=0.39]{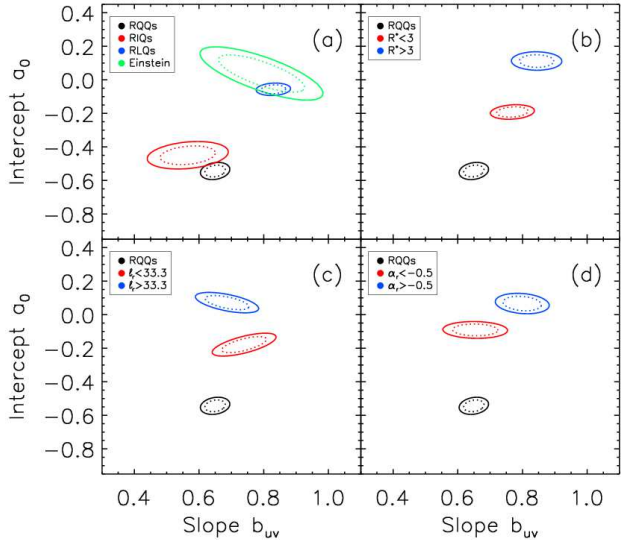} \figcaption{ Joint 90\%
  (solid) and 68\% (dotted) confidence ellipses for fitting X-ray
  luminosity as a (sole) function of optical/UV luminosity for various
  subsamples of RLQs. The model considered is ${\ell}_{\rm x} = a_{\rm
    0} + b_{\rm uv}{\times}{\ell}_{\rm uv}$ (where ${\ell}_{\rm x}$
  and ${\ell}_{\rm uv}$ have been normalized prior to fitting as
  described in $\S$5). RLQs with larger values of radio-loudness or
  luminosity or with flat radio spectra have greater best-fit
  intercepts than those with smaller values of radio-loudness or
  luminosity or with steep radio spectra.}
\end{figure}

\section{Parameterizing the X-ray luminosity of RIQs and RLQs}

We parameterize \hbox{X-ray} luminosity as a sole function of
optical/UV luminosity and as a joint function of optical/UV and radio
luminosity for various groupings of RIQs and RLQs, and also consider
whether an additional dependence upon redshift is required. All
fitting is carried out with the IDL code of Kelly (2007); using the
Astronomy Survival Analysis (ASURV) package (e.g., Isobe \& Feigelson
1990) gives similar results. The potential measurement errors in
optical magnitudes and radio fluxes are generally small (see
references to SDSS and FIRST in $\S$1), and most objects have
sufficient \hbox{X-ray} counts that the uncertainties may be assumed
to be dominated by intrinsic random flux variability (we use
20\%/30\%/40\% or 0.0792/0.114/0.146 in log units for uncertainties on
${\ell}_{\rm r}/{\ell}_{\rm uv}/{\ell}_{\rm x}$, motivated by, e.g.,
$\S3.5$ of Gibson et al.~2008). The luminosities are normalized prior
to fitting as ${\ell}_{\rm r}-33.3, {\ell}_{\rm uv}-30.5, {\ell}_{\rm
  x}-27.0$ (the subtracted values are nearly the medians for the
sample of RLQs). Results are given in Tables 7 and 8 and illustrated
in Figures 9--17. For any given model fit the quoted parameter values
are the median of draws from the posterior distribution and the errors
are credible intervals corresponding to 1$\sigma$. See Appendix D for
additional discussion of fitting methodology.

\subsection{${\ell}_{\rm x}({\ell}_{\rm uv})$}

The first model we consider is \hbox{X-ray} luminosity as a sole
function of optical/UV luminosity, such as is often applied to RQQs
(e.g., Steffen et al.~2006; Just et al.~2007): ${\ell}_{\rm x} =
a_{\rm 0} + b_{\rm uv}\times{\ell}_{\rm uv}$. It is most convenient to
treat ${\ell}_{\rm x}$ as the dependent variable for the purposes of
fitting, given the X-ray limits (and perhaps also most appropriate for
our predominantly optically-selected sample of RIQs and RLQs, whose
X-ray properties were not considered in selection). Since our analysis
is comparative in nature, it suffices to maintain consistency and so
we do not also calculate coefficients treating ${\ell}_{\rm uv}$ as
the dependent variable, nor do we calculate a bisector fit. This
approach also simplifies analysis when additional variables are
considered. We first re-fit the RQQs from Steffen et al.~(2006) using
the same procedure we adopt for analyzing the RIQs and RLQs, to
demonstrate the consistency of this method with previous work. Our
results agree with those of Steffen et al.~(2006) for their Equation
1a (see also Appendix D). We also re-fit the luminous {\it Einstein\/}
RLQs from Worrall et al.~(1987) separately, to assess the influence
this subsample exerts upon the full sample. Finally, we fit the full
sample and then also fit various groupings of RIQs and RLQs. Results
for this model are given on the left side of Table 7 and plotted in
Figure 9.

The general tendencies revealed by Figure~9 are (9a) RIQs have
\hbox{X-ray} luminosities only modestly greater than those of RQQs of
comparable optical/UV luminosities, whereas RLQs become increasingly
\hbox{X-ray} bright relative to comparable RQQs as ${\ell}_{\rm uv}$
increases; (9b) when RLQs are subdivided by radio loudness, RLQs with
$R^{*}>3$ are more \hbox{X-ray} luminous than those with $R^{*}<3$; (9c)
when RLQs are subdivided by radio luminosity, RLQs with ${\ell}_{\rm
  r}>33.3$ are more \hbox{X-ray} luminous than those with ${\ell}_{\rm
  r}<33.3$; (9d) RLQs with flat radio spectra are more \hbox{X-ray}
luminous than those with steep radio spectra, and in particular almost
all of the most X-ray luminous RLQs (with ${\ell}_{\rm x}>28$) have
flat radio spectra. There does not appear to be any grouping of RIQs
or RLQs that contains objects with \hbox{X-ray} luminosities less than
those of comparable RQQs at any optical/UV luminosities; this is
broadly consistent with RIQs and RLQs being similar to RQQs but with
an ``extra'' source of \hbox{X-ray} emission whose strength depends
upon radio properties. Over all groupings and models, there is a
general tendency for the RLQs at the highest optical/UV luminosities
to lie above their best-fit models (to a degree exceeding any possible
slight systematic flattening of the slope due to the fitting method;
see Appendix D), and this structure in the residuals suggests that a
linear fit (to logarithmic quantities) of \hbox{X-ray} luminosity as a
sole function of optical/UV luminosity is not an adequate model even
when applied within subgroups of RLQs, at least for particularly
radio-loud, luminous, or flat-spectrum RLQs.

Joint 68\% and 90\% confidence ellipses for the various fits to this
model are plotted in Figure~10. In all panels the RQQ result is
plotted as a black ellipse for comparison. It can be seen in (10a)
that the confidence region for RIQs is near to that of RQQs; the
modest radio loudness and radio luminosity of RIQs generally do not
appear to enhance substantially their \hbox{X-ray} emission. In
contrast, the confidence region for RLQs is well separated from that
of RQQs, with both a greater model intercept and slope. It can also be
seen that our sample substantially increases the precision with which
the model parameters can be assessed over that provided by previous
studies, such as that of Worrall et al.~(1987) for their {\it
  Einstein\/} sample of RLQs (recall we use 93 of their 114
objects). In (10b), (10c), and (10d) the confidence regions for RLQs
with $R^{*}>3$, ${\ell}_{\rm r}>33.3$, and ${\alpha}_{\rm r}>-0.5$ are
offset from their weaker RLQ counterparts, which are themselves still
fully distinct from RQQs. The increased \hbox{X-ray} brightness for
RLQs with $R^{*}>3$ or ${\ell}_{\rm r}>33.3$ is primarily due to a
larger intercept in the modeled relation. It is possible that flatter
radio spectra objects may have a stronger dependence of ${\ell}_{\rm
  x}$ on ${\ell}_{\rm uv}$, although the 90\% confidence regions
overlap in projection onto the slope variable. The trends displayed in
(10b) and (10d) are qualitatively similar to those shown by Worrall et
al.~(1987) in their Figure~1.

\begin{figure}
\includegraphics[scale=0.47]{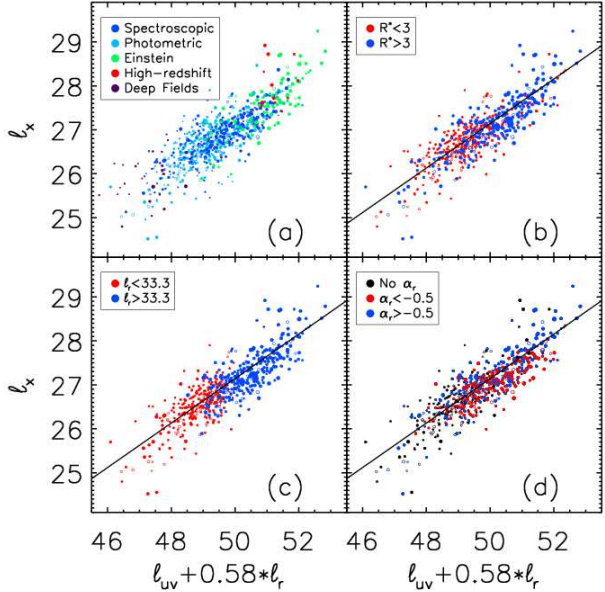} \figcaption{X-ray luminosity
  as a joint function of optical/UV and radio luminosity for various
  sub-samples of RLQs. The model considered is \hbox{${\ell}_{\rm x} =
    a_{\rm 0} + b_{\rm uv}{\times}{\ell}_{\rm uv}+ c_{\rm
      r}{\times}{\ell}_{\rm r}$} (where ${\ell}_{\rm x}$, ${\ell}_{\rm
    uv}$, and ${\ell}_{\rm r}$ have been normalized prior to fitting
  as described in $\S$5). Filled symbols are X-ray detections and
  larger symbols are more radio-loud throughout. The x-axis is based
  on the best-fit model for RLQs, for which $c_{\rm
    r}\simeq0.58{\times}b_{\rm uv}$. The black solid line in panels
  (b), (c), and (d) is the best-fit model for RLQs.}
\end{figure}

\subsection{${\ell}_{\rm x}({\ell}_{\rm uv},{\ell}_{\rm r})$}

\subsubsection{Inclusion of radio luminosity as a fit parameter}

The second parameterization we consider is \hbox{X-ray} luminosity as
a joint function of optical/UV luminosity and radio luminosity:
${\ell}_{\rm x} = a_{\rm 0} + b_{\rm uv}\times{\ell}_{\rm uv} + c_{\rm
  r}\times{\ell}_{\rm r}$. The resulting coefficients obtained from
fitting various groupings of RIQs and RLQs are listed on the right
side of Table 7. The optical/UV luminosity coefficient is now $b_{\rm
  uv} = 0.506$ for RLQs and \hbox{$b_{\rm uv}=0.432/0.439/0.471$} for
highly radio-loud/radio-luminous/flat-spectrum RLQs, compared to the
$b_{\rm uv} = 0.649$ for RQQs; this suggests that the apparently
stronger dependence of RLQ \hbox{X-ray} luminosity upon ${\ell}_{\rm
  uv}$ indicated in the previous model of ${\ell}_{\rm x}({\ell}_{\rm
  uv})$ was actually reflecting the influence of radio luminosity,
which is now explicitly considered. The best-fit model for RIQs
indicates that radio properties do not strongly infuence the
\hbox{X-ray} luminosity of RIQs ($c_{\rm r}$ is formally consistent
with zero). Figure~11 plots ${\ell}_{\rm x}$ versus ${\ell}_{\rm
  uv}+0.58\times{\ell}_{\rm r}$. This choice of variables is motivated
by the coefficients of the ${\ell}_{\rm x}({\ell}_{\rm uv},{\ell}_{\rm
  r})$ model for RLQs, for which $c_{\rm r}$ is
$\simeq0.58{\times}b_{\rm uv}$ (Table 7). Collapsing the ${\ell}_{\rm
  uv}-{\ell}_{\rm r}$ plane to a single joint variable simplifies
presentation of the modeling results and enables ready comparison of
the properties of subgroups of RLQs to those of RLQs as a whole. It
can be seen from (11a) that the {\it Einstein\/} sample of RLQs
dominates the highest luminosity region of the full sample, which also
contains the high-redshift sample objects. Conversely, the lowest
luminosity region of the full sample is strongly influenced by the
deep-field sample objects, although there are many primary sample
photometric quasars in this region as well. In (11b), (11c), and
(11d), data for the same subgroups of RLQs as in Figure~9 are shown
along with the best-fit ${\ell}_{\rm x}({\ell}_{\rm uv},{\ell}_{\rm
  r})$ model for RLQs for comparison. The subgroups of RLQs do not
deviate strongly from the trend for RLQs in general in these
coordinates, although the particularly radio and optical/UV luminous
RLQs from the {\it Einstein\/} sample (along with a few high-redshift
objects) are still excessively \hbox{X-ray} bright.

\begin{figure}
\includegraphics[scale=0.39]{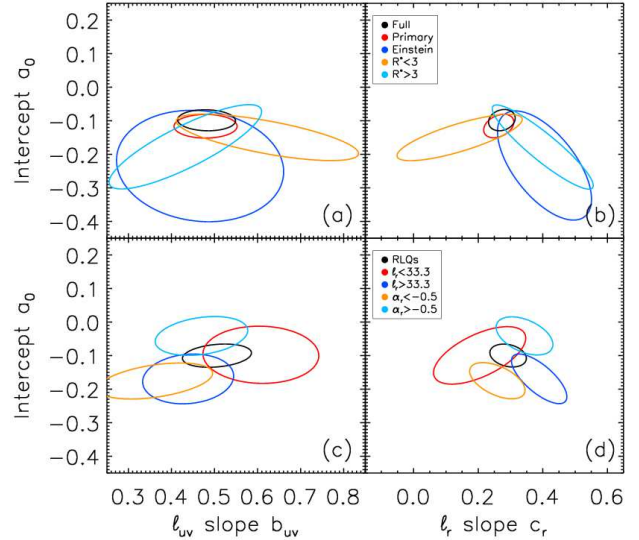} \figcaption{Joint 90\%
  confidence ellipses for fitting X-ray luminosity as a function of
  radio and optical/UV luminosity. The model considered is
  ${\ell}_{\rm x} = a_{\rm 0} + b_{\rm uv}{\times}{\ell}_{\rm uv}+
  c_{\rm r}{\times}{\ell}_{\rm r}$ (where ${\ell}_{\rm x}$,
  ${\ell}_{\rm uv}$, and ${\ell}_{\rm r}$ have been normalized prior
  to fitting as described in $\S$5). The legend in (b) also applies to
  (a) and the legend in (d) also applies to (c). See $\S$5.2 for
  discussion.}
\end{figure}

Joint 90\% confidence ellipses (calculated after collapsing the third
dimension for ease of viewing) for the various fits to this model are
plotted in Figure~12. The full, primary, and {\it Einstein\/} samples
are plotted in (12a) and (12b), as are the subgroups of RLQs with
$R^{*}<3$ and $R^{*}>3$. The parameters for the primary sample of SDSS
RIQs and RLQs are consistent with those of the full sample (not
unexpected, since the primary sample makes up a majority of the full
sample). There is no evidence of a statistically significant
difference in the \hbox{X-ray} luminosity dependence of these
subgroups upon optical/UV luminosity.\footnote{It appears possible
  that the $R^{*}<3$ objects may have a larger $b_{\rm uv}$
  coefficient (and a smaller $c_{\rm r}$ coefficient); this could be
  probed with a larger sample.} There is suggestive support for {\it
  Einstein\/} RLQs and (relatedly) for RLQs with $R^{*}>3$ possessing
a stronger dependence of \hbox{X-ray} luminosity upon radio
luminosity, but the projected confidence ellipses all overlap in
(12b). The joint 90\% confidence ellipses for the ${\ell}_{\rm
  x}({\ell}_{\rm uv},{\ell}_{\rm r})$ model applied to subgroups of
RLQs divided by radio luminosity and by radio spectral index are
plotted in (12c) and (12d), along with the result for RLQs in general
provided for comparison. Flat-spectrum RLQs are \hbox{X-ray} brighter
than steep-spectrum RLQs due to a larger intercept, but have a
consistent dependence of \hbox{X-ray} luminosity upon both optical/UV
and radio luminosity. In possible contrast, RLQs with ${\ell}_{\rm
  r}<33.3$ and those with ${\ell}_{\rm r}>33.3$ share similar best-fit
intercepts, but the more radio-luminous RLQs may have a greater/lesser
dependence of ${\ell}_{\rm x}$ upon ${\ell}_{\rm r}/{\ell}_{\rm uv}$,
although the confidence ellipses overlap in projection onto $c_{\rm
  r}/b_{\rm uv}$.

\subsubsection{A ``radio-adjusted'' ${\ell}_{\rm
  x}({\ell}_{\rm uv})$ relation for RIQs and RLQs}

We consider briefly whether RIQs and RLQs can be treated as similar to
RQQs but with an additional jet-linked contribution to the
\hbox{X-ray} luminosity (which could imply a consistent disk/coronal
structure). If so, then after accounting for the influence of radio
emission, the \hbox{X-ray} luminosity in RIQs and RLQs should be
correlated with optical/UV luminosity through a relation similar to
that for RQQs. RIQs display no significant dependence of \hbox{X-ray}
luminosity upon radio luminosity and may share the same dependence
upon optical/UV luminosity as holds for RQQs, although this is not
strongly constrained in our data.

One method of investigating a ``radio-adjusted'' ${\ell}_{\rm
  x}({\ell}_{\rm uv})$ relation could be to set the radio luminosity
in the best-fit RLQ ${\ell}_{\rm x}({\ell}_{\rm uv},{\ell}_{\rm r})$
model to a value representative of RQQs. This requires an accurate
parameterization of ${\ell}_{\rm r}({\ell}_{\rm uv})$ for RQQs, a
difficult function to evaluate given the inherent radio weakness of
RQQs. A simple model could have radio luminosity proportional to
optical/UV luminosity as ${\ell}_{\rm r} = \alpha +
\beta\times{\ell}_{\rm uv}$; in this case ${\ell}_{\rm x}({\ell}_{\rm
  uv},{\ell}_{\rm r})$ would transform to ${\ell}_{\rm x}({\ell}_{\rm
  uv})$ as (adding in the implicit luminosity normalizations)
${\ell}_{\rm x}-27 = a_{\rm 0}+c_{\rm
  r}\times(\alpha+30.5\beta-33.3)+(b_{\rm uv}+c_{\rm
  r}\beta)\times({\ell}_{\rm uv}-30.5)$. White et al.~(2007) present a
correlation (their Equation 2) that extends to low radio-loudness
values ($R^{*}<1$) and is equivalent\footnote{Their radio-loudness has
  been slightly adjusted as a function of optical luminosity; see
  their Equation~4.} to $\alpha=4.57$ and $\beta=0.85$. Adopting these
values of $\alpha$ and $\beta$, the RLQ fit of ${\ell}_{\rm x} =
-0.100 + 0.506{\ell}_{\rm uv} + 0.292{\ell}_{\rm r}$ becomes
${\ell}_{\rm x} = -0.919 + 0.754{\ell}_{\rm uv}$; this may be compared
with the best-fit RQQ relation of ${\ell}_{\rm x} = -0.545 +
0.649{\ell}_{\rm uv}$ (all fitted coefficients from Table~7). The
slope for RLQs in the radio-adjusted relation is closer to but
slightly larger than that for RQQs (the difference is
$0.105\pm0.049$). Since we have not yet taken the likely beaming of
some fraction of the \hbox{X-ray} emission in RLQs into account, this
result does not mandate that the disk/corona in RLQs is more X-ray
efficient at high optical/UV luminosities than in RQQs. (We
demonstrate in $\S$6 that the RLQ ${\ell}_{\rm x}({\ell}_{\rm
  uv},{\ell}_{\rm r})$ fit can be reproduced assuming a disk/coronal
scaling as in RQQs plus a jet component.) The difference in intercepts
as compared to the RQQ relation may be reflective of greater beaming
of radio emission in the RLQs (although some RQQs may have a boosted
component of radio emission; e.g., Miller et al.~1993; Falcke et
al.~1996), in which case the RQQ relation ought to have the radio
luminosity similarly enhanced prior to substitution for a first-order
comparison. For illustrative purposes, accounting for an additional
beaming enhancement of a factor of 19 (e.g., corresponding to an
inclination of $\simeq12^{\circ}$ with $\gamma=10.5$; see $\S$6) would
change the transformed RLQ intercept to match the RQQ result.

\begin{figure}
\includegraphics[scale=0.39]{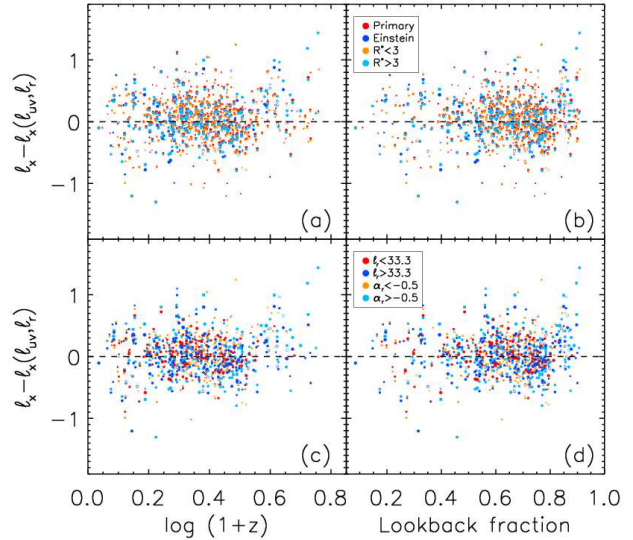}
\figcaption{Residuals for best-fit results for X-ray luminosity
  parameterized by optical/UV and radio luminosity, plotted against
  redshift and fractional lookback time. The legend in (b) also
  applies to (a) and the legend in (d) also applies to (c). A single
  object may be a member of two categories in a given panel, in which
  case the residual values from the separate fits to each category are
  each plotted at the redshift of that object. There is no apparent
  redshift dependence. See $\S$5.3 for discussion.}
\end{figure}

\subsection{${\ell}_{\rm x}({\ell}_{\rm uv},{\ell}_{\rm r}, z)$}

\subsubsection{Inclusion of redshift as a fit parameter}

We now investigate whether there is a dependency upon redshift in
addition to the dependencies upon optical/UV and radio luminosity.
Although the ${\ell}_{\rm x}({\ell}_{\rm uv},{\ell}_{\rm r})$
residuals (see Figure~13) do not show any obvious redshift
dependence,\footnote{The possible tendency for RLQs at high redshift to
  have positive residuals is not necessarily a redshift effect; the
  best-fit ${\ell}_{\rm x}({\ell}_{\rm uv},{\ell}_{\rm r})$ model
  appears to underpredict X-ray emission for particularly radio-loud
  or radio luminous RLQs such as these (see also $\S$2.2.2).} it is
in principle possible that some of the apparent luminosity dependence
might actually be driven by redshift evolution. We test this
possibility by including the redshift dependence as a parameter when
modeling, by fitting ${\ell}_{\rm x}({\ell}_{\rm uv},{\ell}_{\rm r},
z)$, where again the luminosities are normalized prior to fitting. The
redshift dependence is put in terms of $\log (1+z)$ or lookback
fraction [${\tau}_{\rm z}$ = 1$-$age($z$)/age($z=0$)]. The best-fit
model for the full sample is ${\ell}_{\rm x} =
(-0.098\pm0.051)+(0.482\pm0.036)\times{\ell}_{\rm
  uv}+(0.273\pm0.019)\times{\ell}_{\rm
  r}+(0.000\pm0.128)\times{\log}(1+z)$; when RLQs only are considered,
the best-fit model is ${\ell}_{\rm x} =
(-0.109\pm0.052)+(0.503\pm0.042)\times{\ell}_{\rm
  uv}+(0.293\pm0.027)\times{\ell}_{\rm
  r}+(0.025\pm0.131)\times{\log}(1+z)$. When the redshift dependence
is expressed instead in terms of the fractional lookback time
${\tau}_{\rm z}$, the best-fit model for the full sample is
${\ell}_{\rm x} = (-0.064\pm0.072)+(0.488\pm0.034)\times{\ell}_{\rm
  uv}+(0.273\pm0.019)\times{\ell}_{\rm
  r}+(-0.052\pm0.109)\times{\tau}_{\rm z}$; when RLQs only are
considered, the best-fit model is ${\ell}_{\rm x} =
(-0.065\pm0.074)+(0.511\pm0.042)\times{\ell}_{\rm
  uv}+(0.292\pm0.027)\times{\ell}_{\rm
  r}+(-0.053\pm0.111)\times{\tau}_{\rm z}$. In all these cases the
difference between the coefficient for the redshift term and zero is
not statistically significant, and the joint 68\% and 90\% confidence
ellipses with the optical/UV and radio luminosity coefficients include
zero (Figure~14). The redshift coefficients for the
${\log}(1+z)/{\tau}_{\rm z}$ models when fitting the primary sample,
spectroscopic sample, or {\it Einstein\/} sample alone are
$-0.500\pm0.167$/$-0.364\pm0.140$, $-0.351\pm0.245$/$-0.337\pm0.202$,
or $0.775\pm0.443/0.282\pm0.379$, respectively. The indicated redshift
dependence is marginal (3.0/2.6, 1.4/1.7, or 1.7/0.7$\sigma$), and the
direction of the potential trend is inconsistent between the primary
and {\it Einstein\/} samples; recall also that Worrall et al.~(1987),
using independent methods, found no significant redshift dependence
among the {\it Einstein\/} objects. The superior coverage of the
${\ell}-z$ plane provided by the full and RLQ samples leads us to
favor those results.

\begin{figure}
\includegraphics[scale=0.48]{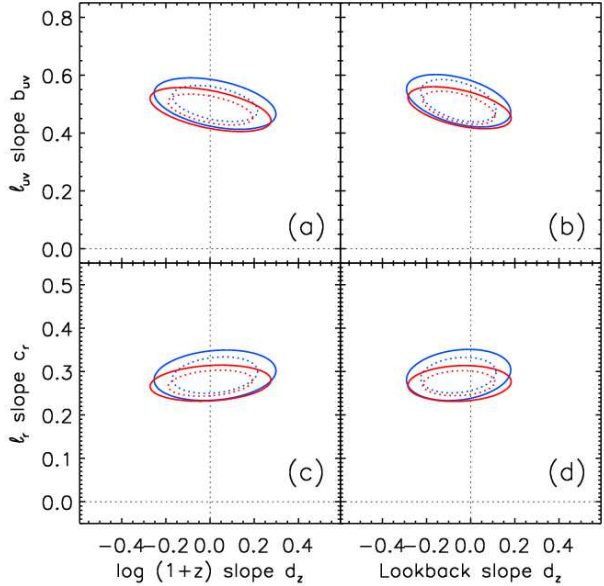} \figcaption{Consideration of
  redshift dependence. The first model investigated is
  \hbox{${\ell}_{\rm x} = a_{\rm 0} + b_{\rm uv}{\times}{\ell}_{\rm
      uv} + c_{\rm r}{\times}{\ell}_{\rm r} + d_{\rm
      z}{\times}{\log}(1+z)$} (where ${\ell}_{\rm x}$, ${\ell}_{\rm
    uv}$, and ${\ell}_{\rm r}$ have been normalized prior to fitting
  as described in $\S$5); the second model investigated replaces
  ${\log}(1+z)$ with the fractional lookback time ${\tau}_{\rm
    z}$. Joint 90\% (solid) and 68\% (dotted) confidence ellipses are
  plotted for the full sample (red) and for the sample of RLQs
  (blue). The difference between the coefficients for redshift
  dependence and zero is not statistically significant.}
\end{figure}

We find no evidence for a dependence upon redshift for the X-ray
emission properties of the full sample or RLQs alone. The degree to
which otherwise comparable RIQs or RLQs at different redshifts could
differ in X-ray luminosity may be constrained via the value and
1$\sigma$ errors on the best-fit coefficient to the redshift term. For
the full sample, the coefficient of $(0.000\pm0.128)\times{\log}(1+z)$
suggests a maximum evolution (1$\sigma$, i.e., using 0.000+0.128 for
the coefficient) in X-ray luminosity between redshift $z=0$ and $z=5$
of ${\ell}_{\rm x,z=0}-{\ell}_{\rm x,z=5}=0.100$, or a ratio in linear
units of 1.26. This suggests that the X-ray luminosity of otherwise
comparable RIQs or RLQs has not changed by more than 30\% over
$z=0-5$. For RLQs only, a similar analysis suggests a maximum change
of $\simlt35$\%. The ${\tau}_{\rm z}$ best-fit coefficients similarly
suggest a maximum redshift-driven change in \hbox{X-ray} luminosity
for the full sample or for RLQs of $\simlt30$\% over $z=0-5$. Although
the fraction of quasars that are RLQs is dependent upon redshift
(e.g., Jiang et al.~2007), the \hbox{X-ray} properties of individual
RIQs and RLQs do not appear to differ strongly when comparing objects
at low versus high redshift. Apparently the cosmic evolution in the
efficiency of generating RLQs does not substantially impact RLQ
structure post-formation.

\begin{figure*}
\includegraphics[scale=0.80]{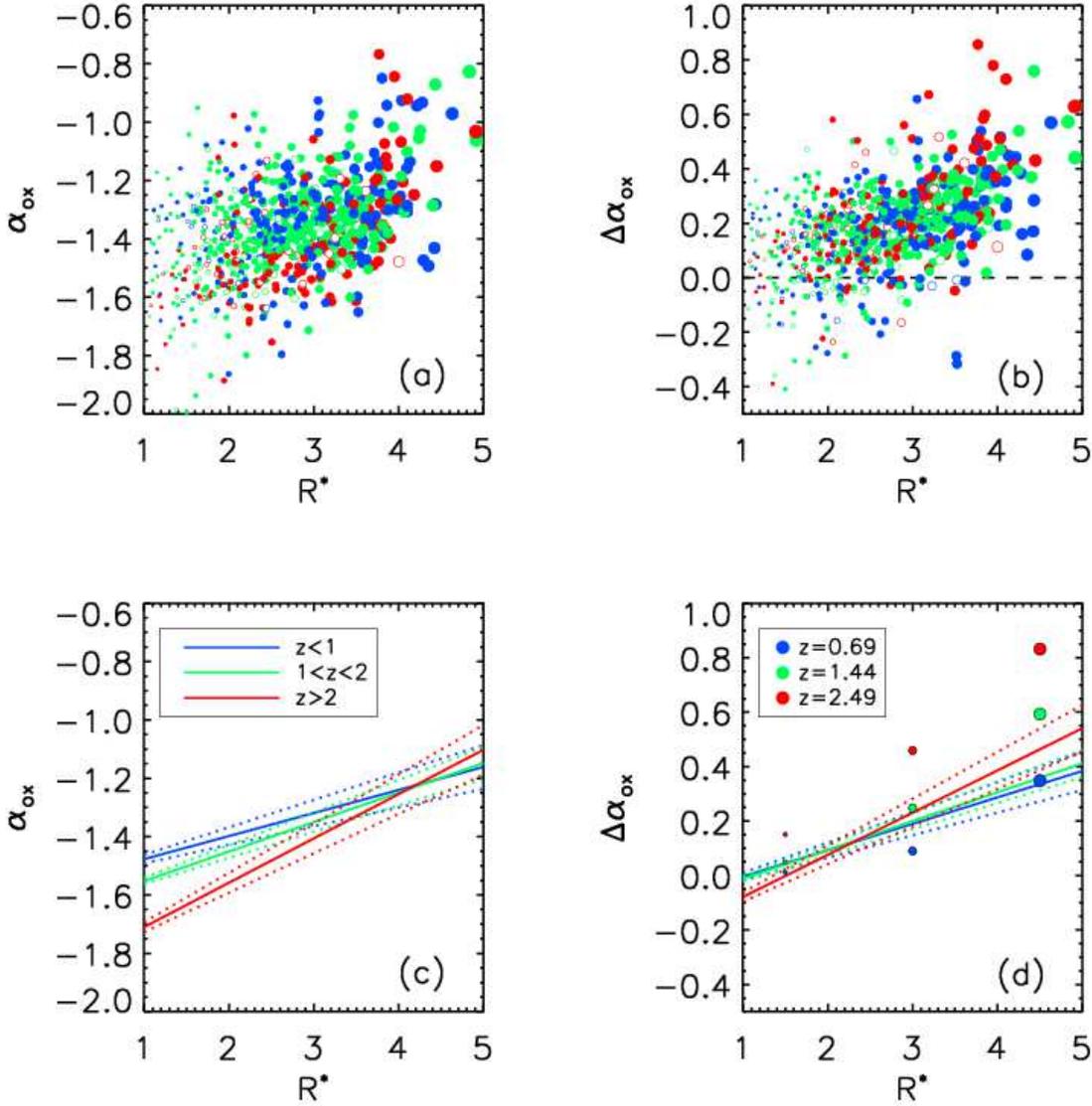}
\figcaption{Consideration of relative X-ray brightness as a function
  of radio-loudness. The full sample of RIQs and RLQs is shown in (a)
  for ${\alpha}_{\rm ox}(R^{*})$ and in (b) for ${\Delta}{\alpha}_{\rm
    ox}(R^{*})$, with blue, green, and red colors indicating low,
  medium, or high redshift ($z<1$, $1<z<2$, and $z>2$,
  respectively). Best-fit trend lines are shown in (c) and (d) with the
  1$\sigma$ errors on the slopes indicated by dashed lines. The
  expected enhancement in X-ray brightness from the $(1+z)^{4}$
  dependence of IC/CMB jet-linked X-ray emission is also plotted in
  (d) for comparison; the set of points at $R^{*}$=1.5, 3, and 4.5
  correspond to models in which the jet-linked fraction of the
  \hbox{X-ray} continuum (prior to considering redshift) is 1\%, 10\%,
  and 99\%, respectively.}
\end{figure*}

\subsubsection{Dependence of relative X-ray brightness on redshift}

An alternative manner of visualizing the potential redshift dependence
of the \hbox{X-ray} luminosity of RIQs and RLQs and of searching for
redshift evolution as a function of radio-loudness is to construct a
multiwavelength color-color plot and group objects within separate
redshift bins. Figure~15a shows ${\alpha}_{\rm ox}$ plotted versus
$R^{*}$ for the full sample divided into low, medium, and high
redshift ($z<1$, $1<z<2$, and $z>2$, respectively). Objects become
increasingly X-ray bright (with less negative values of ${\alpha}_{\rm
  ox}$) as they become increasing radio-loud, as previously shown in
Figure~7, but there is no obvious difference in this trend with
redshift. Fitting ${\alpha}_{\rm ox}(R^{*})$ does suggest that the
slope steepens slightly with redshift, but this is offset by a
decreased intercept (Figure~15c; see also Figure~5 of Lopez et
al.~2006). It is preferable to take the dependence of \hbox{X-ray}
luminosity upon optical/UV luminosity into account when constructing
the optical/UV-to-\hbox{X-ray} color. We calculate
${\Delta}{\alpha}_{\rm ox}={\alpha}_{\rm ox} - {\alpha}_{\rm
  ox}({\ell}_{\rm uv})$ by taking ${\alpha}_{\rm ox}({\ell}_{\rm uv})$
from the Just et al.~(2007) relation for RQQs (their Equation
3). Figure~15b shows ${\Delta}{\alpha}_{\rm ox}$ plotted versus
$R^{*}$ for the full sample grouped into low, medium, and high
redshift bins. There is no obvious stratification by redshift, but
here fitting ${\Delta}{\alpha}_{\rm ox}(R^{*})$ suggests that the
slight increase in slope with increasing redshift may result in a
slight increase in relative \hbox{X-ray} brightness, at least for RLQs
with $R^{*}>2$ (Figure~15d). It is possible that a disproportionate
number of objects in the high-redshift bin with low line-of-sight
inclinations could produce this result.  The best-fit model parameters
are given in Table~8. Two-dimensional KS tests indicate that the
distribution of ${\alpha}_{\rm ox}-R^{*}$ is inconsistent between
different redshift bins (low-mid, mid-high, or low-high), but that of
${\Delta}{\alpha}_{\rm ox}-R^{*}$ is not ($p\simgt0.12$ for all
cases).

It has been suggested that \hbox{X-ray} jets for which inverse Compton
scattering of cosmic microwave background photons (IC/CMB; e.g.,
Tavecchio et al.~2000) provides the primary emission component should
become the dominant source of RLQ \hbox{X-ray} emission at high
redshift (e.g., Rees \& Setti 1968; Schwarz 2002). We estimate the
enhancement in total \hbox{X-ray} luminosity from this process for
sources at redshifts of 0.69/1.44/2.48 (the medians within the
low/medium/high groupings) and for a jet-linked fraction $j$ of the
nuclear X-ray emission (prior to accounting for the redshift
dependence of IC/CMB \hbox{X-ray} emission) of 1\%, 10\%, and
99\%. This is expressed in terms of an increase in
${\Delta}{\alpha}_{\rm ox}$ as $0.384\times\log{[j(1+z)^{4}+(1-j)]}$,
where the coefficient converts the increase in ${\ell}_{\rm x}$ to a
flattening of the optical/UV-to-X-ray spectral slope. The results are
plotted (in increasing order of jet dominance) at radio-loudness
values\footnote{These $R^{*}$ values increase with increasing jet
  dominance, as is qualitatively expected, but the precise numerical
  association between $R^{*}$ and \hbox{X-ray} jet dominance is
  model-dependent (as explored in $\S$6.3) and so the chosen $R^{*}$
  values should be regarded as illustrative.} of 1.5, 3, and 4.5 in
Figure~14d. It can be seen that the $(1+z)^{4}$ dependence of
\hbox{X-ray} jet-linked IC/CMB emission would lead for these
parameters to a stronger splitting with redshift than is observed. We
find that the \hbox{X-ray} luminosities within our sample of RIQs and
RLQs are unlikely to include significant contributions from
\hbox{X-ray} IC/CMB jet emission. In principle it is possible that
extremely bright \hbox{X-ray} jet features could be routinely found
outside of the extraction regions we use to measure the \hbox{X-ray}
luminosity, but to our knowledge such sources are extremely rare. This
is consistent with the general lack of observed high-redshift RLQs in
which the \hbox{X-ray} jet outshines the core (e.g., Bassett et
al.~2004; Lopez et al.~2006).

\subsection{Additional considerations}

We briefly examine four additional topics relevant to the analysis of
luminosity correlations: the impact of the selection method, the
influence of targeted sources, the normality of the variables, and the
potential for spurious or inaccurate results due to luminosity
dispersion effects.

\subsubsection{Optical versus radio selection}

As discussed in $\S$2.1.1, the median properties of a
quasi-radio-selected sample of RIQs and RLQs (constructed from SDSS
spectroscopic quasars by requiring the ``FIRST'' target flag to be
set) are similar to those of ``QSO/HIZ'' targeted SDSS spectroscopic
quasars (with substantial overlap between these samples). The
dependence of \hbox{X-ray} luminosity upon optical/UV and radio
luminosities is also similar for the ``FIRST'' and for the ``QSO/HIZ''
samples, as may be seen from the best-fit relations given in
Table~7. Joint 68\% and 90\% confidence contours from fitting the
``FIRST'' and the ``QSO/HIZ'' samples to ${\ell}_{\rm x} = a_{\rm 0} +
b_{\rm uv}\times{\ell}_{\rm uv} + c_{\rm r}\times{\ell}_{\rm r}$ are
provided in Figure~16, where it can be seen that the parameter values
are consistent. By excluding the small fraction of objects with
${\Delta}(g-i)>1$ we omit strongly dust-reddened RLQs, which may be
associated with young quasars (e.g., Urrutia et al.~2008,
2009). Within the relative color range we accept, there does not
appear to be a difference in the \hbox{X-ray} properties of SDSS RIQs
and RLQs selected by color compared to those selected due to radio
emission. Presumably the \hbox{X-ray} emission mechanisms are likewise
similar, and so the results from $\S$5 and the modeling in $\S$6 are
generally applicable.

\subsubsection{Influence of targeted objects}

Only 14.1\% of RIQs and RLQs within the primary sample were targeted
for \hbox{X-ray} observations ($\S$3.4), with the remainder possessing
serendipitous off-axis coverage. Inclusion of the targeted sources
does not bias the results of the luminosity correlation analysis,
although there are some minor differences in the properties of
targeted and off-axis objects. Compared to those objects observed
serendipitously, the targeted RIQs and RLQs are at somewhat lower
redshifts (median $z$ of 1.05 versus 1.45) but are also brighter
(median $m_{\rm i}$ of 17.94 versus 19.54) and so have modestly higher
optical/UV luminosities (median ${\ell}_{\rm uv}$ of 30.74 versus
30.33). The targeted RIQs and RLQs are also somewhat more radio-loud
(median $R^{*}$ of 3.09 versus 2.44) and $\sim$40\% X-ray brighter
relative to optical/UV luminosity (median ${\ell}_{\rm x}-{\ell}_{\rm
  x,RQQ}$ of 0.53 versus 0.38, with ${\ell}_{\rm x}-{\ell}_{\rm
  x,RQQ}$ as defined in $\S$4). Presumably, targets were
preferentially selected from sources already known to be \hbox{X-ray}
bright.

The results of parameterizing \hbox{X-ray} luminosity for the off-axis
objects are similar to those obtained for the full and primary samples
(Table~7). The differences between the best-fit parameters for the
slopes for the off-axis sample compared to the full or primary samples
are essentially zero.\footnote{Differences between the off-axis and
  full/primary samples are ${\Delta}b_{\rm uv} =
  (-0.045\pm0.044)/(0.023\pm0.048)$ for the ${\ell}_{\rm
    x}({\ell}_{\rm uv})$ model and ${\Delta}b_{\rm uv} =
  (0.046\pm0.051)/(0.048\pm0.052)$, ${\Delta}c_{\rm r} =
  (-0.030\pm0.030)/(-0.020\pm0.031)$ for the ${\ell}_{\rm
    x}({\ell}_{\rm uv},{\ell}_{\rm r})$ model.} Fitting the targeted
objects separately indicates they possess a weaker dependence of
\hbox{X-ray} luminosity upon optical/UV luminosity and perhaps a
stronger dependence upon radio luminosity than do the full or primary
samples, although their parameters are consistent with those of the
spectroscopic sample.

\begin{figure}
\includegraphics[scale=0.39]{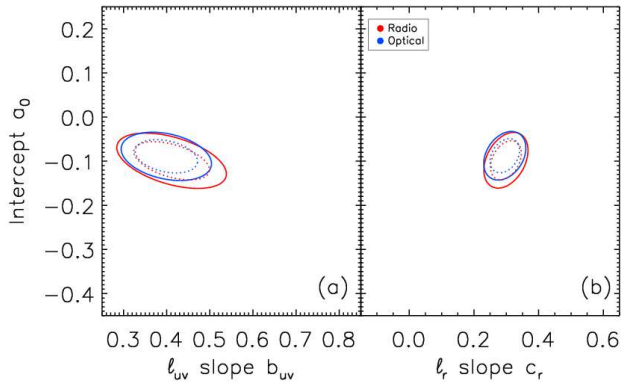} \figcaption{Comparison of a
  quasi-radio-selected (red ellipses) sample of RIQs and RLQs to an
  optically-selected (blue ellipses) sample (both drawn from
  SDSS/FIRST data; there is $\simgt$90\% overlap), illustrated with
  joint 90\% (solid) and 68\% (dotted) confidence ellipses for the
  model \hbox{${\ell}_{\rm x} = a_{\rm 0} + b_{\rm
      uv}{\times}{\ell}_{\rm uv} + c_{\rm r}{\times}{\ell}_{\rm
      r}$}. The axes are scaled to match Figure~12. The RIQs and RLQs
  targeted by SDSS as FIRST sources have properties consistent with
  those targeted due to optical colors.}
\end{figure}

\begin{figure}
\includegraphics[scale=0.39]{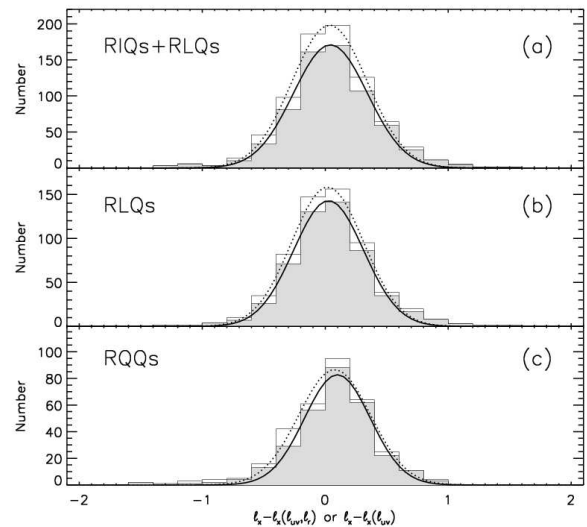}
\figcaption{Evaluation of normality for residuals from fitting X-ray
  luminosity as a joint function of optical/UV and radio luminosity
  (RIQs and RLQs, top panel; RLQs, middle panel; fits from Table~7)
  and as a sole function of optical/UV luminosity (RQQs, bottom panel;
  data from Steffen et al.~2006). The open histograms include limits
  and are fit with Gaussians as indicated by the dotted curves, while
  the filled histograms are detections only and are fit with Gaussians
  as indicated by the solid curves.}
\end{figure}

\subsubsection{Log-normality of residuals}

A presumption that considered variables are normally distributed is
inherent in the chosen method of analysis (and indeed in most
parametric modeling), and can be checked by examining the residuals
from the best-fit models for X-ray luminosity in our sample of RIQs
and RLQs. Histograms of the residual \hbox{X-ray} luminosity for RIQs
and RLQs, RLQs alone, and RQQs are provided in Figure~17. While it is
not clear {\it a priori\/} that the RIQs and RLQs considered here
ought to show normally-distributed residual X-ray luminosity (in
logarithmic units), since there are distinct X-ray emission mechanisms
for RIQs and RLQs (disk/corona and jet-linked) as opposed to a single
dominant X-ray emission mechanism (disk/corona) in RQQs, it may be
seen in Figure~17 that the distributions closely match the overplotted
best-fit Gaussians. Unfortunately the presence of limits in our sample
makes it difficult to utilize standard tests for normality. We examine
the distribution of detected objects and also the distribution of all
objects, treating limits as detections for this calculation only. The
distribution of residual \hbox{X-ray} luminosity is acceptably
characterized as normal, with KS test probabilities $p>0.18$ in all
cases. While this is best regarded as suggestive rather than
conclusive, it does indicate that our methodology is unlikely to
produce biased results due to substantial underlying non-normality.

There appears to be a tail of X-ray weak objects within the RQQ sample
(17c), some of which might be low-redshift BAL RQQs. The sample of
RIQs and RLQs (17a) also appears to show an X-ray weak tail, perhaps
less prominently for RLQs only (17b). These may again be BAL objects,
which decrease in percentage as radio-loudness increases (e.g.,
Shankar et al.~2008). However, in contrast to the situation for RQQs,
there may also be an X-ray bright tail for RIQs and RLQs (for example,
the percentage of objects above 1.645$\sigma$ from the mean is not 5\%
but rather 8\%/8\% for (RIQs+RLQs)/RLQs, calculated using the fitted
values for $\sigma$ and $\mu$; it is 6\% for RQQs). These \hbox{X-ray}
bright objects are generally members of the set of particularly
radio-loud or radio-luminous RLQs, and hence may be jet-dominated at
\hbox{X-ray} frequencies.

\subsubsection{Robustness of luminosity correlations}

The degree of inherent scatter in the luminosities can affect the
results of correlation studies. For example, Yuan et al.~(1998)
demonstrated that when the optical/UV luminosity scatter
${\sigma}_{\rm uv}$ is significantly larger than the X-ray luminosity
scatter ${\sigma}_{\rm x}$, a fit to ${\alpha}_{\rm ox}({\ell}_{\rm
  uv})$ over a limited range of ${\ell}_{\rm uv}$ can indicate an
anti-correlation where none necessarily exists. Relatedly,
${\sigma}_{\rm uv}\gg{\sigma}_{\rm x}$ can give a fitted slope for the
${\ell}_{\rm x}({\ell}_{\rm uv})$ relation less than unity even if
these luminosities are actually proportional (when the considered
luminosity ranges are small). In general, examined luminosity
correlations are significantly less likely to yield spurious or
inaccurate results with a sample spanning a large range in
luminosities relative to the observed dispersion (e.g., see discussion
in $\S$5 of Just et al.~2007), as does our sample.

It is possible to assess the luminosity scatter through consideration
of the standard deviation of the residuals about the ${\ell}_{\rm
  x}({\ell}_{\rm uv}, {\ell}_{\rm r})$ relation. For the primary
sample, these values are 0.34/0.71/1.30 for ${\ell}_{\rm
  x}/{\ell}_{\rm uv}/{\ell}_{\rm r}$, with corresponding 5th--95th
percentile luminosity ranges of 25.81--27.62/29.40--31.26/31.47--34.46
(treating \hbox{X-ray} limits as detections for this calculation
only). This fit is performed treating ${\ell}_{\rm x}$ as the
dependent variable, so the residuals about this fit are larger than
the inherent scatter. We use a Monte Carlo approach to determine the
corresponding ${\sigma}_{\rm x}/{\sigma}_{\rm uv}/{\sigma}_{\rm r}$
values. For simplicity, we simulate ${\ell}_{\rm uv,mod}$ as uniformly
distributed ($n=654$) between 29.50 and 31.25, with ${\ell}_{\rm
  r,mod}$ = ${\ell}_{\rm uv,mod}$ + 2.5 (with random log-normal
scatter of ${\sigma}_{\rm R^{*}}$) and ${\ell}_{\rm x,mod}$ calculated
from ${\ell}_{\rm x}({\ell}_{\rm uv},{\ell}_{\rm r})$. Then random
log-normal scatter ${\sigma}_{\rm x}/{\sigma}_{\rm uv}/{\sigma}_{\rm
  r}$ is added to the model luminosities and the standard deviations
of the residuals and the luminosity ranges are calculated. A good
match to the observed data is provided with ${\sigma}_{\rm
  x}/{\sigma}_{\rm uv}/{\sigma}_{\rm r}$ =
0.28/0.32/0.42,\footnote{These are larger than the uncertainty
  typically associated with variability, plausibly suggesting a spread
  in physical properties among RIQs and RLQs.} and ${\sigma}_{\rm
  R^{*}}=0.55$ gives a radio-loudness standard deviation of
$\simeq$0.75 in agreement with the primary sample.

We assess the impact of such inherent scatter by fitting the simulated
data (including \hbox{X-ray} censoring as described in
$\S$6.1.3). This scatter acts to depress the dependence of
\hbox{X-ray} luminosity upon optical/UV and radio luminosities, so
that fitted values of $b_{\rm uv}$ and $c_{\rm r}$ are slightly lower
than those used in the input ${\ell}_{\rm x}({\ell}_{\rm
  uv},{\ell}_{\rm r})$ relation. The observed best-fit values of
$b_{\rm uv}\simeq0.48$ and $c_{\rm r}\simeq0.26$ for the primary
sample would result in fitted coefficients of 0.39 and 0.23,
respectively, and may themselves be obtained for input coefficients of
$\simeq$0.6 and $\simeq$0.3, respectively. Further, data simulated
with an input model of ${\ell}_{\rm x}({\ell}_{\rm uv},{\ell}_{\rm
  r})$ that is then fit as ${\ell}_{\rm x}({\ell}_{\rm uv})$ produce
fitted values for $b_{\rm uv}$ greater than the input coefficient
(e.g., fitted $b_{\rm uv}=0.67\pm0.03$ for input $b_{\rm uv}/c_{\rm
  r}$=0.6/0.3), as suggested in $\S$5.2.1.

A contribution to the \hbox{X-ray} luminosity linked to the radio
luminosity appears to be mandated for RIQs and RLQs. For the derived
inherent scatter, it is not possible to reproduce the observed
best-fit ${\ell}_{\rm x}({\ell}_{\rm uv},{\ell}_{\rm r})$ relation for
the primary sample with an input model using ${\ell}_{\rm
  x}({\ell}_{\rm uv})$ as for RQQs. With the luminosity ranges of our
primary sample, the $b_{\rm uv}\simeq0.65$ observed for RQQs may
result from an input coefficient of $\simeq$0.85 (1.0 is excluded;
see, e.g., $\S$3.5 of Strateva et al.~2005). Data simulated with an
input model of ${\ell}_{\rm x}\propto0.85\times{\ell}_{\rm uv}$ that
is then fit as ${\ell}_{\rm x}({\ell}_{\rm uv},{\ell}_{\rm r})$
produce fitted coefficients of $b_{\rm uv}=0.56\pm0.03$ and $c_{\rm
  r}=0.09\pm0.02$. If ${\ell}_{\rm r}$ is initially taken directly
from ${\ell}_{\rm uv}$ (i.e., ${\sigma}_{R^{*}}=0$), the fitted
coefficients are $b_{\rm uv}=0.47\pm0.03$ and $c_{\rm r}=0.22\pm0.03$,
but the scatter in radio-loudness is then significantly lower than
observed. (Using ${\ell}_{\rm r}=\alpha+\beta\times{\ell}_{\rm uv}$
with $\alpha=4.57$ and $\beta=0.85$ as in $\S$5.2.2 gives similar
results.)

We note for completeness that more extreme values of ${\sigma}_{\rm
  x}/{\sigma}_{\rm uv}/{\sigma}_{\rm r}$ could distort luminosity
correlations (e.g., for ${\sigma}_{\rm x}/{\sigma}_{\rm
  uv}/{\sigma}_{\rm r}$ = 0.28/0.64/0.42 or 0.28/0.32/0.84 rather than
the derived 0.28/0.32/0.42, input parameters of $b_{\rm uv}/c_{\rm
  r}$=0.6/0.3 give fitted coefficients of 0.22/0.33 or 0.56/0.15,
respectively). However, as demonstrated above, these considerations do
not apply to our sample. Additionally, since much of our analysis is
concerned with comparing the results of (similarly conducted) fits to
different groupings of RIQs and RLQs, any modest systematic skewing of
correlations does not impact our conclusions regarding the relative
differences in how \hbox{X-ray} luminosity is dependent on optical/UV
and radio luminosities within these subgroups.

\section{A physical model for X-ray emission in RIQs and RLQs}

It is of interest to evaluate the physical basis for the correlations
discussed in $\S$5. It is widely theorized that the increasing X-ray
brightness of RIQs and RLQs with increasing radio loudness or
luminosity is driven by a source of nuclear X-ray emission that is
directly or indirectly powered by the radio jet (e.g., see discussion
in $\S$1), but the precise nature of this linkage is not clearly
understood. We make use of previous results from the literature to
simulate a population of RIQs and RLQs with radio and optical
properties consistent with observations, and then test competing
models for the X-ray emission through comparing the properties of the
simulated data sets to observations. The structure and parameters of
this modeling are given in Table~9.

We adopt a physical model that contains emission contributions from
the core and lobes at radio frequencies (making the common assumption
that the core is dominated by the small-scale radio jet, but see also
Bell \& Comeau 2010), from the disk and small-scale jet at optical/UV
frequencies, and from the disk/corona and an additional ``jet-linked''
component at X-ray frequencies. Most of the parameters for the radio
and optical emission components in this model are fixed by prior work;
likely values for a few free parameters were determined through
comparison to our sample data (see Table~9 for details). The X-ray
disk/corona emission is presumed to scale with the optical/UV disk
emission as established for RQQs, as seems reasonable based on the
results of $\S$4 and $\S$5. The Doppler beaming factor is
$\delta={\gamma}^{-1}(1-\beta\cos\theta)^{-1}$ (e.g., Worrall \&
Birkinshaw~2006).

We consider three possibilities for the X-ray jet-linked emission: 
\begin{itemize}
\item{In model A, the \hbox{X-ray} jet-linked emission is proportional to the
intrinsic radio-jet emission (prior to applying beaming) and is also
itself unbeamed.}
\item{In model B, the X-ray jet-linked emission shares the
beaming factor $\delta$ that applies to the radio-jet emission.}
\item{In model C, the X-ray jet-linked emission has a lower bulk Lorentz factor
${\gamma}_{\rm x}$ (and thus a lower beaming factor ${\delta}_{\rm x}$
at low inclinations) than the radio-jet emission.}
\end{itemize} 
The models A, B, and C correspond in a general sense to cases in which 
the jet-linked X-ray emission originates in an additional accretion flow 
structure, or co-spatially with the radio-jet emission, or within the small-scale 
jet but predominantly in a less-relativistic region, respectively. Note that 
we are not attempting to model the \hbox{X-ray} emission in 
individual sources (such an approach is unproductive with these data). 
We instead conduct a statistical study to examine different plausible and 
representative physical scenarios, and present the consequent
implications including, for example, the implied fraction of RIQs and
RLQs for which the X-ray emission is jet-dominated.

In this section, we first describe the various components of the basic
model, then match the simulated luminosities to the observed primary
sample data, and then consider additional observational constraints
upon the models and determine the dominant source of X-ray emission in
each case.

\subsection{The model components}
\subsubsection{Radio emission: core and lobes}

We follow the general unification scheme described by Jackson \& Wall
(1999), in which FR~I and low-excitation emission-line FR~II radio
galaxies are the parent population of BL Lacs, while high-excitation
emission-line FR~II radio galaxies are the parent population of RLQs
(cf.~Donoso et al.~2009), but update the beaming model and luminosity
function to reflect more recent consensus. The radio source population
is presumed to be described by the luminosity function presented by
Willott et al.~(2001), which is based on low-frequency (151~MHz and
178~MHz) data and is thus relatively unbiased toward beaming, and is
computed at luminosities sufficiently high that contamination from
star-forming galaxies is negligible (Willott et al.~2001 estimate it
to be only $\sim$2\% even at their 0.1 Jy limit). The simulated
objects include radio galaxies and quasars; we select for the latter
simply by requiring the inclination to be $\theta<60^{\circ}$ (we also
require $\theta>5.8^{\circ}$ to remove highly beamed objects, which
our sample attempts to exclude\footnote{An indication that
  $\sim5^{\circ}$ is reasonable is provided by applying the
  orientation measure from Wills \& Brotherton~(1995) to the most
  radio-loud group of primary sample sources. When we also model the
  optical/UV emission ($\S$6.1.2) we find that a cutoff within
  $4^{\circ}-7^{\circ}$ is required to match the observed
  distributions of both ${\ell}_{\rm r}$ and ${\ell}_{\rm uv}$. The
  adopted $\theta>5.8^{\circ}$ helps provide a simulated ${\ell}_{\rm
    x}({\ell}_{\rm uv},{\ell}_{\rm r})$ relation ($\S$6.1.3) matching
  that observed for the primary sample.}). A sample of RIQs and RLQs
is synthesized with redshift and luminosity distributions drawn from
the luminosity function, and with randomly assigned orientations
(uniform in $\sin{\theta}$). The intrinsic core prominence (the ratio
of core-to-lobe radio flux at low frequencies; i.e., unaffected by
beaming) is taken from the Bayesian modeling of FR~II sources carried
out by Mullin \& Hardcastle (2009) and is simulated including
intrinsic scatter in core power based on their best-fit model; we also
take the typical bulk Lorentz factor $\gamma=10.5$ for core emission
from their work. The parameters we adopt are those from Table~5 of
Mullin \& Hardcastle (2009) for the model excluding low-excitation
emission-line objects.

The observed radio characteristics for a given simulated source are
calculated for the inclination of that source. The core emission from
the small-scale jet is boosted by ${\delta}^{2-{\alpha}_{\rm r}}$
(e.g., Worrall \& Birkinshaw~2006). The 1.4~GHz flux densities for the
lobes and core of each simulated source are determined assuming
${\alpha}_{\rm r}=-0.9$ for lobe emission and ${\alpha}_{\rm r}=-0.3$
for core emission (consistent with the methodology used in $\S$3.2 to
calculate ${\ell}_{\rm r}$ for the sample sources from the observed
FIRST flux densities). A limit of 1~mJy was imposed to match the FIRST
catalog detection limit, and components with simulated flux densities
below this limit were dropped from further consideration. Faint and
diffuse lobes may not always register as FIRST catalog sources; for
only $\sim$5\% of the primary sample objects with lobes is the
integrated lobe flux below 2.7~mJy, so we additionally do not include
any contribution from lobes with simulated fluxes below 2.7~mJy in the
calculated radio luminosities.

\subsubsection{Disk-dominated optical/UV emission}

Optical emission from RLQs can be generated by quasi-thermal emission
from the accretion disk and by nonthermal (e.g., synchrotron) emission
from the small-scale jet (e.g., Wills et al.~1995). In our sample, the
ordinary equivalent widths of the broad emission lines place an upper
limit upon the degree to which a featureless jet-linked component can
contribute to the optical/UV emission. However, it is not possible in
practice to determine simply the fraction of optical/UV emission that
is disk-linked in individual objects from either the optical/UV
spectra\footnote{Such an approach would have to account for the known
  trends in emission-line strength with luminosity (Baldwin effects;
  e.g., Baldwin 1977; Osmer \& Shields~1999) that occur even absent
  optical jet emission, but there is still significant scatter in
  broad line strength between otherwise similar objects and so the
  accuracy of this method is fundamentally limited for individual
  objects.} or from the radio multifrequency data.\footnote{In
  principle, the observed radio core flux could be extrapolated to the
  optical/UV band using the measured radio spectral index. However,
  even ignoring the uncertainty in the spectral index, any intervening
  spectral break (which is common, even in a simple synchrotron
  context; e.g., Worrall \& Birkinshaw 2006) essentially destroys the
  accuracy of the extrapolation.}  The observational indications that
disk emission remains dominant in broad-line RIQs and RLQs therefore
guide our models but do not provide sufficient motivation to fix
simply the fraction of optical/UV emission that is disk-linked.

We estimate the jet-linked optical/UV component from the jet radio
emission presuming ${\alpha}_{\rm ro}=-0.8$, reflective of the
tendency for the synchrotron spectrum of jet emission to steepen at
higher frequencies (e.g., Worrall \& Birkinshaw 2006).  The
disk-linked component is dependent upon physical processes not
directly incorporated into our modeling (most notably the accretion
rate); fortunately, there are radio characteristics which are thought
to be likewise sensitive to such processes, and so it is possible to
estimate the optical disk luminosity from the already-modeled radio
properties. In particular, Willott et al.~(1999) find a correlation
between low-frequency radio power and narrow-line emission in radio
galaxies which they argue indicates an underlying dependence upon
accretion rate (to feed the jet and furnish sufficient ionizing
photons incident upon the narrow line region; Willott et al.~1999 also
comment that direct illumination from the jet or jet-cloud collisions
are generally of secondary importance but are relevant in some
individual objects). Motivated by such relationships, we set the disk
to have a monochromatic luminosity at rest-frame 2500~\AA~that is a
fixed fraction of the intrinsic (unbeamed) jet monochromatic
luminosity at rest-frame 5~GHz. A log offset of $-$1.8, or a fraction
of 1.6\%, provides a reasonable match to observed ${\ell}_{\rm uv}$;
values of $-1.7$ to $-2.0$ (2--1\%) are viable. The correlation
between low-frequency radio power and narrow-line emission observed by
Willott et al.~(1999) has a 1$\sigma$ scatter of 0.5 dex; we introduce
into the model a similar spread in disk optical/UV emission at a given
intrinsic jet luminosity by adding random normal scatter with
$\sigma=0.5$ dex to the disk optical/UV emission. We emphasize that
this methodology is not reflective of a direct physical link between
jet and disk emission but rather captures the influence of additional
processes, particularly the accretion rate, on both jet and disk
emission.

For most of the simulated sources the disk emission dominates over
that from the jet at optical/UV wavelengths (by factors of a few to
several hundred), and it is only for particularly low-inclination
objects that the jet emission contributes significantly to the
optical/UV emission. This is consistent with the observed optical/UV
spectra for our sample RIQs and RLQs.

\subsubsection{Dual X-ray emission components}

{\it Chandra\/} and {\it XMM-Newton\/} observations of FR~II radio
galaxies show that their \hbox{X-ray} spectra often contain two
components, which can be interpreted as emission from both a
disk/corona and a jet-linked component (e.g., Evans et al.~2006;
Hardcastle et al.~2009). The X-ray spectra of powerful RLQs, on the
other hand, can typically be fit with a single power-law model; this
suggests they are dominated by jet emission, generally inferred to be
inverse Compton radiation (e.g., Belsole et al.~2006). Guided by prior
work and the results discussed in $\S$4 and $\S$5, we include in our
model X-ray emission from a disk/corona and also from the small-scale
jet. Our sample includes objects of modest radio loudness and
luminosity, including RIQs; since these observed properties are
dependent on intrinsic power but also viewing angle, we expect our
sample to include objects at inclinations intermediate between radio
galaxies and luminous RLQs, allowing us to investigate in our
simulations the increasing jet contribution to the X-ray continuum as
inclination decreases (with scatter reflective of intrinsic variance
in unbeamed radio core luminosities).

\begin{figure}
\includegraphics[scale=0.41]{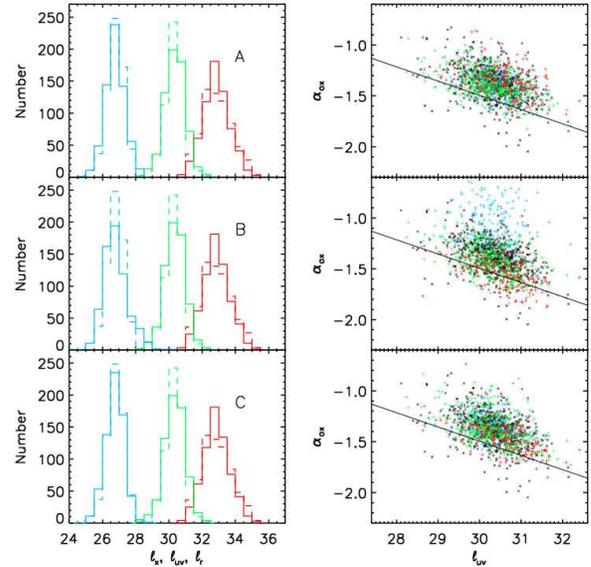} \figcaption{\small Simulated
  populations of RIQs and RLQs compared with the observed primary
  sample. The models differ principally in the degree of beaming of
  the jet-linked X-ray emission: A is unbeamed, B is beamed with
  ${\gamma}_{\rm x}=10.5$ as for the radio jet, and C is beamed to a
  lesser degree with ${\gamma}_{\rm x}=2.0$. The left-hand side shows
  histograms of radio (red), optical/UV (green), and X-ray (blue)
  luminosity, for observations (dashed) and simulation (solid). The
  right-hand side shows the ${\alpha}_{\rm ox}-{\ell}_{\rm uv}$
  relation for the primary sample (black crosses) and for the
  simulation, with objects color-coded by inclination (blue is
  $5.8^{\circ}<\theta<10^{\circ}$, green is
  $10^{\circ}<\theta<20^{\circ}$, and red is
  $20^{\circ}<\theta<60^{\circ}$). The solid line is the RQQ relation
  from Just et al.~(2007).}
\end{figure}

It is not possible to leave both the disk/corona and jet-linked X-ray
emission as simultaneously free parameters in the absence of
additional observational input (such as high-quality X-ray spectra)
that most of our sources lack. We assume the X-ray emission associated
with the disk/corona is related to the optical disk emission in the
same manner as for RQQs (note that this does not necessarily require
identical accretion structure) and can therefore be calculated using
Equation~7 from Just et al.~(2007) and the simulated optical/UV disk
emission. Random log normal scatter with $\sigma=0.3$ dex is added to
the disk/coronal \hbox{X-ray} luminosities to mimic the scatter in the
RQQ relation.

The jet-linked\footnote{The use of the term ``jet-linked'' with
  respect to model A should be understood to refer to the correlation
  between X-ray and intrinsic (unbeamed) radio-jet emission; it is
  possible but not required that this X-ray component is produced in a
  jet.} X-ray emission is unlikely to be synchrotron emission from the
same population of electrons as generates the radio-jet emission (in
contrast to the apparent situation for lower power FR~I jets; e.g.,
Chiaberge et al.~2000), as this process does not appear to generate
sufficient \hbox{X-ray} emission to match observations (e.g., Landt et
al.~2008). Inverse Compton processes must be considered; sources of
seed photons include emission from the jet (self-Compton, or SSC),
radiation from the central engine (external Compton, or EC), and the
cosmic microwave background (IC/CMB). The IC/CMB model has a strong
dependence upon redshift, which conflicts with the lack of redshift
dependence observed in our sample (see $\S$5.3), and in any case is
unlikely to produce as many seed photons as can the nucleus on the
relevant parsec-scale or smaller distances (e.g., Schwartz
2002). Comparison of the SSC and EC processes suggests the latter
dominates this close to the central engine (e.g., Sokolov \& Marscher
2005).

The precise physical parameters governing X-ray jet emission (such as
the size of the emission region or the magnetic-field strength) likely
vary significantly from object to object. Since we are interested in
general trends rather than the specific details of a given source, we
assume a ``standard'' intrinsic ratio between radio and X-ray
jet-linked emission, which we set to obtain consistency with the
observed primary sample X-ray luminosities. The intrinsic jet-linked
X-ray emission is then modified in models B and C for the observer
based on the bulk velocity and inclination of the jet, with presumed
dominant EC emission boosted by ${\delta}^{3-2{\alpha}_{\rm x}}$ (an
additional factor of $1-\alpha$ beyond the radio synchrotron emission;
e.g., Dermer 1995), with ${\alpha}_{\rm x}=-0.3$ for model B taken to
match the radio value and ${\alpha}_{\rm x}=-0.5$ for model C taken to
match the typical energy index of RLQs (e.g., Page et
al.~2005). \hbox{X-ray} censoring is added to the simulated sample in
quasi-random fashion, with fainter objects more likely to be labeled
as upper limits. For simplicity, this is accomplished through matching
the \hbox{X-ray} detection fraction within the top/bottom half of the
primary sample ranked by ${\ell}_{\rm x}$.  We randomly select
88\%/77\% of objects with simulated \hbox{${\ell}_{\rm x}>26.8/<26.8$}
as detections, and then the simulated detection rate matches that of
the primary sample overall and within each \hbox{X-ray} luminosity
bin.

\subsection{Comparison to observed luminosities}

In principle, this modeling process will produce a simulated
population of RIQs and RLQs that can be directly compared to the
observed primary sample data. However, the nature of our sample
introduces a few complicating effects that need to be considered. One
important parameter is the limiting magnitude for inclusion within the
primary sample. Since we prefer to retain as many observed objects as
possible, we do not impose a magnitude limit on the observed
objects. However, only $\sim$5\% of primary sample RIQs and RLQs have
$m_{\rm i}>19.95/20.13$ and $m_{\rm i}>20.60/21.08$ for $z<1/z\ge1$
spectroscopic and photometric quasars, respectively, and so these
values are used as effective magnitude cutoffs for the
simulations. The numbers of simulated ``spectroscopic'' and
``photometric'' objects (distinguished in the model only by the
applied magnitude cutoffs) are set to match those of the primary
sample.

\begin{figure*}
\includegraphics[scale=0.80]{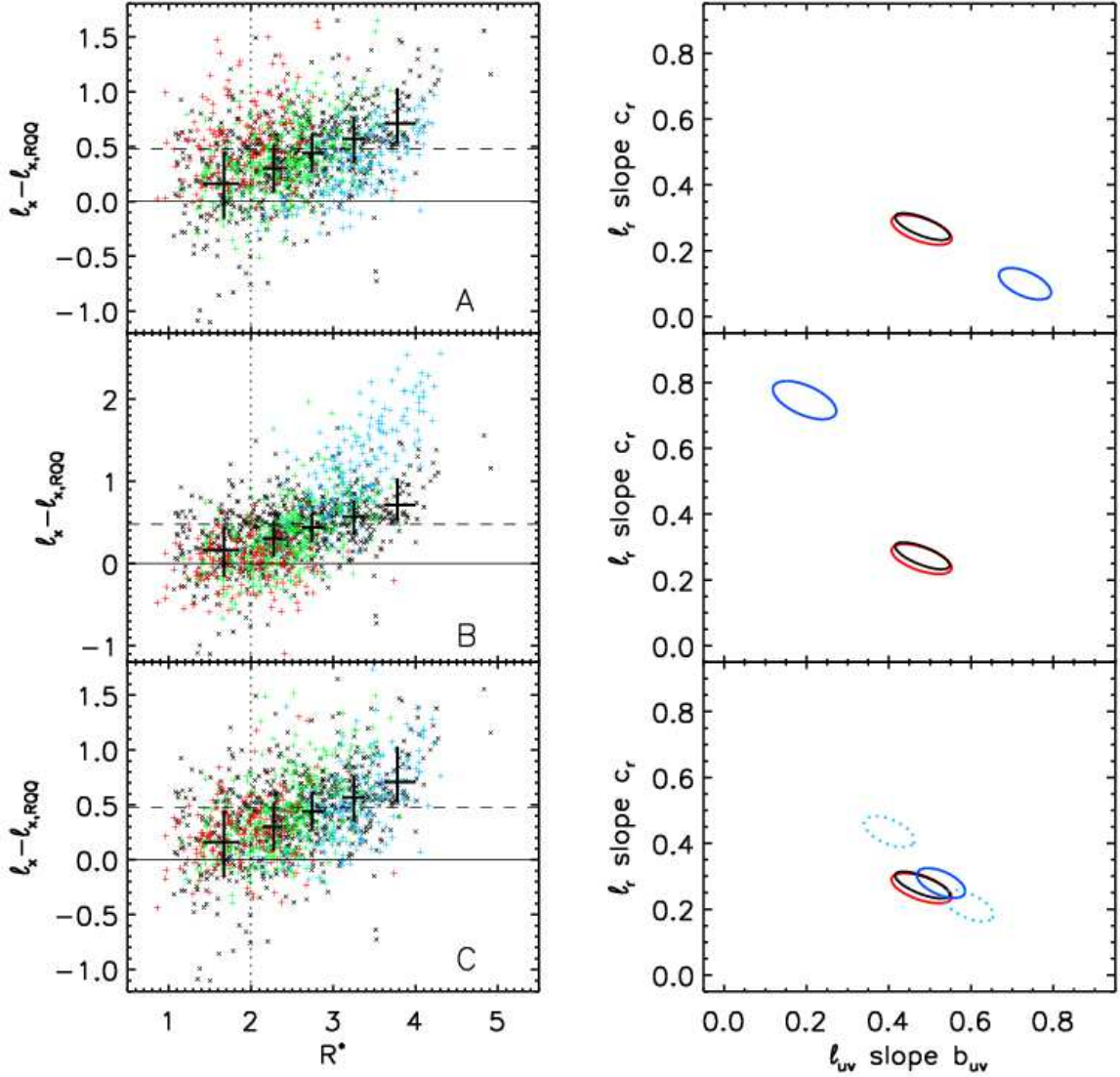} \figcaption{\small Simulated
  populations of RIQs and RLQs compared with the observed primary
  sample. The models differ principally in the degree of beaming of
  the jet-linked X-ray emission: A is unbeamed, B is beamed with
  ${\gamma}_{\rm x}=10.5$ as for the radio jet, and C is beamed to a
  lesser degree with ${\gamma}_{\rm x}=2.0$. The left-hand side shows
  ``excess'' X-ray luminosity relative to comparable RQQs (see
  Figure~7a) for the observed sample (small black crosses) and for the
  simulation, with objects color-coded by inclination (blue is
  $5.8^{\circ}<\theta<10^{\circ}$, green is
  $10^{\circ}<\theta<20^{\circ}$, and red is
  $20^{\circ}<\theta<60^{\circ}$). The large black crosses give the
  median and interquartile range for the primary sample. The
  right-hand side shows joint 90\% confidence ellipses for X-ray
  luminosity fit as a function of optical/UV and radio luminosity, for
  the primary sample (red), the full sample (black), and each model
  (blue). The dotted blue ellipses are for models C1.5 and C3.0 with
  ${\gamma}_{\rm x}=1.5$ and 3.0. See $\S$6.3 for discussion.}
\end{figure*}

The normalization (i.e., number of objects) of the simulated
population should be able to be set solely based on the sky coverage
fraction of the observed sample, but this is difficult to evaluate for
the various X-ray missions and their complex instrumental fields of
view. Matched total numbers of simulated and sample objects are not
critical for conducting a comparison (it is the distribution of
luminosities that is of interest); nonetheless, it is convenient for
examining the results of the simulations if the simulated population
is of the same size as the observed sample. Since we are interested in
how X-ray luminosity is related to radio and optical/UV luminosities
and unconcerned with the sky density of (selected) RIQs and RLQs, we
simulate a large population and then draw from it to obtain an equal
number of simulated and observed objects. We also simulate variability
due to the radio/optical observations being carried out
non-simultaneously with the \hbox{X-ray} observations by adding random
normal scatter to the simulated ${\ell}_{\rm r}/{\ell}_{\rm uv}$
values (after calculating all interdependent quantities) with
amplitudes of 0.0792/0.114 (the same uncertainties as used with
fitting; see $\S$5), but inclusion of this minor effect does not
significantly impact the modeling.

Another relevant effect is the redshift-dependent efficiency of the
SDSS color-selection targeting algorithm and of the photometric quasar
classification algorithm. The median redshift of the simulated
population is 1.79/1.88 for the spectroscopic/photometric magnitude
cutoffs, higher than that of the observed primary sample (for which
the median redshift is 1.41). Although the redshift dependencies of
quasar color-selection techniques presumably explain most of the
discrepancy between the observed and simulated median redshifts, there
are a few more factors that could act to decrease the median redshift
of the observed primary sample. For example, there is a bias toward
lower redshifts for the minority of sources that were \hbox{X-ray}
targeted (rather than serendipitous) and a possible bias against
spectroscopic identification of fainter objects for which measuring
broad emission lines is more difficult. Fortunately, our goal of
evaluating \hbox{X-ray} luminosity as a function of radio and
optical/UV luminosity does not depend on redshift (e.g., see $\S$5.3),
and so it is not necessary to account for all the various selection
effects that may influence the difference in redshift distribution. We
match in redshift and number simultaneously by drawing from the large
simulated population those objects with redshifts close to the
observed sample. Note that the underlying basic model could be
falsified at this stage were the simulated radio luminosities
(specified primarily by the adopted luminosity function, the adopted
intrinsic jet/lobe properties, and the random inclination) to disagree
with observation. However, a KS test gives a probability of
\hbox{$p=0.21$}, indicating that the simulated and observed samples
cannot be considered to differ significantly in their radio luminosity
distributions. The median radio luminosity for the primary sample is
${\ell}_{\rm r}=32.83$ with a standard deviation of 0.89, while that
of the simulated sample is ${\ell}_{\rm r}=32.80$ with a standard
deviation of 0.79. It would also be possible at this stage to falsify
the nature of the presumed correlation between intrinsic radio jet
power and optical/UV disk luminosity if the distribution of simulated
optical/UV luminosities were to disagree with observation (the median
can always be made to agree by adjusting the ratio, but the
distribution reflects the lack of beaming in the optical/UV
luminosities, in contrast to the radio). However, a KS test indicates
that the simulated and observed samples are not inconsistent
($p=0.41$) in their optical/UV luminosity distributions. The median
optical/UV luminosity for the primary sample is ${\ell}_{\rm
  uv}=30.36$ with a standard deviation of 0.57, while that of the
simulated sample is ${\ell}_{\rm uv}=30.33$ with a standard deviation
of 0.63. The radio-loudness distributions are also consistent (KS
probability $p=0.15$) with observed/simulated median values of
2.50/2.46 and standard deviations of 0.74/0.70. These radio and
optical/UV properties are the same for each of models A, B, and C.

For any of models A, B, or C to constitute viable descriptions of the
X-ray emission from RIQs and RLQs, it is necessary (but not
sufficient) that the distribution of simulated \hbox{X-ray}
luminosities not disagree with observation. The \hbox{X-ray}
disk/corona emission is completely specified by the optical/UV disk
emission, and so it is the additional jet-linked \hbox{X-ray} emission
that must be accurately modeled. The median value of the overall
simulated \hbox{X-ray} luminosity can always be made to agree with
observations by adjusting the ratio between the jet-linked
\hbox{X-ray} component and the intrinsic radio-jet luminosity, but the
distribution is strongly influenced by the degree to which the
jet-linked emission is beamed. The log offsets between intrinsic
radio-jet luminosity (prior to applying beaming) and intrinsic
\hbox{X-ray} luminosity of the jet-linked component (prior to applying
beaming to models B and C) required to match simulated with observed
median \hbox{X-ray} luminosities are $-5.35, -7.25,$ and $-7.25$ for
models A, B, and C, respectively. Random log-normal scatter is added
to these offsets; we find empirically that a somewhat better match to
the ${\ell}_{\rm x}$ scatter observed for the primary sample is
obtained if the magnitude of this scatter is greater at lower
simulated redshifts (0.7/0.3 for $z<1/z\ge1$), but this could
potentially be related to the increased fraction of \hbox{X-ray}
bright targeted objects at lower redshifts within the primary sample
($\S$5.4.2), so we do not ascribe physical significance to this
result. The \hbox{X-ray} bulk Lorentz factor in model C is
${\gamma}_{\rm x}=2.0$; for reference, this corresponds to
${\delta}_{\rm x}=3.6$ for an inclination of $\theta=5^{\circ}$
(rather than ${\delta}=11.4$ as with the ${\gamma}=10.5$ that applies
to the radio jet). For model C, \hbox{X-ray} bulk Lorentz factors of
${\gamma}_{\rm x}\simeq1-3$ can also match the primary sample
\hbox{X-ray} luminosity distributions, but as shown in $\S$6.3,
${\gamma}_{\rm x}=2.0$ produces the best agreement to the observed
${\ell}_{\rm x}({\ell}_{\rm uv},{\ell}_{\rm r})$ relation. KS tests
(treating upper limits as detections, but the simulated censoring
matches the data) give probabilities of $p=0.37$ and $p=0.30$ for
models A and C, respectively, indicating that the simulated and
observed samples are not inconsistent in their \hbox{X-ray} luminosity
distribution for these models. However, the KS test probability for
model B is $p=0.0052$; the simulated distribution of \hbox{X-ray}
luminosities differs significantly from that observed. In particular,
a tail to large \hbox{X-ray} luminosity values in the simulated
population for model B does not agree with the data, which appear to
show at most a modest skew toward higher \hbox{X-ray} luminosities
(see also $\S$5.4.3). This tail in the model is more pronounced at
\hbox{X-ray} than radio frequencies because the dependence upon the
$\delta$ beaming factor is greater (by $1-\alpha$ in the exponent) for
the \hbox{X-ray} emission. (Model B would be an even worse match to
the primary sample if the ${\ell}_{\rm x}$ medians were equalized.)
These results may be seen in the left-hand panels of Figure~18.

All three models obviously produce (by design) enhanced \hbox{X-ray}
emission relative to RQQs of comparable optical/UV luminosity, as can
be seen in the right-hand panels of Figure~18 and as given in
Table~10. However, the manner in which they do so differs. In these
figures the simulated objects are color-coded by inclination (blue for
$5.8^{\circ}<\theta<10^{\circ}$, green for
$10^{\circ}<\theta<20^{\circ}$, and red for
$20^{\circ}<\theta<60^{\circ}$). For model A, there is essentially no
inclination dependence to the X-ray excess, since the X-ray luminosity
is unbeamed. For model B, there is a strong inclination dependence,
with the lowest inclination RLQs being extremely X-ray bright (to a
degree that does not match observations). Note that these simulated
low-inclination objects are still expected to be disk-dominated at
optical/UV frequencies (as also applies to models A and C) and are
thus not blazars, so the comparison to the observed primary sample
(from which all identified blazars have been removed) is properly
matched. For model C, there is a definite but modest dependence upon
inclination. Simultaneous consideration of the distribution of radio,
optical/UV, and \hbox{X-ray} luminosities, following the analysis
performed in $\S$4 and $\S$5 for observed RIQs and RLQs, provides
additional insight helpful to evaluating the feasibility of the three
models.

\subsection{Modeling results}

We now assess the accuracy with which these models (generated as
described in $\S$6.1, $\S$6.2, and Table~9) reproduce the observed
dependence of \hbox{X-ray} luminosity upon radio and optical/UV
luminosities. We first examine how the ``excess'' X-ray luminosity
(X-ray luminosity minus that of a RQQ of matched optical/UV
luminosity) for the simulated RIQs and RLQs depends upon radio
loudness for the three models. In Figure~19, left-hand side, the
simulated objects are again color-coded by inclination, and the
primary sample data are plotted along with the median and
interquartile range measurements from Figure~7a. It may be observed
that model A does not appear to capture accurately the rise in X-ray
brightness with increasing radio loudness. This is because in model A
the X-ray luminosity is unbeamed, whereas the radio luminosity is
beamed, and so the most radio-luminous and radio-loud objects tend to
have low inclinations (are color-coded blue) but are not \hbox{X-ray}
brighter than higher inclination objects. Conversely, model B
overpredicts the degree to which \hbox{X-ray} brightness depends on
increasing radio loudness, or inclination. It is only for model C that
the X-ray brightness compared to RQQs may be seen to increase with
radio loudness in a manner analogous to that observed.

Modeling X-ray luminosity as a joint function of optical/UV and radio
luminosity, as done for the observed RIQs and RLQs in $\S$5.2, makes
clear the differing manner in which the three models are dependent on
optical/UV and radio luminosity (see Table~10). The right-hand side of
Figure~19 shows the joint 90\% confidence ellipses for the primary
(red) and for the full (black) observed samples, along with the
calculated result for each model (blue). For model A, the X-ray
luminosity is strongly dependent on optical/UV luminosity but only
weakly dependent on radio luminosity. The best-fit relation for the
population simulated with model A is ${\ell}_{\rm x} =
-0.142+0.730\times{\ell}_{\rm uv}+0.100\times{\ell}_{\rm r}$. This
result reflects the underlying dependence of both the optical disk
emission (and thereby the disk/corona \hbox{X-ray} component) and the
``jet-linked'' X-ray component on the intrinsic (unbeamed) radio-jet
luminosity. The parameterization of \hbox{X-ray} luminosity upon
optical/UV and radio luminosity in model A is inconsistent with the
fits to the primary and full samples. For model B, the X-ray
luminosity is strongly dependent upon radio luminosity but only weakly
dependent on optical/UV luminosity. This result is due to the
increased dominance of the \hbox{X-ray} continuum by jet-linked
emission for a large fraction of the simulated sources, particularly
those at low inclinations. The best-fit relation for the population
simulated with model B is ${\ell}_{\rm x} =
0.099+0.195\times{\ell}_{\rm uv}+0.746\times{\ell}_{\rm r}$. This
reflects the mutual beaming of the radio and \hbox{X-ray} jets, but
these parameters are also inconsistent with observations. For model C,
the X-ray luminosity is dependent upon both optical/UV and radio
luminosities in a manner similar to that observed, as illustrated by
the overlapping confidence ellipses. The best-fit relation for the
population simulated with model C is ${\ell}_{\rm x} =
-0.098+0.526\times{\ell}_{\rm uv}+0.279\times{\ell}_{\rm r}$. This
result reflects the milder beaming with ${\gamma}_{\rm x}=2.0$ that
places model C in some sense intermediate between model A and model
B. For comparison, model C with ${\gamma}_{\rm x}=1.5/3.0$ (models
C1.5/C3; best-fit relations given in Table~10 and shown as dotted blue
ellipses in Figure~19) bracket the ${\gamma}_{\rm x}=2.0$ result but
do not provide as close a match to observations.

\begin{figure}
\includegraphics[scale=0.41]{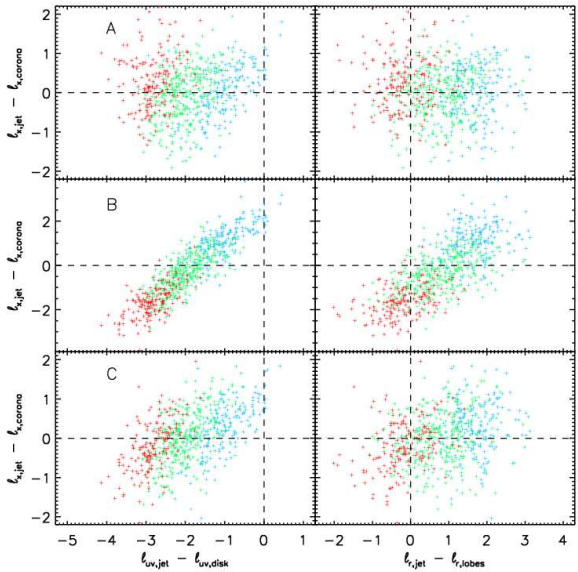} \figcaption{\small Simulated
  populations of RIQs and RLQs. The models differ principally in the
  degree of beaming of the jet-linked X-ray emission: A is unbeamed, B
  is beamed with ${\gamma}_{\rm x}=10.5$ as for the radio jet, and C
  is beamed to a lesser degree with ${\gamma}_{\rm x}=2.0$. In all
  panels objects are color-coded by inclination (blue is
  $5.8^{\circ}<\theta<10^{\circ}$, green is
  $10^{\circ}<\theta<20^{\circ}$, and red is
  $20^{\circ}<\theta<60^{\circ}$). The left-hand side shows X-ray
  luminosity components versus optical/UV luminosity components, while
  the right-hand side shows X-ray luminosity components versus radio
  luminosity components.}
\end{figure}

The simulated populations may be utilized to evaluate the dominant
emission component at each frequency, and in particular to examine the
degree to which X-ray luminosity is dominated by the jet-linked
component, as a function of inclination, for the different models (see
Figure~20). In nearly all cases, the optical/UV luminosity is
disk-dominated. For most sources, and for all the low-inclination
sources, the radio luminosity is jet-dominated. In each of models A,
B, and C, the X-ray luminosity is dominated by disk/corona emission
for a sizeable fraction of objects; there is agreement that despite
the greater median X-ray luminosities of RIQs and RLQs relative to
RQQs, there are many individual objects for which a disk/corona
analogous to that in RQQs is sufficient to produce the majority of
observed \hbox{X-ray} emission. For models B and C, the degree to
which the X-ray emission is jet-dominated is correlated with the
degree to which the radio emission is jet-dominated (to a lesser
degree in model C, for which the beaming factor is lower). For the
preferred model C, the majority of objects with inclinations less than
$20^{\circ}$ have X-ray luminosities dominated by jet-linked emission.

Despite the general nature of the scenarios considered under models A,
B, and C, these simulations provide insight into the degree to which
the ``excess'' X-ray luminosity in RIQs and RLQs is jet-linked and
beamed. It seems likely that some beaming must be present, as model A
does not satisfactorily match observations; this conclusion is
consistent with the results of previous work (e.g., Browne \& Murphy
1987; Shastri et al. 1993). It also appears that the jet-linked
\hbox{X-ray} emission is beamed to a lesser degree than affects the
radio jet, as model B does not satisfactorily match observations; this
result has also been suggested by other work (e.g., Dou \&
Yuan~2008). The success of model C suggests that the majority of the
jet-linked X-ray emission is not produced cospatially with the radio
jet emission, but likely primarily originates in a somewhat slower jet
region. The preferred model C implies that high signal-to-noise X-ray
spectroscopy of RIQs or RLQs established (e.g., from radio properties)
as possessing intermediate inclinations should result in some such
objects showing X-ray spectra in which an unabsorbed
disk/corona-linked component and a jet-linked component are
simultaneously apparent. This is consistent with observational results
(although additional \hbox{X-ray} studies would help establish this
point); for example, Sambruna et al.~(2006) describe three RLQs with
concave broken power law \hbox{X-ray} spectra, which they attribute to
dual jet/disk-corona components.

Future work with significantly larger samples and denser coverage of
the ${\ell}-z$ plane\footnote{We are currently carrying out {\it
    Swift\/} snapshot observations of high-luminosity RLQs (Jianfeng
  Wu et al., in preparation) that will help fill in the ${\ell}-z$
  plane.}  could potentially investigate whether the small-scale jet
has velocity structure (as in a fast-spine/slow-sheath model; see,
e.g., Jester et al.~2006b), most simply through considering two X-ray
jet components with differing bulk Lorentz factors. Modeling
deficiencies that are unlikely to impact these results but could be
addressed to improve the precision of comparison to data include
limited consideration of selection effects and ad-hoc normalization of
simulated sample size.

\subsection{Alternative model parameters}

Although we have not conducted an exhaustive search of the
multidimensional parameter space, the presented models (A, B, C, C1.5,
and C3.0) together cover a large range of plausible possibilities, and
from this set model C best matches observations. It is of interest to
evaluate whether a modified set of models would still yield
qualitatively similar results. Here, we briefly describe the effects
of altering selected model parameters in a reasonable fashion. Minor
adjustments to the log offsets between intrinsic radio-jet luminosity
and intrinsic jet-linked X-ray luminosity are generally made without
comment below; other model parameters are as given in $\S$6.1,
$\S$6.2, and Table~9, unless otherwise stated.

The credible interval for $\gamma$ from the adopted Mullin \&
Hardcastle (2009) model ranges from 3.20 to 14.05. The higher $\gamma$
produces an acceptable match to the primary sample radio and
optical/UV luminosities (KS probabilities of 0.59/0.26/0.13 for
${\ell}_{\rm r}/{\ell}_{\rm uv}/R^{*}$) and can match the X-ray
luminosity of models A and C but not B (KS probabilities of
\hbox{0.21/$<0.01$/0.49} for A/B/C). The lower $\gamma$ does not
easily give a good match to both the radio and optical/UV
luminosities. For a ratio of optical/UV disk emission to unbeamed
radio emission of $-2.0$ (logarithmic units) and a minimum inclination
of 4$^{\circ}$, the KS probabilities are 0.059/0.075/0.0076 for
${\ell}_{\rm r}/{\ell}_{\rm uv}/R^{*}$. The simulated median X-ray
luminosities may be matched to observations with resulting KS
probabilities of 0.35/$<0.01$/0.10 for models A/B/C. In this scenario
the majority of the simulated objects have inclinations with
${\theta}>20^{\circ}$, but the \hbox{X-ray} brightness of the
lower-inclination objects cause the increase in ``excess'' X-ray
luminosity with increasing radio-loudness for model B and (to a lesser
degree) model C to be more rapid than observed. For model C under this
scenario, the dependence of \hbox{X-ray} luminosity upon optical/UV
and radio luminosity is weaker and stronger ($b_{\rm uv}=0.31$ and
$c_{\rm r}=0.55$), respectively, than holds for the primary sample.

Willott et al.~(2001) consider two additional models for their radio
luminosity function in addition to the version we use (see their
$\S$3.2). If instead a population of RIQs and RLQs is simulated based
on their model which contains a steeper high-redshift decay (their
Equation~11), then an acceptable match to the primary sample may be
obtained (KS probabilities of 0.24/0.12/0.42 for ${\ell}_{\rm
  r}/{\ell}_{\rm uv}/R^{*}$) with a ratio of optical/UV disk emission
to unbeamed radio emission of $-1.9$. The KS probabilities for
matching simulated versus observed X-ray luminosity are
0.065/$<0.01$/0.35 for models A/B/C. Model B is again not a good
match. If instead the Willott et al.~(2001) model which becomes
constant at high redshift (their Equation~12) is used to simulate a
population of RIQs and RLQs, the match to the primary sample is again
acceptable (KS probabilities of 0.082/0.19/0.26 for ${\ell}_{\rm
  r}/{\ell}_{\rm uv}/R^{*}$). The KS probabilities for matching
simulated versus observed X-ray luminosity are 0.10/$<0.01$/0.20 for
models A/B/C. Model B remains a poor match. Additionally, the median
simulated redshift (prior to matching redshifts to the primary sample)
is 1.95/2.22 for the spectroscopic/photometric magnitude cutoffs,
higher than that of the primary sample (1.41) and also higher than
that simulated with the version of the Willott et al.~(2001) model we
use (1.79/1.88).

We also consider a scenario in which the X-ray jet emission is beamed
as ${\delta}^{2-{\alpha}_{\rm x}}$ (similarly to the radio jet
emission) rather than as ${\delta}^{3-2{\alpha}_{\rm x}}$. This might
apply if, for example, the X-ray jet emission were to be dominated by
synchrotron emission from a secondary population of high-energy
electrons. This change only affects models B and C. Maintaining
\hbox{${\alpha}_{\rm x}=-0.3/-0.5$} and adjusting the X-ray jet
offsets to $-6.45/-6.50$, the KS probabilities are 0.065/0.76 for
models B/C. With less extreme X-ray beaming model B can match the
observed distribution of X-ray luminosities. However, model B still
predicts a greater fraction of X-ray-bright objects than is observed,
and the increase in ``excess'' X-ray luminosity with increasing
radio-loudness is also more rapid than observed. Model C remains a
superior explanation of the data. For model C under this scenario, the
dependence of \hbox{X-ray} luminosity upon optical/UV and radio
luminosity is moderately stronger and weaker ($b_{\rm uv}=0.58$ and
$c_{\rm r}=0.23$), respectively, than holds for the primary sample.

\section{Summary}

The primary results of our analysis of the nuclear \hbox{X-ray}
properties of RIQs and RLQs are the following:

1. Sample: We have compiled a sample of 188 RIQs and 603 RLQs
(primarily by matching optically selected SDSS quasars to the FIRST
survey, taking into account extended radio emission) with high-quality
archival X-ray coverage by {\it Chandra\/}, {\it XMM-Newton\/}, or
{\it ROSAT\/}. The full sample is almost unbiased with respect to
\hbox{X-ray} properties, has a high (85\%) detection rate, and can be
utilized to investigate the nature and origin of X-ray emission in
RIQs and RLQs. The sample size is significantly larger than earlier
studies of RLQs and provides superior coverage of the
luminosity-redshift plane.

2. Trends: We calculate the ratio of X-ray luminosity in RIQs and RLQs
relative to that of comparable RQQs and determine how this ratio
increases with both radio loudness and luminosity. This ratio of
``excess'' X-ray luminosity ranges from $\sim$0.7--2.8 for RIQs
through the canonical $\sim$3 for RLQs to $\simgt$10 for strongly
radio-loud ($R^{*}>4$) or luminous (${\ell}_{\rm r}>35$) objects. We
also carry out fits to the X-ray luminosity as a function of both
optical and radio luminosity, which are useful for determining the
``typical'' X-ray luminosity for any given RIQ or RLQ. The
\hbox{X-ray} emission in RIQs is not significantly dependent upon
radio luminosity. We quantify the manner in which RLQs become more
X-ray luminous for large $R^{*}$ or ${\ell}_{\rm r}$ or for flat radio
spectra. Finally, we do not find any significant redshift dependence
in the properties of RIQs and RLQs (implying, e.g., that IC/CMB
jet-linked emission does not contribute substantially to the nuclear
\hbox{X-ray} continuum).

3. Models: We conduct Monte Carlo simulations based on a low-frequency
radio luminosity function to which we add a randomly inclined
relativistic jet. The optical disk emission is successfully modeled as
correlated with the intrinsic (unbeamed) radio-jet luminosity, and the
X-ray emission contains both disk/corona and jet-linked components. We
consider three models for the non-coronal core X-ray luminosity
component, and conclude that the jet-linked X-ray emission is likely
beamed but with a lesser bulk Lorentz factor than applies to the
radio-jet emission. The alternative possibilities of unbeamed
\hbox{X-ray} emission and of \hbox{X-ray} emission with
$\gamma\sim10.5$ as for the radio jet do not appear to match
adequately the observed data. For the preferred model, the radio
emission is mostly jet dominated, the optical/UV emission is almost
exclusively disk dominated, and the X-ray emission is split between
disk/corona and jet-linked components with the jet becoming
increasingly dominant at low inclinations.

\acknowledgments

We thank the anonymous referee for many constructive suggestions that
improved this paper. We thank B.~Luo, D.~A.~Rafferty, and Y.~Xue for
assistance with photometric redshifts and for making their catalogs
available prior to publication. We thank M.~Eracleous and B.~Kelly for
useful discussions. BPM, WNB, and DPS thank NASA ADP grant NNX10AC99G
for support; DPS also thanks NSF grant AST06-07634. Funding for the
SDSS and SDSS-II has been provided by the Alfred P. Sloan Foundation,
the Participating Institutions, the National Science Foundation, the
U.S. Department of Energy, the National Aeronautics and Space
Administration, the Japanese Monbukagakusho, the Max Planck Society,
and the Higher Education Funding Council for England. The SDSS Web
Site is http://www.sdss.org/.


\section*{A. Details of primary sample radio/optical/X-ray matching process}

We here describe the details of the selection, screening, and matching
process used to generate the spectroscopic sample. Many of these steps
are identically used in the construction of the photometric
sample. Together, the spectroscopic and photometric samples make up
the primary sample of RIQs and RLQs used throughout this paper.

The SDSS quasar survey is $\sim$95\% complete for unresolved sources
(for objects categorized as quasars on the basis of optical spectra)
and $\sim$89\% complete overall (Vanden Berk et al.~2005); the DR5
quasar catalog of Schneider et al.~(2007) consists of quasars with
strong broad emission lines (of width $>$1000~km~s$^{-1}$) and by
design does not include BL Lacs with featureless spectra (e.g.,
Plotkin et al.~2008, 2010). It does, however, include broad-absorption
line (BAL) quasars, and it is now clear (e.g., Becker et al.~2000,
2001; Menou et al.~2001) that a small fraction ($\simlt$8\% at high
radio luminosities; e.g., Shankar et al.~2008) of RLQs have BALs and
are, like BAL RQQs, relatively \hbox{X-ray} weak due to intrinsic
absorption (e.g., Brotherton et al.~2005; Gibson et al.~2008; Miller
et al.~2009). Since we are interested in intrinsic emission and not
absorption, we exclude from our sample any objects included in the
SDSS BAL catalog of Gibson et al.~(2009) and also any RLQs we are
aware of as possessing BALs not observable in the SDSS spectrum (e.g.,
J100726.10+124856.2, the $z=0.24$ object PG~1004+130 identified as a
BAL RLQ by Wills et al.~1999). Through this process 22 BAL quasars are
identified and dropped from further consideration. Contamination from
low-redshift ($z<1.7$) RLQs not identifiable as having BALs due to
SDSS bandpass limitations (e.g., spectroscopic quasars lacking
coverage of the \hbox{1400--1550~\AA}~region, which includes the
strong C~IV BAL transition) is estimated to be $\sim$2\% of the entire
sample,\footnote{This estimate assumes $\sim$8\% (e.g., Shankar et
  al.~2008) of the 210 spectroscopic RIQs and RLQs with $z<1.7$ are
  BAL quasars, out of a total sample of 791 objects; BAL contamination
  within the photometric sample is addressed separately in $\S$2.1.2.}
and a small number of \hbox{X-ray} weak outliers would not materially
impact our results below. Objects with apparent heavy intrinsic
reddening are sometimes also associated with \hbox{X-ray} absorption
(e.g., Hall et al.~2006). The quantity ${\Delta}(g-i)$ (plotted in
Figure 2) is the $g-i$ color of an object (corrected for Galactic
extinction) less the median $g-i$ quasar color at the redshift of that
object; negative ${\Delta}(g-i)$ values correspond to objects bluer
than the norm and positive ${\Delta}(g-i)$ values indicate redder
colors (e.g., Richards et al.~2001). For reference, the median
relative color for our sample of spectroscopic SDSS RIQs and RLQs is
${\Delta}(g-i)=0.07$ (there is a general tendency for RLQs to have
somewhat redder colors than RQQs; e.g., Labita et al.~2008). Almost
all non-reddened objects have ${\Delta}(g-i)<1$ while intrinsically
reddened (and potentially absorbed) objects form a tail in the
distribution that extends past this value (e.g., Hall et al.~2006), so
objects with ${\Delta}(g-i)>1$ are culled from the sample. This cut
removes 8 objects that would otherwise be included (an additional 4 of
the already excluded BAL quasars also have
${\Delta}(g-i)>1$). Finally, 3 objects possess radio spectra
(described in $\S$3.3) indicating they are GHz-peaked spectrum (GPS;
e.g., O'Dea 1998) sources, which are generally considered to be young
sources and have X-ray properties that are not necessarily
representative of RLQs in general (e.g., Siemiginowska et al.~2008);
these GPS sources are therefore culled and not considered further.

Extended radio sources (with jets or lobes) may have separate entries
in the FIRST source catalog for each detected component, and
lobe-dominated sources do not always possess a detected core
component. It is therefore necessary to match to the entire
environment surrounding each quasar to recover all associated emission
(e.g., Best et al.~2005; Lu et al.~2007). Very few radio sources that
lack a core component have lobe-to-lobe angular sizes greater than
90$''$ (e.g., de Vries et al.~2006), so we search for radio emission
within 90$''$ of each optically-selected quasar. We considered as
candidates for radio-detected quasars those objects with a radio
source within 2$''$, referred to hereafter as the core, or two or more
radio sources within 2$''< \theta < 90''$ of which at least one pair
has component angular orientations relative to the core differing by
more than 90$^{\circ}$ (where a 180$^{\circ}$ separation would
indicate components on directly opposite sides of the core),
classified as potential lobes pending additional review. The potential
lobes were then matched to the SDSS photometric catalog, and if they
had an optical counterpart (unless that apparent counterpart was a
spectroscopically-classified star) they were flagged as intruding
background sources and not considered further in determining the radio
characteristics of that quasar; if this process left only one
potential lobe and no core then the object was eliminated as not
radio-detected. All objects with remaining potential lobes were then
examined visually to screen further for misclassifications (for
example, unrelated background double-lobed radio sources are readily
identifiable by eye as intruding but are difficult to categorize
correctly automatically). A few cases in which an obviously-associated
extended component had been mistakenly flagged as background due to a
randomly coincident SDSS optical source were also corrected. Visual
screening also identified four objects with likely associated radio
components outside the 90$''$ consideration radius
(J093200.08+553347.4, J112956.53+364919.2, J094745.14+072520.6, and
J142735.60+263214.5, with maximal lobe offset from the core of 93$''$,
99$''$, 117$''$, and 148$''$, respectively); since we wish to take
into account all lobe emission from objects already included in the
sample, these components were added to the total radio fluxes for
these objects. Visual screening also confirmed that the FIRST catalog
contains entries for all obvious components in all sources included in
the sample [with only one exception: we added a core component to
  J170441.37+604430.5 (3C~351) based on examination of the FIRST and
  other radio maps]. The FIRST survey is not particularly sensitive to
large-scale regions of low surface brightness, so diffuse lobes may
not be detected or may have their fluxes underestimated (e.g., Lu et
al.~2010). The NRAO VLA Sky Survey (NVSS; Condon et al.~1998) provides
lower angular resolution coverage that complements the high angular
resolution imaging of FIRST; 1.4~GHz fluxes from the NVSS catalog were
used for a handful of obviously extended FIRST sources.

The 312 RIQs and RLQs in the spectroscopic sample all possess
sensitive \hbox{X-ray} coverage from {\it Chandra\/}, {\it
  XMM-Newton\/}, or {\it ROSAT\/}. For {\it Chandra\/}, we checked all
public non-grating ACIS-S or ACIS-I observations with exposures longer
than 1~ks and off-axis angles of less than 12$'$ (note that each one
of the ACIS CCDs covers an area of approximately 8.4$'$ by
8.4$'$). For {\it XMM-Newton\/}, we considered only those observations
with exposure times greater than 1~ks and off-axis angles of less than
15$'$. For {\it ROSAT\/}, we initially considered unfiltered PSPC
observations with exposure times greater than 2~ks and off-axis angles
less than 40$'$. All \hbox{X-ray} observations were examined to screen
out cases in which the quasar did not fall on the detector, was too
close to another bright source, was located within an instrumental
artifact, or was otherwise unsuitable for further analysis. Due to
their greater \hbox{X-ray} sensitivity, {\it Chandra\/} or {\it
  XMM-Newton\/} observations were prioritized over {\it ROSAT\/}
observations. In cases where both {\it Chandra\/} and {\it
  XMM-Newton\/} provide coverage, the observation with the greatest
\hbox{X-ray} signal-to-noise ratio (estimated based on exposure time,
off-axis angle, detector efficiency, and average background) was
used. {\it ROSAT\/} observations were ranked by \hbox{X-ray}
signal-to-noise ratio, and lower-quality observations were
discarded. The threshold value for discarding {\it ROSAT\/}
observations was chosen so as to provide a large sample size while
maintaining a relatively high detection fraction. In cases where
multiple {\it ROSAT\/} observations were available, the observation
with the greatest \hbox{X-ray} signal-to-noise ratio was used. Despite
their lower sensitivity, the {\it ROSAT\/} observations are useful for
this project, as they provide a large area of sky coverage beyond that
available solely from {\it Chandra\/} and {\it XMM-Newton\/}
observations. Note that this method is unbiased with respect to
intrinsic quasar properties (including \hbox{X-ray} brightness) as it
is based only on the expected quality of the \hbox{X-ray}
observations, so it should not lead to any biases or systematic errors
in our analyses.

\section*{B. Using GALEX data to improve photometric redshifts}

We here describe the process through which GALEX observations were
used to identify and discard 34 RIQs and RLQs with likely incorrect
photometric redshifts and a further 25 objects for which the accuracy
of the photometric redshifts could not be checked (see also
$\S$2.1.2). 

Out of the 406 photometric RIQs and RLQs under consideration prior to
incorporation of GALEX data, 151 (37\%) have $z_{\rm phot}>1.9$, of
which 121 have GALEX coverage. From these 121, we use a color-color
cut and a joint-detection cut to identify 43 objects with questionable
photometric redshifts. For these 43 objects, 37/29/23 (86\%/67\%/53\%)
are flagged by the color-color/joint-detection/both methods (so 14/6,
or 33\%/14\%, are flagged exclusively by the
color-color/joint-detection method). From these 43 objects, 9 have
spectroscopic redshifts (in all cases confirming the photometric
redshift was incorrect) and are retained with corrected redshifts and
luminosities. The remaining 34 objects are discarded. Of the 30
objects with $z_{\rm phot}>1.9$ lacking GALEX coverage, 5 have
spectroscopic redshifts (in all cases indicating the photometric
redshift is correct); the remaining 25 objects are conservatively
discarded, sacrificing some completeness to minimize potential
contamination. Additional details are provided below.

The color-color index $(m_{\rm NUV}-m_{\rm r})-2.5(m_{\rm g}-m_{\rm
  r})$ cleanly separates $z_{\rm spec}\le1$ from $z_{\rm spec}\ge1.9$
objects (D.~W.~Hogg 2009, personal communication). We flag the 37
objects with photometric redshifts $z_{\rm phot}\ge1.9$ but $(m_{\rm
  NUV}-m_{\rm r})-2.5(m_{\rm g}-m_{\rm r})<1.5$, of which we estimate
$\sim$31 are actually low-redshift quasars. From the catalog of SDSS
DR3 quasars matched to GALEX data provided by Trammell et al.~(2007),
of the quasars with both FUV and NUV detections the fraction with
$z_{\rm spec}>1.9$ is 6.9\% (or 7.7\%/8.6\% for objects with $m_{\rm
  i}>19/20$; the median $m_{\rm i}$ for the photometric sample is
20.0). In contrast, within the primary sample, $\simeq$23\% of the
quasars with both FUV and NUV detections have $z_{\rm phot}\ge1.9$. We
flag the 29 objects with NUV and FUV detections and $z_{\rm
  phot}\ge1.9$, of which we estimate $\sim$18 are actually
low-redshift quasars. Most (23/29) of the objects with both FUV and
NUV detections are already flagged for rejection by the color-color
cut, so this process finds 6 new objects with questionable photometric
redshifts. After combining methods we generate a rejection list of 43
objects with photometric redshifts $z\ge1.9$ but GALEX properties more
characteristic of low-redshift quasars; we estimate $\sim$(30--35)/43
are indeed low-redshift, and only $\sim$10\% of genuine $z\ge1.9$
quasars are lost through these cuts. The GALEX detection rate of
low-redshift SDSS quasars is sufficiently high (Trammell et al.~2007)
that virtually all interlopers with GALEX coverage are expected to be
identifiable through this process.

We checked the rejection list of 43 objects (with photometric
redshifts $z\ge1.9$ but GALEX properties more characteristic of
low-redshift quasars) for SDSS DR6 or NED spectroscopic coverage,
finding 9 objects with spectroscopic redshifts available for
comparison. From these, 7 objects have incorrect photometric redshifts
$z\ge1.9$ but spectroscopic redshifts $z\le1$ (of which 6/5/4 objects
were flagged for rejection by color-color/joint-detection/both cuts.)
The other 2 objects also have (less dramatically) incorrect
photometric redshifts, with spectroscopic redshifts $\sim$0.5 below
the photometric estimate (these objects are flagged for rejection by
both methods). No objects with spectroscopically confirmed photometric
redshifts of $z>1.9$ are on this rejection list. We correct the
redshifts (and luminosities) of these 9 objects with spectroscopic
coverage and then remove them from the rejection list, which therefore
consists of 34 objects. Of the 30 objects with photometric redshifts
$z\ge1.9$ that lack GALEX coverage, 5 also have spectroscopic
redshifts and are retained, while the remaining 25 (of which $\sim$7
are likely actually low-redshift) are conservatively added to the
rejection list. After discarding objects on the rejection list, the
updated candidate list of photometric RIQs and RLQs contains 347
($=406-34-25$) objects, within which the remaining fraction with this
type of redshift misindentification is only $\sim$1.5\%.

\section*{C. Notes on individual deep-field objects}

Here we present brief commentary on selected RIQs and RLQs from the
CDF-N and the E-CDF-S; the interested reader is referred to the
provided references for additional detail.

{\it 123538.51+621643.0\/} is included in the {\it XMM-Newton\/}
Bright Serendipitous Survey (Della Ceca et al. 2004). Galbiati et
al.~(2005) describe it as a non-blazar AGN (with $R^{*}=1.22$ and
${\alpha}_{\rm r}=2.5$ calculated from 1.4 to 8.5 GHz) and find that
the X-ray spectrum can be adequately fit by a power-law model with
$\Gamma=1.96^{+0.08}_{-0.05}$ and no intrinsic absorption.

{\it 123649.62+620737.8\/} is near the borderline of our X-ray
hardness-ratio cut and may be mildly obscured, although it is
nevertheless relatively X-ray bright. It is classified as a type-2 AGN
by, e.g., Padovani et al.~(2004), using the definition of Szokoly et
al.~(2004). Trouille et al.~(2008) characterize the optical spectrum
as showing high excitation lines (i.e., not an obvious broad-line
object), but its ${\Delta}(g-i)$ value is $-0.17$, bluer than the
typical quasar at that redshift.

{\it 123746.63+621739.0\/} is listed in the radio catalog of Richards
et al.~(1998) as having a radio flux density of 11.1 $\mu$Jy at 8.5
GHz, and is also in the radio catalog of Richards et al.~(1999) with a
flux density of 998 $\mu$Jy at 1.4 GHz, in agreement with the flux
density from Biggs \& Ivison~(2006). The optical counterpart has
quasar-like colors but lacks an optical spectrum that would permit
definitive classification.

{\it 033115.03$-$275518.5\/} has a lobe-dominated radio morphology and
shows double-peaked optical broad-line emission structure. Details may
be found in Luo et al.~(2009).

{\it 033124.86$-$275207.1\/} has two other optical objects within
$2.5''$, but the selected counterpart is secure. It is described in
Rovilos et al.~(2007) as a type-1 AGN based on its soft X-ray
spectrum.

{\it 033139.49$-$274119.6\/} is listed in the ACTA radio catalog of
Norris et al.~(2006) as having a 1.4~GHz flux density of 0.2 mJy, in
agreement with the flux density of 206 $\mu$Jy from Miller et
al.~(2008). This object is detected in the {\it Spitzer\/} Wide-Area
Infrared Extragalactic Survey (Lonsdale et al.~2003) and has optical
quasar-like colors but lacks an optical spectrum that would permit
definitive classification.

{\it 033208.66$-$274734.4\/} is a bright RIQ with a broad-line optical
spectrum (e.g., Tozzi et al.~2009). Lehmer et al.~(2005) describe the
X-ray spectrum (containing $\sim$20000 counts) as having an effective
photon index of $\Gamma=1.73$ (calculated from the hardness ratio),
and Steffen et al.~(2006) quote a radio-loudness of $R^{*}=1.14$.

{\it 033210.91$-$274414.9\/} is a RLQ with a broad-line optical
spectrum (e.g., Tozzi et al.~2009). Wang et al.~(2007) find that the
X-ray spectrum (containing $\sim$1000 counts) can be well-fit by a
power-law model with $\Gamma=1.88\pm0.09$ and no intrinsic absorption.

{\it 033211.63$-$273725.9\/} has COMBO-17 colors consistent with those
of a template quasar spectrum. Lehmer et al.~(2005) describe the X-ray
spectrum as having an effective photon index of $\Gamma=1.44$
(calculated from the hardness ratio), and Steffen et al.~(2006) quote
a radio-loudness of $R^{*}=1.29$. This source is detected in the {\it
  Spitzer\/} Wide-Area Infrared Extragalactic Survey.

{\it 033227.01$-$274105.0\/} is a broad-line quasar (e.g., Tozzi et
al.~2009). Galbiati et al.~(2005) quote $R^{*}=2.05$ and
${\alpha}_{\rm r}=-0.53$ and fit the X-ray spectrum with a power-law
plus blackbody model ($\Gamma=2.04\pm0.15$ and $kT=0.15\pm0.06$) with
no intrinsic absorption.

\section*{D. Verification of fitting methodology}

We fit \hbox{X-ray} luminosity as a function of various parameters
using the IDL code of Kelly (2007), which takes into account censoring
of the dependent variable and uncertainties in all variables. This is
an advance over methods that consider either censoring (e.g., ASURV)
or errors (e.g., the IDL program fitexy) but not both
simultaneously. Here, we illustrate the fitting technique through
application to the RQQ sample of Steffen et al.~(2006), for which
results using an alternative method (ASURV) have already been
reported.

In this sample of RQQs the uncertainties are dominated by intrinsic
variability (see, e.g., $\S$3.5 of Gibson et al.~2008), as is also the
case for the RIQs and RLQs that form the primary focus of this
work. The uncertainties we use throughout are based on observed data,
but we investigate here the degree to which the fitting results are
sensitive to alternative values for the uncertainties. Figure~21a
shows ${\ell}_{\rm x}({\ell}_{\rm uv})$ computed for RQQs for our
standard errors (solid line), for a fit resulting from doubling the
standard errors (dashed line), and for a fit resulting from halving
the standard errors (dotted line). The coefficients for the standard
error fit are ${\ell}_{\rm x} =
(-0.545\pm0.023)+(0.649\pm0.021)\times{\ell}_{\rm uv}$; for the double
errors fit they are ${\ell}_{\rm x} =
(-0.540\pm0.023)+(0.673\pm0.022)\times{\ell}_{\rm uv}$; for the half
errors fit they are ${\ell}_{\rm x} =
(-0.546\pm0.023)+(0.644\pm0.021)\times{\ell}_{\rm uv}$. When comparing
the coefficient values it must be kept in mind that these coefficients
are probabilistically derived, so a large number of fits would be
required to quantify most accurately the impact of using different
error values. However, it is clear that varying the uncertainties
within reasonable bounds does not dramatically alter the parameters of
the best-fit line (i.e., the credible intervals of the parameters
overlap).

\begin{figure}
\includegraphics[scale=0.85]{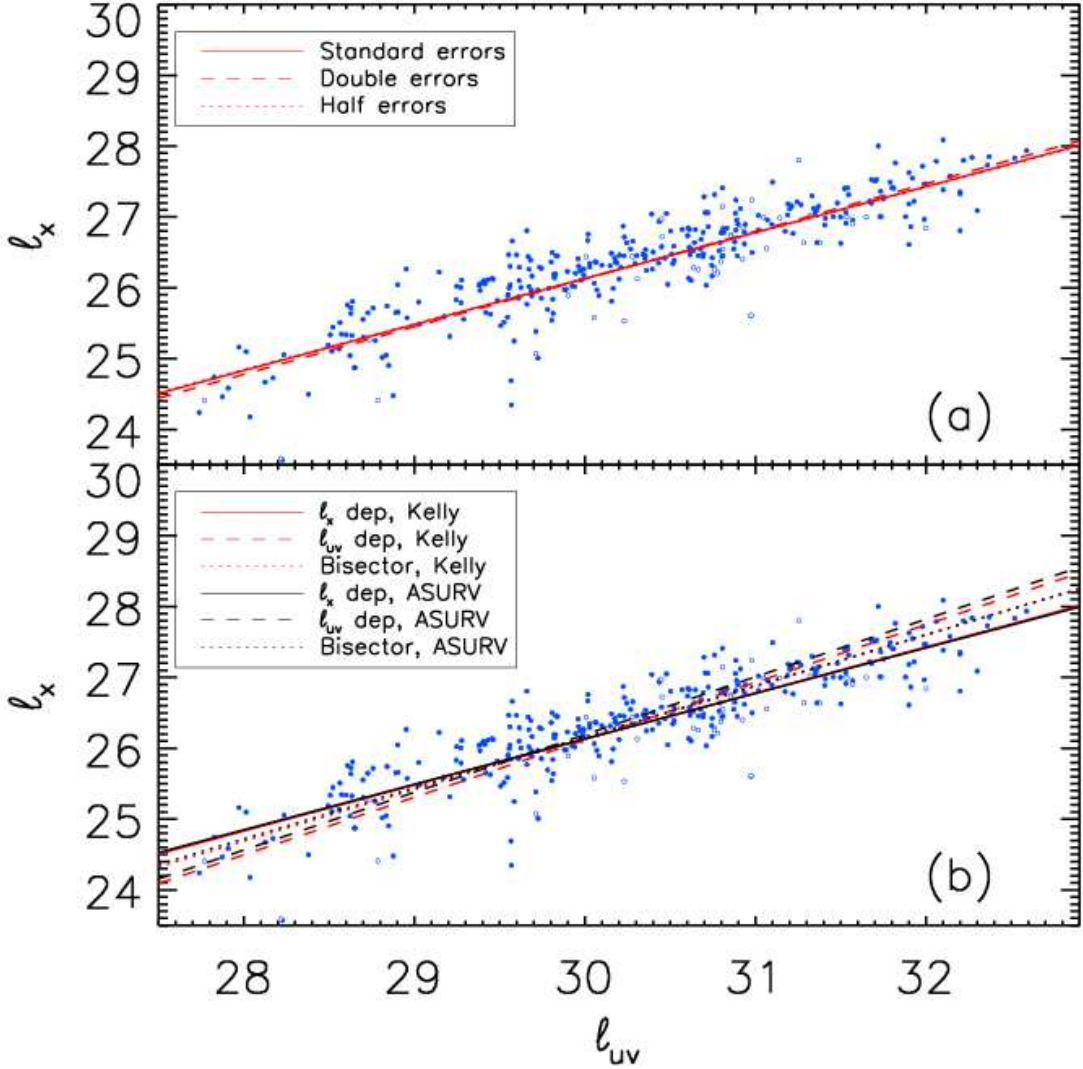}
\figcaption{\small Demonstration of fitting methodology for
  ${\ell}_{\rm x}({\ell}_{\rm uv})$ applied to RQQs. The top panel
  illustrates that varying the uncertainties in the variables within a
  reasonable range has only a minimal impact upon the best-fit
  correlation. The bottom panel shows that using the code of Kelly
  (2007) provides results consistent with those given by ASURV, and
  illustrates the effect of considering ${\ell}_{\rm uv}$ rather than
  ${\ell}_{\rm x}$ as the dependent variable, or of using the bisector
  line. See Appendix~D for details.}
\end{figure}

The best-fit model obtained through using the Kelly (2007) code agrees
with that calculated by Steffen et al.~(2006) using ASURV and treating
${\ell}_{\rm x}$ as the dependent variable (their Equation 1a), which
is ${\ell}_{\rm x} = (-0.546\pm0.023)+(0.642\pm0.021)\times{\ell}_{\rm
  uv}$ using our normalization convention. As described in $\S$5.1, it
seems reasonable on both practical and statistical grounds to consider
${\ell}_{\rm x}$ as the dependent variable throughout. However, other
approaches are possible; we investigate the effect of considering
${\ell}_{\rm uv}$ as the dependent variable and of using the bisector
best-fit line. When fitting ${\ell}_{\rm uv}({\ell}_{\rm x})$ and
inverting the results, the slope will generally be steeper than for
the fit conducted with ${\ell}_{\rm x}$ as the dependent variable. A
complicating factor is that it is no longer possible to consider the
censoring in ${\ell}_{\rm x}$. However, treating all upper limits as
\hbox{X-ray} detections, we find ${\ell}_{\rm x} =
-0.475+0.812\times{\ell}_{\rm uv}$, which may be compared with the
Steffen et al.~(2006) result (their Equation 1b) of ${\ell}_{\rm x} =
-0.400+0.815\times{\ell}_{\rm uv}$. For reference, the best-fit lines
calculated using the IDL program fitexy are ${\ell}_{\rm x} =
-0.500+0.695\times{\ell}_{\rm uv}$ and ${\ell}_{\rm x} =
-0.491+0.736\times{\ell}_{\rm uv}$ treating ${\ell}_{\rm x}$ and
${\ell}_{\rm uv}$ in turn as the dependent variable. Calculating the
bisector using Table~1 of Isobe et al.~(1990) gives ${\ell}_{\rm x} =
-0.493+0.727\times{\ell}_{\rm uv}$ for our fits, or ${\ell}_{\rm x} =
-0.478+0.721\times{\ell}_{\rm uv}$ from Steffen et al.~(2006; their
Equation 1c). The bisector slope is slightly higher, but it may be
seen in Figure~21b that the change in the distance from the best-fit
line for any given point is at most $\sim$0.2 over the span of the
variables. When we discuss ``outliers'' or ``positive residuals'' we
are referring to greater distances from the best-fit line. We note
that the best-fit lines may not always appear to run through the
highest density of points when upper limits are present; for example,
the RIQ fit to ${\ell}_{\rm x}({\ell}_{\rm uv})$ in Figure~9a may
appear different than would be the case were all of the upper limits
detections.

As a final consistency check, we also performed every fit given in
Table~7 with ASURV.\footnote{For computational purposes, for the full,
  primary, off-axis, and RLQ subsamples only (which each contain
  $>$500 objects) we randomly selected $\sim$500 objects to fit with
  ASURV.} In all cases, the parameters agreed (in the sense that the
1$\sigma$ intervals overlapped) with those computed using the code of
Kelly~(2007). The median differences between the best-fit parameters
obtained using the code of Kelly~(2007) and those obtained with ASURV
(and the standard deviations thereof) for the ${\ell}_{\rm
  x}({\ell}_{\rm uv})$ model are $-0.0009~(0.0085)$ for $a_{\rm 0}$
and $0.0262~(0.0082)$ for $b_{\rm uv}$; without considering the
uncertainties, ASURV apparently computes slightly flatter slopes for
these datasets (recall that the typical values for $b_{\rm uv}$ are
0.6--0.8 in these fits, so this systematic difference is 3--4\%). The
median differences (and standard deviations) for the ${\ell}_{\rm
  x}({\ell}_{\rm uv},{\ell}_{\rm r})$ model are $0.0006~(0.0049)$ for
$a_{\rm 0}$, $0.0200~(0.0130)$ for $b_{\rm uv}$, and
$-0.0088~(0.0094)$ for $c_{\rm r}$. For illustrative purposes, if the
Kelly~(2007) code is run with an uncertainty on ${\ell}_{\rm uv}$ of
0.05 rather than 0.114 (keeping the uncertainties on ${\ell}_{\rm r}$
and ${\ell}_{\rm x}$ at 0.0792 and 0.146), the average change in
$b_{\rm uv}/c_{\rm r}$ is $-0.022/0.0079$, producing coefficients
matching the ASURV results (with a similar impact on $b_{\rm uv}$ in
the ${\ell}_{\rm x}({\ell}_{\rm uv})$ relation).

We conclude that using only ${\ell}_{\rm x}$ as the dependent variable
is acceptable for our comparative study and that using the IDL code of 
Kelly~(2007) to take both censoring and uncertainties into account 
is advantageous. 

\clearpage

\begin{deluxetable}{p{90pt}rrrrrrrrrrrrr}
\tablecaption{Primary sample RIQs and RLQs}
\tabletypesize{\scriptsize}
\tablewidth{19.0cm}

\tablehead{ \colhead{Name\tablenotemark{a}} & \colhead{$z$} & \colhead{$m_{\rm i}$} &
  \colhead{${\Delta}(g-i)$} & \colhead{${\ell}_{\rm r}$\tablenotemark{b}} &
  \colhead{${\ell}_{\rm uv}$} & \colhead{${\ell}_{\rm x}$} &
  \colhead{Det\tablenotemark{c}} & \colhead{$R^{*}$} &
  \colhead{${\alpha}_{\rm ox}$} & \colhead{${\alpha}_{\rm r}$\tablenotemark{d}} &
  \colhead{${\ell}_{\rm r,core}$} & \colhead{SDSS\tablenotemark{e}} &
  \colhead{X-ray\tablenotemark{f}}}

\startdata

000442.18$+$000023.3 &1.008 &19.02 &$-$0.02 &31.92 &30.28 &26.91 &1 &1.64 &$-$1.30 & 0.00 &  31.92 &S &X \\
000622.60$-$000424.4 &1.038 &19.49 & 0.21 &34.97 &30.05 &27.29 &1 &4.91 &$-$1.06 &$-$0.72 &  34.97 &S &C \\
001130.40$+$005751.8 &1.492 &20.08 &$-$0.05 &33.96 &30.13 &27.17 &1 &3.84 &$-$1.13 &$-$0.22 &  33.96 &S &X \\
004230.40$-$092202.7 &2.085 &19.97 & 0.01 &34.01 &30.51 &26.31 &1 &3.50 &$-$1.61 & 0.00 &  34.01 &P &c \\
004413.72$+$005141.0 &0.941 &18.44 & 0.06 &32.64 &30.42 &26.61 &1 &2.22 &$-$1.46 & 0.00 &$<$31.29 &S &X \\
005009.81$-$003900.6 &0.728 &19.89 & 0.63 &31.68 &29.36 &26.25 &1 &2.32 &$-$1.20 & 0.00 &  31.68 &S &C \\
005905.50$+$000651.6 &0.719 &17.46 &$-$0.07 &34.42 &30.62 &27.37 &1 &3.80 &$-$1.24 &$-$0.40 &  34.42 &S &C \\
012401.76$+$003500.9 &1.850 &20.23 & 0.26 &34.12 &30.26 &26.90 &1 &3.86 &$-$1.29 &$-$0.70 &  34.12 &S &C \\
012528.84$-$000555.9 &1.077 &16.47 &$-$0.09 &34.59 &31.36 &27.73 &1 &3.23 &$-$1.39 & 0.22 &  34.59 &S &r \\
012734.57$-$000523.8 &1.598 &20.37 & 0.14 &33.32 &30.04 &26.74 &1 &3.28 &$-$1.27 & 0.00 &  33.32 &S &r \\

\enddata

\tablecomments{Table 1 is provided in its entirety in the online
  edition of the journal. A portion is shown here for guidance as to
  its form and content. Some values here and in the following tables
  are given to precision greater than the uncertainty to avoid future
  accumulated round-off error.}
\tablenotetext{a}{Name is J2000 from optical RA/Dec and is SDSS DR5
  for spectroscopic quasars or SDSS DR6 for photometric quasars.}
\tablenotetext{b}{The radio luminosity ${\ell}_{\rm r}$ is calculated
  from both core and lobe components as described in $\S$3.2, while
  ${\ell}_{\rm r,core}$ is for the core only (limits indicate no FIRST
  core and are calculated for 1~mJy).}
\tablenotetext{c}{1 = X-ray detection; 0 = X-ray upper limit.}
\tablenotetext{d}{${\alpha}_{\rm r}$ is calculated between 1.4 and 5
  GHz where Green Bank fluxes are available and is otherwise based on
  low-frequency (Texas and Westerbork) surveys; zero-valued entries
  indicate a lack of multi-frequency radio data.}
\tablenotetext{e}{S = spectroscopic, from the DR5 quasar catalog of
  Schneider et al.~(2007); P = photometric, from the DR6 quasar
  catalog of Richards et al.~(2009).}  
\tablenotetext{f}{C/c = {\it Chandra\/}; X/x = {\it XMM-Newton\/}; r =
  {\it ROSAT\/}. Lowercase indicates our measurements, uppercase from
  catalogs as described in $\S$3.4.}

\end{deluxetable}

\begin{deluxetable}{rrrrrrrrr}
\tablecaption{RIQs and RLQs selected from deep surveys}
\tabletypesize{\scriptsize}
\tablewidth{13.0cm}

\tablehead{ \colhead{Name\tablenotemark{a}} & \colhead{$z$} &
  \colhead{$m_{\rm i}$} & \colhead{${\ell}_{\rm r}$} &
  \colhead{${\ell}_{\rm uv}$} & \colhead{${\ell}_{\rm
      x}$\tablenotemark{b}} & \colhead{$R^{*}$} &
  \colhead{${\alpha}_{\rm ox}$} & \colhead{Ref\tablenotemark{c}}}

\startdata
\cutinhead{Color selected}

033115.03$-$275518.5&    1.368 &  21.43 & 32.71& 29.60& 26.48 &  3.11 &$-$1.20 &   1 \\
033124.86$-$275207.1&    1.328 &  21.09 & 33.18& 29.31& 25.72 &  3.87 &$-$1.38 &     \\
033139.49$-$274119.6&    2.215 &  23.12 & 31.36& 29.29&$<$25.04& 2.06 &$<-$1.63 &     \\
033208.66$-$274734.4&    0.543 &  18.78 & 31.03& 29.26& 25.94 &  1.77 &$-$1.27 &   2 \\ 
033210.91$-$274414.9&    1.605 &  22.95 & 32.26& 29.10& 26.08 &  3.16 &$-$1.16 &   3 \\
033211.63$-$273725.9&    1.636 &  18.99 & 32.36& 30.72& 26.85 &  1.64 &$-$1.49 &     \\
033227.01$-$274105.0&    0.737 &  19.09 & 32.31& 29.95& 26.34 &  2.36 &$-$1.39 &   3 \\
033302.67$-$274823.1&    3.021 &  23.96 & 31.38& 29.31& 26.51 &  2.06 &$-$1.08 &     \\
033310.20$-$274841.9&    1.034 &  22.66 & 32.68& 28.58& 25.71 &  4.10 &$-$1.10 &   3 \\
123529.38$+$621256.4&    2.413 &  23.21 & 31.24& 29.37& 26.11 &  1.88 &$-$1.25 &   5 \\
123538.51$+$621643.0&    0.712 &  19.70 & 31.50& 29.53& 26.50 &  1.97 &$-$1.16 &   5 \\
123649.62$+$620737.8&    1.610 &  23.51 & 31.25& 28.85& 26.17 &  2.40 &$-$1.03 &   6 \\
123746.63$+$621739.0&    2.316 &  23.36 & 32.12& 29.25&$<$25.20& 2.87 &$<-$1.56 &     \\
095958.54$+$015254.6&    1.019 &  21.86 & 31.02& 28.56& 24.80 &  2.46 &$-$1.44 &     \\
100046.92$+$020726.6&    1.210 &  21.70 & 31.81& 28.97& 25.37 &  2.85 &$-$1.38 &     \\
100114.86$+$020208.9&    0.969 &  20.63 & 31.97& 29.31& 26.00 &  2.66 &$-$1.27 &   8 \\
                                                                                 
\cutinhead{Broad-line selected}                        

033225.17$-$274218.8&    1.617 &  23.37 & 30.70& 28.82& 25.78 &  1.87 &$-$1.17 &   4 \\
123704.11$+$620755.4&    1.253 &  22.07 & 30.44& 29.11& 25.54 &  1.32 &$-$1.37 &   7 \\
123707.51$+$622148.0&    1.451 &  23.04 & 30.45& 28.92& 26.20 &  1.53 &$-$1.04 &   7 \\
095821.65$+$024628.2&    1.403 &  19.05 & 32.36& 30.48& 26.75 &  1.88 &$-$1.43 &   8 \\
095838.47$+$022439.3&    1.161 &  21.86 & 30.34& 28.71& 26.23 &  1.63 &$-$0.95 &   8 \\
095908.32$+$024309.6&    1.317 &  18.43 & 33.32& 30.66& 27.16 &  2.66 &$-$1.35 &   8 \\
095921.31$+$024412.4&    1.004 &  20.46 & 31.57& 29.42& 25.53 &  2.15 &$-$1.49 &   8 \\
100114.86$+$020208.8&    0.969 &  21.00 & 31.97& 29.27& 26.18 &  2.70 &$-$1.19 &   8 \\
100129.83$+$023239.0&    0.826 &  20.54 & 30.36& 28.80& 25.54 &  1.55 &$-$1.25 &   8 \\
100205.03$+$023731.5&    0.519 &  19.05 & 30.68& 29.25& 26.23 &  1.43 &$-$1.16 &   8 \\
100213.42$+$023351.7&    1.143 &  21.76 & 30.25& 28.94& 25.68 &  1.32 &$-$1.25 &   8 \\
100228.82$+$024016.9&    3.144 &  21.27 & 31.56& 29.69& 26.76 &  1.87 &$-$1.12 &   8 \\
100230.06$+$014810.4&    0.626 &  19.65 & 30.22& 28.30& 25.51 &  1.92 &$-$1.07 &   8 \\
100240.93$+$023448.4&    1.677 &  21.98 & 30.74& 29.35& 25.89 &  1.40 &$-$1.33 &   8 \\
100249.33$+$023746.5&    2.124 &  19.75 & 32.84& 30.62& 26.37 &  2.22 &$-$1.63 &   8 \\

\enddata

\tablenotetext{a}{Name is J2000 from optical RA/Dec.}

\tablenotetext{b}{X-ray luminosities are from {\it Chandra\/}
  observations except for the broad-line selected COSMOS objects, for
  which the X-ray luminosities are from {\it XMM-Newton\/}
  observations. See $\S$2.2.3 for details.}

\tablenotetext{c}{References for spectroscopic redshifts: 1 = Luo et
  al.~(2009); 2 = Mignoli et al.~(2005); 3 = Silverman et al., in
  prep; 4 = Szokoly et al.~(2004); 5 = Barger et al.~(2003); 6 =
  Barger et al.~(2008); 7 = Trouille et al.~(2008); 8 = Trump et
  al.~(2009). Unlabled redshift values are photometric from Luo et
  al.~(2010; CDF-S \hbox{X-ray} detections); Rafferty et al.~(2010;
  \hbox{X-ray} limits and E-CDF-S detections); Xue et al.~(2010;
  CDF-N); and Ilbert et al.~(2009; COSMOS).}

\end{deluxetable}

\begin{deluxetable}{p{90pt}rrrcrrrrrr}
\tablecaption{Sample characteristics}
\tabletypesize{\scriptsize}
\tablewidth{14.0cm}

\tablehead{ & \multicolumn{3}{c}{X-ray sources} & &
  \multicolumn{6}{c}{Median properties} \\ \colhead{Sample} &
  \colhead{$N$} & \colhead{$N_{\rm det}$} & \colhead{100$\frac{N_{\rm
        det}}{N}$} & & \colhead{$z$} & \colhead{${\ell}_{\rm r}$} &
  \colhead{${\ell}_{\rm uv}$} & \colhead{${\ell}_{\rm x}$} &
  \colhead{$R^{*}$} & \colhead{${\alpha}_{\rm ox}$}}

\startdata

\cutinhead{Full, primary, and supplemental samples}

Full                        & 791 &  674 &   85 &~~&   1.40 &  32.95 &  30.42 &  26.79 &   2.59 &  $-$1.40  \\[+4pt]  
Primary                     & 654 &  547 &   84 &~~&   1.41 &  32.83 &  30.36 &  26.73 &   2.50 &  $-$1.40  \\       
~~~Spectroscopic            & 312 &  274 &   88 &~~&   1.32 &  33.00 &  30.49 &  26.83 &   2.46 &  $-$1.41  \\       
~~~~~~QSO/HIZ               & 200 &  186 &   93 &~~&   1.20 &  33.14 &  30.67 &  26.97 &   2.50 &  $-$1.41  \\       
~~~Photometric              & 342 &  273 &   80 &~~&   1.47 &  32.72 &  30.23 &  26.64 &   2.52 &  $-$1.39  \\       
~~~Off-axis                 & 562 &  456 &   81 &~~&   1.45 &  32.73 &  30.33 &  26.70 &   2.44 &  $-$1.40  \\       
~~~Targeted                 &  92 &   91 &   99 &~~&   1.05 &  33.65 &  30.74 &  27.04 &   3.09 &  $-$1.39  \\[+4pt] 
Supplemental                & 137 &  127 &   93 &~~&   1.32 &  34.33 &  30.96 &  27.38 &   3.11 &  $-$1.36  \\       
~~~{\it Einstein\/}         &  93 &   85 &   91 &~~&   1.02 &  34.63 &  31.15 &  27.55 &   3.44 &  $-$1.39  \\       
~~~High-$z$                 &  13 &   13 &  100 &~~&   4.31 &  34.69 &  31.42 &  27.70 &   2.90 &  $-$1.34  \\       
~~~Deep fields              &  31 &   29 &   94 &~~&   1.32 &  31.50 &  29.27 &  26.04 &   2.06 &  $-$1.26  \\[+3pt]         
FIRST                       & 180 &  164 &   91 &~~&   1.27 &  33.16 &  30.73 &  26.99 &   2.41 &  $-$1.43  \\       
\cutinhead{RIQs and RLQs}                                                                                            
RIQs                        & 188 &  142 &   76 &~~&   1.42 &  32.05 &  30.41 &  26.50 &   1.67 &  $-$1.50  \\       
RLQs                        & 603 &  532 &   88 &~~&   1.38 &  33.30 &  30.44 &  26.89 &   2.88 &  $-$1.37  \\       
\cutinhead{Groupings of RLQs}                                                                                            
$R^{*}<3$                   & 343 &  291 &   85 &~~&   1.42 &  32.87 &  30.34 &  26.69 &   2.47 &  $-$1.41  \\       
$R^{*}\ge3$                 & 260 &  241 &   93 &~~&   1.33 &  34.08 &  30.59 &  27.16 &   3.46 &  $-$1.31  \\       
${\ell}_{\rm r}<33.3$       & 301 &  251 &   83 &~~&   1.26 &  32.68 &  30.11 &  26.53 &   2.50 &  $-$1.37  \\       
${\ell}_{\rm r}\ge33.3$     & 302 &  281 &   93 &~~&   1.63 &  34.04 &  30.82 &  27.22 &   3.31 &  $-$1.37  \\       
${\alpha}_{\rm r}<-0.5$     & 181 &  160 &   88 &~~&   1.31 &  33.59 &  30.48 &  26.93 &   3.18 &  $-$1.38  \\       
${\alpha}_{\rm r}\ge-0.5$   & 221 &  207 &   94 &~~&   1.26 &  33.66 &  30.61 &  27.11 &   3.10 &  $-$1.34  \\       

\enddata

\tablecomments{The columns $N$, $N_{\rm det}$, and 100$\frac{N_{\rm
      det}}{N}$ give the number of sources, the number of X-ray
  detected sources, and the percentage of sources with X-ray
  detections, respectively. Radio, optical/UV, and X-ray monochromatic
  luminosities ${\ell}_{\rm r}$, ${\ell}_{\rm uv}$, and ${\ell}_{\rm
    x}$ are expressed as logarithmic quantities with units of
  erg~s$^{-1}$~Hz$^{-1}$ at rest-frame 5~GHz, 2500~\AA, and 2~keV,
  respectively. The radio loudness is $R^{*} = {\ell}_{\rm
    r}-{\ell}_{\rm uv}$ and the optical/UV-to-X-ray spectral slope is
  ${\alpha}_{\rm ox} = 0.384\times({\ell}_{\rm x}-{\ell}_{\rm
    uv})$. RIQs have $1{\leq}R^{*}$$<2$ and RLQs have
  $R^{*}\ge2$. Medians that include limits were determined with
  ASURV. Details regarding the various samples are given in $\S$2.}
\end{deluxetable}

\begin{deluxetable}{p{50pt}rrrrrr}
\tablecaption{RIQ and RLQ properties as a function of ${\ell}_{\rm uv}$}
\tabletypesize{\scriptsize}
\tablewidth{11.0cm}

\tablehead{ & \multicolumn{6}{c}{${\ell}_{\rm uv}$ bin range}
  \\ \colhead{Statistic} & \colhead{$<$29.5} & \colhead{29.5--30} &
  \colhead{30--30.5} & \colhead{30.5--31} & \colhead{31--31.5} &
  \colhead{$\ge$31.5}}

\startdata
\cutinhead{X-ray sources}
$N$                         &   69 &   119 &    257 &    210 &     92 &     44 \\
$N_{\rm det}$               &   57 &    98 &    217 &    176 &     86 &     40 \\
100$\frac{N_{\rm det}}{N}$  &   83 &    82 &     84 &     84 &     93 &     91 \\
\cutinhead{Redshift}                                  
Mean    & 0.87 &  1.04 &   1.36 &   1.72 &   2.09 &   2.71 \\
25th \% & 0.47 &  0.74 &   1.05 &   1.09 &   1.27 &   2.01 \\
50th \% & 0.71 &  1.03 &   1.39 &   1.70 &   1.99 &   2.44 \\
75th \% & 1.15 &  1.33 &   1.69 &   2.20 &   2.90 &   3.52 \\
\cutinhead{${\ell}_{\rm uv}$}                                  
Mean    &29.12 & 29.80 &  30.27 &  30.73 &  31.21 &  31.78 \\
25th \% &28.85 & 29.67 &  30.16 &  30.61 &  31.11 &  31.60 \\
50th \% &29.27 & 29.82 &  30.28 &  30.73 &  31.19 &  31.73 \\
75th \% &29.38 & 29.92 &  30.40 &  30.85 &  31.31 &  31.91 \\
\cutinhead{${\ell}_{\rm x}$}                                  
Mean    &25.77 & 26.35\tablenotemark{a}&  26.67 &  26.97 &  27.52 &  27.83 \\
25th \% &25.38 & 26.06 &  26.42 &  26.72 &  27.27 &  27.48 \\
50th \% &25.75 & 26.38 &  26.72 &  27.03 &  27.53 &  27.71 \\
75th \% &26.23 & 26.57 &  26.94 &  27.21 &  27.78 &  28.13 \\
\cutinhead{${\alpha}_{\rm ox}$}                                           
Mean    &$-$1.28 & $-$1.32 &  $-$1.39 &  $-$1.45 &  $-$1.42 &  $-$1.52 \\
25th \% &$-$1.46 & $-$1.43 &  $-$1.46 &  $-$1.52 &  $-$1.53 &  $-$1.62 \\
50th \% &$-$1.25 & $-$1.32 &  $-$1.38 &  $-$1.44 &  $-$1.42 &  $-$1.54 \\
75th \% &$-$1.16 & $-$1.25 &  $-$1.28 &  $-$1.34 &  $-$1.31 &  $-$1.45 \\
\cutinhead{$R^{*}$}                                  
Mean    & 2.66 &  2.52 &   2.51 &   2.67 &   2.96 &   2.67 \\
25th \% & 1.92 &  2.03 &   1.97 &   2.06 &   2.33 &   2.19 \\
50th \% & 2.66 &  2.43 &   2.47 &   2.64 &   3.14 &   2.77 \\
75th \% & 3.27 &  2.90 &   3.01 &   3.37 &   3.71 &   3.24 \\
\enddata

\tablecomments{Properties here are for the full sample of RIQs and
  RLQs. Quantities and units are as defined in Table 3. The
  ${\ell}_{\rm uv}$ bins are plotted in Figure 3.}

\tablenotetext{a}{Estimate of the mean is biased because it was
  computed by ASURV treating the censored first point in this bin as a
  detection.}

\end{deluxetable}

\begin{deluxetable}{p{50pt}rrrrrcrrrrr}
\tablecaption{Excess X-ray luminosity as a function of $R^{*}$ and ${\ell}_{\rm r}$: full sample}
\tabletypesize{\scriptsize}
\tablewidth{18.0cm}

\tablehead{ & \multicolumn{5}{c}{$R^{*}$ bin range} &
  \multicolumn{5}{c}{${\ell}_{\rm r}$ bin range}
  \\ \colhead{Statistic} & \colhead{$<$2.0} & \colhead{2.0--2.5} &
  \colhead{2.5--3.0} & \colhead{3.0--3.5} & \colhead{$\ge3.5$} & &
  \colhead{$<32.2$} & \colhead{32.2--32.9} & \colhead{32.9--33.6} &
  \colhead{33.6--34.3} & \colhead{$\ge34.3$}}

\startdata         
$N$                         &    188 &    175 &    168 &    141 &    119 &~~~&  176 &    198 &    187 &    108 &    122  \\
$N_{\rm det}$               &    142 &    145 &    146 &    128 &    113 &~~~&  138 &    156 &    163 &    102 &    115  \\
100$\frac{N_{\rm det}}{N}$  &     76 &     83 &     87 &     91 &     95 &~~~&   78 &     79 &     87 &     94 &     94  \\
\cutinhead{${\ell}_{\rm x}-{\ell}_{\rm x,RQQ}$}
Mean    &   0.11 &  0.29 &    0.41 &    0.58 &     0.77 &~~~&    0.22 &   0.22 &    0.42 &   0.55 &     0.76 \\
Error   &   0.04 &  0.03 &    0.03 &    0.03 &     0.05 &~~~&    0.04 &   0.04 &    0.03 &   0.04 &     0.04 \\
25th \% &$-$0.16 &  0.08 &    0.26 &    0.35 &     0.53 &~~~& $-$0.13 &   0.05 &    0.22 &   0.34 &     0.51 \\
50th \% &   0.16 &  0.30 &    0.44 &    0.57 &     0.71 &~~~&    0.29 &   0.29 &    0.41 &   0.53 &     0.67 \\
75th \% &   0.45 &  0.52 &    0.62 &    0.77 &     1.03 &~~~&    0.59 &   0.52 &    0.62 &   0.72 &     0.94 \\
\cutinhead{$R^{*}$}                                                                                                               
Mean    &   1.62 &  2.26 &    2.75 &    3.25 &     3.85 &~~~&    1.88 &   2.27 &    2.77 &   3.15 &     3.61 \\
Error   &   0.02 &  0.01 &    0.01 &    0.01 &     0.03 &~~~&    0.04 &   0.04 &    0.04 &   0.05 &     0.04 \\
25th \% &   1.42 &  2.14 &    2.62 &    3.11 &     3.62 &~~~&    1.52 &   1.88 &    2.50 &   2.88 &     3.35 \\
50th \% &   1.67 &  2.28 &    2.74 &    3.25 &     3.78 &~~~&    1.86 &   2.29 &    2.79 &   3.20 &     3.65 \\
75th \% &   1.86 &  2.39 &    2.88 &    3.38 &     4.00 &~~~&    2.14 &   2.58 &    3.07 &   3.48 &     3.87 \\
\cutinhead{${\ell}_{\rm r}$}                                                             
Mean    &  31.98 & 32.62 &   33.14 &   33.75 &    34.38 &~~~&   31.69 &  32.54 &   33.23 &  33.91 &    34.77 \\
Error   &   0.05 &  0.05 &    0.05 &    0.06 &     0.07 &~~~&    0.03 &   0.01 &    0.01 &   0.02 &     0.03 \\
25th \% &  31.56 & 32.17 &   32.79 &   33.34 &    33.93 &~~~&   31.44 &  32.36 &   33.07 &  33.72 &    34.46 \\
50th \% &  32.05 & 32.58 &   33.18 &   33.70 &    34.51 &~~~&   31.82 &  32.54 &   33.23 &  33.92 &    34.72 \\
75th \% &  32.36 & 32.95 &   33.47 &   34.23 &    34.94 &~~~&   32.03 &  32.70 &   33.39 &  34.06 &    34.97 \\
\enddata

\tablecomments{The ``excess'' X-ray luminosity is defined as
  ${\ell}_{\rm x}-{\ell}_{\rm x,RQQ}$, where ${\ell}_{\rm x,RQQ} =
  0.709{\ell}_{\rm uv}+4.822$ from Just et al.~(2007). Other
  quantities and units are as defined in Table 3. Note that the first
  $R^{*}$ bin is the sample of RIQs. The inner bins in $R^{*}$
  increase by linear factors of $\sim$3; the inner bins in
  ${\ell}_{\rm r}$ increase by linear factors of $\sim$5. These values
  are plotted in Figure~7.}

\end{deluxetable}

\begin{deluxetable}{p{50pt}rrrrrcrrrrr}
\tablecaption{Excess X-ray luminosity as a function of $R^{*}$ and ${\ell}_{\rm r}$: primary sample}
\tabletypesize{\scriptsize}
\tablewidth{18.0cm}

\tablehead{ & \multicolumn{5}{c}{$R^{*}$ bin range} &
  \multicolumn{5}{c}{${\ell}_{\rm r}$ bin range}
  \\ \colhead{Statistic} & \colhead{$<$2.0} & \colhead{2.0--2.5} &
  \colhead{2.5--3.0} & \colhead{3.0--3.5} & \colhead{$\ge3.5$} & &
  \colhead{$<32.2$} & \colhead{32.2--32.9} & \colhead{32.9--33.6} &
  \colhead{33.6--34.3} & \colhead{$\ge34.3$}}

\startdata         
$N$                         &    171 &    154 &    147 &    111 &     71 &~~~&  154 &    187 &    172 &     90 &     51  \\
$N_{\rm det}$               &    125 &    126 &    128 &    102 &     66 &~~~&  118 &    145 &    150 &     84 &     50  \\
100$\frac{N_{\rm det}}{N}$  &     73 &     82 &     87 &     92 &     93 &~~~&   77 &     78 &     87 &     93 &     98  \\
\cutinhead{${\ell}_{\rm x}-{\ell}_{\rm x,RQQ}$}  
Mean    &   0.07 &   0.30 &    0.42 &    0.56 &   0.67 &~~~&   0.20 &   0.21 &  0.44 &  0.54 &   0.70 \\
Error   &   0.04 &   0.03 &    0.03 &    0.03 &   0.06 &~~~&   0.04 &   0.04 &  0.03 &  0.04 &   0.04 \\
25th \% &$-$0.20 &   0.12 &    0.26 &    0.34 &   0.53 &~~~&$-$0.13 &   0.03 &  0.23 &  0.34 &   0.52 \\
50th \% &   0.13 &   0.30 &    0.44 &    0.54 &   0.65 &~~~&   0.23 &   0.29 &  0.41 &  0.53 &   0.65 \\
75th \% &   0.41 &   0.52 &    0.62 &    0.74 &   0.97 &~~~&   0.57 &   0.52 &  0.63 &  0.71 &   0.87 \\  
\cutinhead{$R^{*}$}                                                                           
Mean    &   1.62 &   2.26 &    2.74 &    3.25 &   3.82 &~~~&   1.86 &   2.25 &  2.77 &  3.21 &   3.66 \\
Error   &   0.02 &   0.01 &    0.01 &    0.01 &   0.03 &~~~&   0.04 &   0.04 &  0.04 &  0.05 &   0.05 \\
25th \% &   1.41 &   2.14 &    2.61 &    3.12 &   3.61 &~~~&   1.51 &   1.87 &  2.50 &  2.95 &   3.45 \\
50th \% &   1.67 &   2.28 &    2.74 &    3.25 &   3.77 &~~~&   1.84 &   2.28 &  2.80 &  3.27 &   3.70 \\
75th \% &   1.84 &   2.39 &    2.87 &    3.38 &   3.92 &~~~&   2.11 &   2.56 &  3.07 &  3.49 &   3.85 \\  
\cutinhead{${\ell}_{\rm r}$}                                
Mean    &  32.06 &  32.57 &   33.08 &   33.58 &  34.09 &~~~&  31.78 &  32.54 & 33.23 & 33.89 &  34.61 \\
Error   &   0.04 &   0.04 &    0.04 &    0.06 &   0.09 &~~~&   0.03 &   0.01 &  0.02 &  0.02 &   0.04 \\
25th \% &  31.69 &  32.17 &   32.76 &   33.26 &  33.66 &~~~&  31.55 &  32.36 & 33.07 & 33.70 &  34.38 \\
50th \% &  32.07 &  32.54 &   33.13 &   33.61 &  34.32 &~~~&  31.87 &  32.52 & 33.22 & 33.90 &  34.57 \\
75th \% &  32.38 &  32.92 &   33.41 &   34.04 &  34.67 &~~~&  32.04 &  32.70 & 33.39 & 34.04 &  34.74 \\
\enddata

\tablecomments{Quantities defined as in Table 5.}

\end{deluxetable}

\begin{deluxetable}{p{90pt}rrcrrr}
\tablecaption{Correlations with X-ray luminosity}
\tabletypesize{\scriptsize}
\tablewidth{14.5cm}

\tablehead{ & \multicolumn{2}{c}{${\ell}_{\rm x} = a_{\rm 0} +
    b_{\rm uv}\times{\ell}_{\rm uv}$} & & \multicolumn{3}{c}{${\ell}_{\rm x} =
    a_{\rm 0} + b_{\rm uv}\times{\ell}_{\rm uv} + c_{\rm r}\times{\ell}_{\rm r}$}
  \\ \colhead{Sample} & \colhead{$a_{\rm 0}$} & \colhead{$b_{\rm uv}$} & &
  \colhead{$a_{\rm 0}$} & \colhead{$b_{\rm uv}$} & \colhead{$c_{\rm r}$}}

\startdata
\cutinhead{Full, primary, and supplemental samples}
Full & $-$0.144$^{+0.017}_{-0.016}$ &  0.789$^{+0.026}_{-0.025}$ &~~~& $-$0.098$^{+0.014}_{-0.015}$& 0.482$^{+0.031}_{-0.031}$& 0.273$^{+0.019}_{-0.019}$ \\[+3pt]
Primary & $-$0.190$^{+0.018}_{-0.018}$ &  0.721$^{+0.031}_{-0.031}$ &~~~& $-$0.116$^{+0.016}_{-0.016}$& 0.480$^{+0.033}_{-0.034}$& 0.263$^{+0.021}_{-0.021}$ \\[+3pt]
~~~Spectroscopic & $-$0.193$^{+0.024}_{-0.024}$ &  0.640$^{+0.042}_{-0.043}$ &~~~& $-$0.103$^{+0.021}_{-0.020}$& 0.380$^{+0.041}_{-0.042}$& 0.296$^{+0.025}_{-0.024}$ \\[+3pt]
~~~~~~QSO/HIZ  & $-$0.168$^{+0.030}_{-0.030}$ &  0.651$^{+0.049}_{-0.048}$ &~~~& $-$0.088$^{+0.025}_{-0.025}$& 0.400$^{+0.048}_{-0.047}$& 0.294$^{+0.030}_{-0.031}$ \\[+3pt]            
~~~Photometric  & $-$0.162$^{+0.027}_{-0.028}$ &  0.816$^{+0.050}_{-0.048}$ &~~~& $-$0.111$^{+0.026}_{-0.027}$& 0.611$^{+0.057}_{-0.058}$& 0.214$^{+0.035}_{-0.034}$ \\[+3pt]           
~~~Off-axis  & $-$0.208$^{+0.020}_{-0.020}$ &  0.744$^{+0.036}_{-0.036}$ &~~~& $-$0.123$^{+0.019}_{-0.020}$& 0.528$^{+0.039}_{-0.040}$& 0.243$^{+0.023}_{-0.025}$ \\[+3pt]           
~~~Targeted  & $-$0.039$^{+0.045}_{-0.043}$ &  0.553$^{+0.064}_{-0.065}$ &~~~& $-$0.089$^{+0.035}_{-0.037}$& 0.335$^{+0.059}_{-0.059}$& 0.320$^{+0.044}_{-0.044}$ \\[+3pt]           
Supplemental  & $+$0.028$^{+0.044}_{-0.044}$ &  0.823$^{+0.045}_{-0.044}$ &~~~& $-$0.017$^{+0.041}_{-0.041}$& 0.491$^{+0.079}_{-0.079}$& 0.253$^{+0.051}_{-0.052}$ \\[+3pt]             
~~~{\it Einstein\/} & $+$0.039$^{+0.072}_{-0.075}$ &  0.793$^{+0.087}_{-0.087}$ &~~~& $-$0.233$^{+0.072}_{-0.077}$& 0.472$^{+0.089}_{-0.090}$& 0.407$^{+0.066}_{-0.068}$ \\[+3pt]
FIRST  & $-$0.196$^{+0.035}_{-0.034}$ &  0.668$^{+0.062}_{-0.063}$ &~~~& $-$0.098$^{+0.028}_{-0.030}$& 0.409$^{+0.060}_{-0.059}$& 0.297$^{+0.032}_{-0.032}$ \\[+3pt]
\cutinhead{RQQs, RIQs, and RLQs}
RQQs & $-$0.545$^{+0.023}_{-0.023}$ &  0.649$^{+0.021}_{-0.021}$ &~~~& \nodata & \nodata & \nodata \\[+3pt]
RIQs & $-$0.447$^{+0.037}_{-0.038}$ &  0.567$^{+0.056}_{-0.057}$ &~~~& $-$0.477$^{+0.212}_{-0.215}$& 0.588$^{+0.181}_{-0.178}$& $-$0.025$^{+0.175}_{-0.177}$ \\[+3pt]
RLQs & $-$0.057$^{+0.017}_{-0.017}$ &  0.831$^{+0.026}_{-0.025}$ &~~~& $-$0.100$^{+0.016}_{-0.016}$& 0.506$^{+0.037}_{-0.038}$& 0.292$^{+0.027}_{-0.026}$ \\[+3pt]
\cutinhead{Groupings of RLQs}
$R^{*}<3$ & $-$0.191$^{+0.020}_{-0.020}$ &  0.771$^{+0.032}_{-0.032}$ &~~~& $-$0.149$^{+0.032}_{-0.032}$& 0.620$^{+0.100}_{-0.100}$& 0.144$^{+0.088}_{-0.091}$ \\[+3pt]
$R^{*}\ge3$ & $+$0.112$^{+0.026}_{-0.025}$ &  0.844$^{+0.036}_{-0.034}$ &~~~& $-$0.177$^{+0.057}_{-0.058}$& 0.432$^{+0.081}_{-0.085}$& 0.398$^{+0.073}_{-0.070}$ \\[+3pt]
${\ell}_{\rm r}<33.3$ & $-$0.180$^{+0.032}_{-0.032}$ &  0.739$^{+0.045}_{-0.046}$ &~~~& $-$0.098$^{+0.040}_{-0.040}$& 0.610$^{+0.061}_{-0.062}$& 0.205$^{+0.065}_{-0.065}$ \\[+3pt]
${\ell}_{\rm r}\ge33.3$ & $+$0.070$^{+0.028}_{-0.028}$ &  0.687$^{+0.045}_{-0.044}$ &~~~& $-$0.170$^{+0.034}_{-0.035}$& 0.439$^{+0.047}_{-0.048}$& 0.389$^{+0.040}_{-0.040}$ \\[+3pt]
${\alpha}_{\rm r}<-0.5$ & $-$0.091$^{+0.023}_{-0.023}$ &  0.652$^{+0.045}_{-0.045}$ &~~~& $-$0.175$^{+0.025}_{-0.025}$& 0.363$^{+0.061}_{-0.061}$& 0.259$^{+0.041}_{-0.041}$ \\[+3pt]
${\alpha}_{\rm r}\ge-0.5$ & $+$0.066$^{+0.029}_{-0.029}$ &  0.800$^{+0.039}_{-0.040}$ &~~~& $-$0.041$^{+0.026}_{-0.026}$& 0.471$^{+0.052}_{-0.050}$& 0.342$^{+0.039}_{-0.040}$ \\[+3pt]
\enddata

\tablecomments{Fitting is done with the IDL code of Kelly (2007),
  which utilizes Bayesian techniques that incorporate both errors and
  upper limits. Errors are assumed to be dominated by intrinsic random
  flux variability; see, e.g., $\S3.5$ of Gibson et al.~(2008). The
  luminosities are normalized prior to fitting as ${\ell}_{\rm
    r}-33.3, {\ell}_{\rm uv}-30.5, {\ell}_{\rm x}-27.0$. For any given
  model fit the quoted parameter values are the median of draws from
  the posterior distribution and the errors are 1$\sigma$. These
  results are plotted in Figures 9--13.}

\end{deluxetable}

\begin{deluxetable}{p{50pt}rrrrrr}
\tablecaption{${\alpha}_{\rm ox}$ and ${\Delta}{\alpha}_{\rm ox}$ as functions of $R^{*}$}
\tabletypesize{\scriptsize}
\tablewidth{12.0cm}

\tablehead{\colhead{Sample} & \colhead{N} &
  \colhead{$z$\tablenotemark{a}} & \colhead{$R^{*}$\tablenotemark{a}}
  & \colhead{${\alpha}_{\rm ox}/{\Delta}{\alpha}_{\rm
      ox}$\tablenotemark{a}} & \colhead{Intercept $a$} &
  \colhead{Slope $b$}}

\startdata

\cutinhead{${\alpha}_{\rm ox} = a + b\times{R^{*}}$}
$z<1$  & 213 & 0.69& 2.68& $-$1.33& $-$1.556$^{+0.043}_{-0.043}$ &  0.079$^{+0.015}_{-0.015}$ \\[+3pt]
$1<z<2$& 406 & 1.44& 2.54& $-$1.39& $-$1.653$^{+0.029}_{-0.029}$ &  0.101$^{+0.011}_{-0.011}$ \\[+3pt]
$z>2$  & 172 & 2.49& 2.56& $-$1.49& $-$1.862$^{+0.048}_{-0.048}$ &  0.152$^{+0.017}_{-0.017}$ \\[+3pt]

\cutinhead{${\Delta}{\alpha}_{\rm ox} = a + b\times{R^{*}}$}
$z<1$  & 213& 0.69& 2.68& 0.18& $-$0.101$^{+0.040}_{-0.041}$ &  0.097$^{+0.014}_{-0.014}$ \\[+3pt]
$1<z<2$& 406& 1.44& 2.54& 0.17& $-$0.118$^{+0.027}_{-0.027}$ &  0.106$^{+0.010}_{-0.010}$ \\[+3pt]
$z>2$  & 172& 2.49& 2.56& 0.16& $-$0.234$^{+0.047}_{-0.048}$ &  0.155$^{+0.017}_{-0.017}$ \\[+3pt]

\enddata

\tablenotetext{a}{Median values from within the subsample. The first
  three rows give ${\alpha}_{\rm ox}$ values and the last three rows
  give ${\Delta}{\alpha}_{\rm ox}$ values, where
  ${\Delta}{\alpha}_{\rm ox}$ corresponds to ${\alpha}_{\rm ox} -
  {\alpha}_{\rm ox,RQQ}$ with ${\alpha}_{\rm ox,RQQ} =
  -0.140\times{\ell}_{\rm uv}+2.705$ (Equation 3 from Just et
  al.~2007; see also $\S4$).}

\tablecomments{Fitting as described for Table 7. Uncertainties for the
  fitting were propagated from those assumed for luminosities. }

\end{deluxetable}

\begin{deluxetable}{p{60pt}p{80pt}l}
\tablecaption{Description of model components}
\tabletypesize{\scriptsize}
\tablewidth{19cm}

\tablehead{\colhead{Parameter} & \colhead{Value\tablenotemark{a}} & \colhead{Comment\tablenotemark{b}}}

\startdata
$z$                                         &   Quasi-random          &  From W01 RLF, then use values close to primary sample to account for redshift selection effects.         \\
$\theta$                                    &   Random          &  Drawn from uniform distribution in $\sin{\theta}$.                           \\
${\theta}_{\rm min}$                        &   $5.8^{\circ \dagger}$   &  Minimum permitted inclination; values between $4^{\circ}-7^{\circ}$ are viable.               \\
\cutinhead{Radio}
$\rho$, ${\sigma}_{\rm \rho}$, $\gamma$     &   $-5.61, 1.38, 10.49$ &  From Table 5 of MH09. Both $\rho$ and ${\sigma}_{\rm \rho}$ are expressed as natural logs.\\
${\ell}_{\rm r,lobes}$                      &   Random          &  From W01 RLF, then shifted to 5~GHz using ${\alpha}_{\rm r}=-0.9$  \\
${\ell}_{\rm r,jet0}$                       &   Calculated      &  ${\ell}_{\rm r,jet0}$ from ${\ell}_{\rm r,lobes}$ and $\rho$ with scatter from ${\sigma}_{\rm \rho}$. \\
${\ell}_{\rm r,jet}$                        &   Calculated      &  ${\ell}_{\rm r,jet}$ = ${\ell}_{\rm r,jet0}\times{\delta}^{2-{\alpha}_{\rm r}}$, $\delta = {\gamma}^{-1}(1-\beta\cos{\theta})^{-1}$, ${\alpha}_{\rm r}=-0.3$. \\
$f_{\rm r,min}$                             &   1, 2.7 mJy      &  Flux density limit at 1.4~GHz for core (FIRST 5$\sigma$), effective limit for lobes ($\simlt5$\% of lobes below).  \\
\cutinhead{Optical}  
${\rho}_{\rm dj0}$, ${\sigma}_{\rm dj0}$    &   $-1.8^{\dagger}, 0.5$     &  Logarithmic ratio of disk to intrinsic (unbeamed) radio jet; correlation motivated by W99 results. \\
${\ell}_{\rm uv,disk}$                      &   Calculated      &  ${\ell}_{\rm uv,disk}={\ell}_{\rm r,jet0}+{\rho}_{\rm dj0}$ with scatter from ${\sigma}_{\rm dj0}$ \\
${\ell}_{\rm uv,jet}$                       &   Calculated      &  Extrapolated from ${\ell}_{\rm r,jet}$ using ${\alpha}_{\rm ro}=-0.8$. Observations restrict optical jet flux. \\
$m_{\rm i,max}$                             &   19.95/20.13, 20.60/21.08  &  Cutoff $m_{\rm i}$ for $z<1/z\ge1$; from depth ($\simlt5$\% objects above) of spectroscopic, photometric samples. \\
\cutinhead{X-ray}
${\ell}_{\rm x,corona}$                     &   Calculated      &  Determined from ${\ell}_{\rm uv,disk}$ following the RQQ relation of J07. \\
${\rho}_{\rm j,A}$, ${\rho}_{\rm j,B}$, ${\rho}_{\rm j,C}$       & $-5.35^{\dagger}, -7.25^{\dagger}, -7.25^{\dagger}$ &  Logarithmic ratio of X-ray jet-linked emission to intrinsic radio jet for models A, B, and C. \\
${\sigma}_{\rm dx},{\sigma}_{\rm j}$         & 0.3, $0.7^{\dagger}/0.3^{\dagger}$      & Scatter in ${\ell}_{\rm x,corona}$ (from RQQ ${\ell}_{\rm x}({\ell}_{\rm uv})$ relation) and in ${\rho}_{\rm j}$ for $z<1/z\ge1$ for models A, B, and C. \\ 
${\gamma}_{\rm x,A}$, ${\gamma}_{\rm x,B}$, ${\gamma}_{\rm x,C}$ & 1.0, 10.49, 2.0$^{\dagger}$     &  Bulk Lorentz factor governing beaming of X-ray jet-linked emission for models A, B, and C. \\
${\ell}_{\rm x,jet}$                        &   Calculated      &  Determined from ${\ell}_{\rm r,jet0}$ and ${\rho}_{\rm j}$, with boosting of ${\delta}_{\rm x}^{3-2{\alpha}_{\rm x}}$, ${\alpha}_{\rm x}=-0.3/-0.5$ for B/C.\\

\enddata

\tablenotetext{a}{Only those numbers indicated with a dagger symbol
  were varied to best match the simulated to the observed data; all
  other numbers were taken from indicated references or fixed by
  observations.}

\tablenotetext{b}{References here are W01: Willott et al.~(2001);
  MH09: Mullin \& Hardcastle 2009; W99: Willott et al.~(1999); J07:
  Just et al.~(2007).}

\tablecomments{See $\S$6 for modeling details.}

\end{deluxetable}

\begin{deluxetable}{p{60pt}rrrcrrr}
\tablecaption{Model X-ray luminosities and ${\ell}_{\rm x}({\ell}_{\rm uv},{\ell}_{\rm r})$}
\tabletypesize{\scriptsize}
\tablewidth{13cm}

\tablehead{ & \multicolumn{3}{c}{X-ray luminosities} & &
  \multicolumn{3}{c}{${\ell}_{\rm x} = a_{\rm 0} + b_{\rm
      uv}\times{\ell}_{\rm uv} + c_{\rm r}\times{\ell}_{\rm r}$}
  \\ \colhead{Sample} & \colhead{${\ell}_{\rm x}$} &
  \colhead{$\sigma$} & \colhead{KS $p$\tablenotemark{a}} & & \colhead{$a_{\rm 0}$} &
  \colhead{$b_{\rm uv}$} & \colhead{$c_{\rm r}$}}

\startdata

Primary   & 26.80 & 0.54 & \nodata &~~~& $-0.116^{+0.016}_{-0.016}$ & $0.480^{+0.033}_{-0.034}$ & $0.263^{+0.021}_{-0.021}$ \\[+3pt]
Model A   & 26.74 & 0.58 & 0.37    &~~~& $-0.142^{+0.016}_{-0.017}$ & $0.730^{+0.030}_{-0.029}$ & $0.100^{+0.022}_{-0.022}$ \\[+3pt]
Model B   & 26.72 & 0.75 &$<0.01$  &~~~& $+0.099^{+0.020}_{-0.020}$ & $0.195^{+0.035}_{-0.035}$ & $0.746^{+0.027}_{-0.027}$ \\[+3pt]
Model C   & 26.75 & 0.55 & 0.30    &~~~& $-0.098^{+0.016}_{-0.016}$ & $0.526^{+0.028}_{-0.027}$ & $0.279^{+0.021}_{-0.021}$ \\[+3pt]
Model C1.5& 26.78 & 0.55 & 0.76    &~~~& $-0.105^{+0.017}_{-0.016}$ & $0.593^{+0.028}_{-0.029}$ & $0.210^{+0.022}_{-0.022}$ \\[+3pt]
Model C3.0& 26.78 & 0.59 & 0.10    &~~~& $-0.000^{+0.016}_{-0.017}$ & $0.400^{+0.028}_{-0.028}$ & $0.435^{+0.022}_{-0.022}$ \\[+3pt]

\enddata

\tablenotetext{a}{Kolmogorov-Smirnov test comparing distribution of
  model ${\ell}_{\rm x}$ values to that of the primary sample
  (computed treating observed and simulated upper limits as
  detections, as are the median ${\ell}_{\rm x}$ values here). The
  probability $p$-values given indicate that for all but Model B the
  simulated X-ray luminosities are not inconsistent with the observed
  X-ray luminosities.}

\tablecomments{Models are described in $\S$6. Fitting as described in
  Table~7. Upper limits are included in the simulated sample at a rate
  consistent with those observed in the primary sample. }

\end{deluxetable}

\end{document}